\documentclass[numbers]{assets/matterlab}

\usepackage{tabularx}
\usepackage{booktabs}
\usepackage{multirow}
\usepackage{graphicx}
\usepackage{array}
\usepackage{ragged2e}
\usepackage{float}
\usepackage{tipa}
\newcolumntype{Y}{>{\RaggedRight\arraybackslash}X} 
\newcolumntype{M}{>{\centering\arraybackslash}X}

\usepackage[most]{tcolorbox}
\usepackage{caption} %
\newtcolorbox{agentoutputbox}[2]{%
  enhanced,
  colback=gray!10,
  colframe=black!70,
  arc=3mm,
  boxrule=0.8pt,
  left=4mm,
  right=4mm,
  top=7mm,   %
  bottom=3mm,
  width=1.0\linewidth,
  fontupper=\footnotesize
    \setlength{\parskip}{0pt}
    \setlength{\abovedisplayskip}{0pt}
    \setlength{\belowdisplayskip}{0pt}
    \setlength{\abovedisplayshortskip}{0pt}
    \setlength{\belowdisplayshortskip}{0pt},
  overlay={%
    \node[anchor=south west] (agenticon)
      at ([yshift=2pt]frame.north west)
      {\includegraphics[width=0.6cm]{#1}};

    \node[anchor=south west, xshift=2pt]
      at (agenticon.south east)
      {\footnotesize\textbf{#2}};
  },
}

\definecolor{agentgray}{RGB}{245,245,245}
\newcommand{\kcal}{kcal\,mol$^{-1}$}

\usepackage{microtype}
\usepackage{graphicx}
\usepackage{booktabs}
\usepackage{float}
\usepackage{xurl}
\usepackage{hyperref}
\usepackage{titlesec}

\usepackage{amsmath}
\usepackage{amssymb}
\usepackage{mathtools}
\usepackage{amsthm}
\usepackage{bm}

\usepackage[noabbrev,nameinlink]{cleveref}

\usepackage{amsmath,amsfonts,bm}
\usepackage{makecell}

\def\eqref#1{equation~\ref{#1}}

\def\1{\bm{1}}

\DeclareMathAlphabet{\mathsfit}{\encodingdefault}{\sfdefault}{m}{sl}
\SetMathAlphabet{\mathsfit}{bold}{\encodingdefault}{\sfdefault}{bx}{n}

\usepackage{orcidlink}
\usepackage{comment}
\usepackage[version=4]{mhchem}
\usepackage{longtable}
\usepackage{array}
\renewcommand{\arraystretch}{1.7}
\usepackage{multirow}           
\usepackage{amsfonts}           
\usepackage{nicefrac}
\usepackage{duckuments}         
\usepackage{thmtools,thm-restate}
\usepackage{enumitem}
\usepackage{xcolor,colortbl}
\usepackage{tikz}
\usepackage{tikz-cd}
\usepackage{caption}
\usepackage{subcaption}
\usepackage{listings}
\usepackage{gensymb}
\usepackage{textcomp} 
\usepackage[inkscapelatex=false]{svg}

\newcommand{\addressCHEM}{Department of Chemistry, University of Toronto,  80 St. George St., Toronto, ON M5S 3H6, Canada}
\newcommand{\addressAC}{Acceleration Consortium, 700 University Ave., Toronto, ON M7A 2S4, Canada}
\newcommand{\addressCS}{Department of Computer Science, University of Toronto, 40 St George St., Toronto, ON M5S 2E4, Canada}
\newcommand{\addressVECTOR}{Vector Institute for Artificial Intelligence, W1140-108 College St., Schwartz Reisman Innovation Campus, Toronto, ON M5G 0C6, Canada}
\newcommand{\addressMSE}{Department of Materials Science \& Engineering, University of Toronto, 184 College St., Toronto, ON M5S 3E4, Canada}
\newcommand{\addressCHEMENG}{Department of Chemical Engineering \& Applied Chemistry, University of Toronto, 200 College St., Toronto, ON M5S 3E5, Canada}
\newcommand{\addressCIFAR}{Canadian Institute for Advanced Research (CIFAR), 661 University Ave., Toronto, ON M5G 1M1, Canada}
\newcommand{\addressNVIDIA}{NVIDIA, 431 King St. W \#6th, Toronto, ON M5V 1K4, Canada}

\newcommand{\addressMS}{Institute of Medical Science, 1 King's College Circle, Medical Sciences Building, Room 2374, Toronto, ON M5S 1A8, Canada}

\usepackage{cinzel}
\usepackage{bibunits}

\title{El Agente Quntur: A research collaborator agent for quantum chemistry}

\author[1,6,\dagger,\orcidlink{0000-0003-1267-3627}]{Juan B. P\'erez-S\'anchez}
\author[1,6,\dagger,\orcidlink{0009-0007-9131-4468}]{Yunheng Zou}
\author[1,6,\orcidlink{0000-0003-1156-3012}]{Jorge A. Campos-Gonzalez-Angulo}
\author[1,6, \ddagger\orcidlink{0000-0002-5398-4192}]{Marcel M\"uller}
\author[1,\ddagger,\orcidlink{0000-0003-0240-8782}]{Ignacio Gustin}
\author[1,6,\ddagger,\orcidlink{0000-0003-3647-5000}]{Andrew Wang}
\author[5,\S]{Han Hao}
\author[1,6,\S, \orcidlink{0000-0002-0802-9559}]{Tsz Wai Ko}
\author[1,6,\S]{Changhyeok Choi}
\author[5,\S]{Eric S. Isbrandt}

\author[1,2,6,\S,\orcidlink{0000-0002-2927-4975}]{Mohammad Ghazi Vakili} 
\author[1,\S,\orcidlink{0009-0009-7887-5468}]{Hanyong Xu} 
\author[2,5]{Chris Crebolder}
\author[*,2,7, \orcidlink{0000-0002-8446-7956}]{Varinia Bernales}
\author[*,1,2,3,4,5,6,7,8,9,\orcidlink{0000-0002-8277-4434}]{Al\'an Aspuru-Guzik}

\affiliation[1]{\addressCHEM}
\affiliation[2]{\addressCS}
\affiliation[3]{\addressMSE}
\affiliation[4]{\addressCHEMENG}
\affiliation[5]{\addressAC}
\affiliation[6]{\addressVECTOR}
\affiliation[7]{\addressCIFAR}
\affiliation[8]{\addressNVIDIA}
\affiliation[9]{\addressMS}
\affiliation[*]{corresponding authors}

\contribution[\dagger,\ddagger,\S]{These authors contributed equally}

\abstract{

Quantum chemistry is a foundational tool for chemistry, materials science, computational biology, and others. Despite its power, its application remains restricted to experts due to methodological complexity, software heterogeneity, and the need for informed interpretation. To bridge this accessibility gap, we introduce {\cinzel El Agente Quntur}, a hierarchical, multi-agent system designed as a research collaborator for computational quantum chemistry. {\cinzel Quntur} follows three strategies: i) reasoning-driven decisions rather than hard-coded procedural policies, ii) general and composable actions, and iii) guided deep research integrating quantum-chemical reasoning with software logic and syntax. Instantiated in {\cinzel ORCA}, {\cinzel Quntur} reasons over software documentation and literature to plan, execute, adapt, and analyze \textit{in silico} chemistry experiments. Benchmarking against Claude Code in the same software environment shows that {\cinzel Quntur}'s domain-specific agent architecture more faithfully translates open-ended chemical requests into defensible end-to-end computational protocols. We outline advances, bottlenecks, and a roadmap toward an autonomous end-to-end computational chemistry research agent.

}

\date{\today}
\correspondence{\email{varinia@bernales.org}, and \email{alan@aspuru.com}}

\begin{document}

\maketitle

\newpage 

Computational chemistry significantly advances the understanding of chemical systems by providing an approximate digital twin of their behaviour. This is achieved by computing the approximate behaviour of electrons and nuclei in molecules and materials. In particular, quantum chemistry calculations have become essential tools in modern drug discovery, offering molecular-level insights that are unattainable through classical computational methods. By solving the Schr\"odinger equation---albeit approximately due to efficiency and computational complexity constraints---these techniques elucidate molecular electronic structures, energies, and binding affinities critical for understanding drug-target interactions \cite{GomezBombarelli2016Virtual,Lin2020,Breijyeh2021}. A variety of computational methods, including molecular docking and pharmacophore modeling, are implemented to optimize drug design and enhance the identification of potential drug candidates \cite{Cova2019}. Beyond drug discovery, quantum chemistry methods play a vital role in catalyst design and in the exploration of various disciplines, including chemistry, biology, and materials science, fundamentally shaping our understanding of biomolecular functions and material structures. Quantum chemistry approaches illuminate structure-property relationships in organic electronics, enabling rational, systematic design of future high-performance materials \cite{Hachmann2011The,Hachmann2014Lead,Huggins2019Biomolecular}.

The computational implementation of quantum mechanical methods continues to evolve, with developments such as real-space density functional theory on graphical processing units (GPUs) significantly accelerating calculations \cite{Andrade2013RealSpace,Olivares-Amaya2010Accelerating}. These advances in computational efficiency enable the application of quantum chemistry to increasingly complex systems, including enzymatic reaction mechanisms, where quantum-mechanical cluster models can yield highly accurate results that reproduce experimental product distributions and activation free energies \cite{Visser2024Tutorial}. Furthermore, quantum chemistry has provided fundamental insights into biological processes, such as the role of quantum coherence in chromophoric energy transport within photosynthetic protein complexes, including the Fenna-Matthews-Olson complex, where coherent dynamics contribute approximately 10\% to energy-transfer efficiency \cite{Rebentrost2009Role}. The continued evolution of quantum chemistry, including the potential integration of quantum computing to accelerate calculations, promises transformative impacts on personalized medicine and the targeting of previously undruggable therapeutic targets \cite{Niazi2025Quantum}.

Given the maturity of the field, applying quantum chemistry in practice is far from a simple matter of pressing a button. Each of the numerous electronic structure methods and software packages has its own conventions and limitations; therefore, accurate results require a deep understanding of the theory and the specific code being used \cite{Huber2021Common}.
Different programs use different algorithms and very different input and output formats, making it difficult, even for experts, to master more than a few \cite{Aspuru-Bernales2026Therise}.
In practice, many computational setups and parameter choices are guided by human experts' experience rather than by automated rules.

Computational chemists often design workflows to encode efficient and reliable sequences of calculations by pre-selecting methods, parameters, and convergence strategies appropriate for specific classes of problems \cite{Wu2025A, Curtarolo2012Aflow, Pizzi2016Aiida, Larsen2017The, Alegre2023AQME, Hicks2024ChemCloud}. However, these workflows are severely limited. One limitation is operational complexity: running a calculation often involves scripting multiple steps, selecting computational resources, and actively monitoring progress.
Typical failures include convergence issues in geometry or self-consistent field iterations, or exhaustion of allocated wall time on a computing cluster.
Handling these failures often requires manual intervention, such as adjusting convergence parameters, restarting jobs with different guesses, or increasing resources.
Another limitation is the methodological uncertainty:
A user must choose an appropriate level of theory (Hartree-Fock, correlated wavefunction theory, density functional theory, etc.) and a suitable basis set for each problem \cite{Cohen2012Challenges, Bursch2022Best}.
No single choice is universally optimal, and different options can give substantially different results. In practice, scientists rely on benchmarking studies or chemical intuition to select appropriate methods, and there is no fully algorithmic way to guarantee reliability in every case. Finally, once a calculation finishes, its raw outputs (energies, optimized geometries, orbital populations, spectral and electrical properties, etc.) must be analyzed to extract chemical meaning. Deciding which results are chemically significant and how they influence one's hypotheses typically requires human insight.
For example, when comparing different methods, recognizing that a discrepancy in activation energies is meaningful may depend on knowledge of experimental uncertainties or the chemical context.

These limitations of automation through traditional workflows motivate a fundamentally new approach: the development of agentic AI systems for research \cite{Aspuru-Bernales2026Therise,Wei2025aiscienceagenticscience}.
AI agents are currently being designed to autonomously drive scientific inquiry by designing and executing wet-lab and computational experiments, or generating potentially novel materials with applications across various domains. One example in chemistry is {\cinzel Coscientist}, in which a team of large language models (LLMs) plans complex synthesis pathways by browsing the literature, parsing hardware documentation, and directly controlling robotic laboratory hardware \cite{Boiko2023Autonomous}. Similarly, {\cinzel ChemCrow} autonomously plans and executes syntheses (e.g., of organocatalysts) and iteratively corrects its procedures \cite{Bran2024Augmenting}. In parallel, some of us developed {\cinzel Organa}, an LLM-integrated robotic assistant that combines natural-language interaction, perception-driven decision-making, task and motion planning, and parallel execution to autonomously perform diverse chemical procedures in the wet lab \cite{Kourosh2025Organa}. More recently, there has been a surge of agent-based interfaces for computational chemistry, spanning a diverse set of complexity, autonomy, and architectural choices \cite{Ansari2024Agent,McNaughton2024CACTUS,Ruan2024LLMRDF,Dai2024Autonomous,Sprueill2024ChemReasoner,Song2025ChemAgents,Chen2025chemminerlargelanguagemodel,Tang2025chemagent,Yang2025moosechem,Yamada2025aiscientistv2, Mitchener2025kosmos,Liu2025moosechem3,Li2025chemhas,Yu2025chemtoolagent,Li2025labutopia,Ma2025GVIM,Kim2025mtmol,Xinhao2025CSstep,Callahan2025agenticCRAGMoW,Chen2025chemistx,Inizan2025MOFGen,Ghafarollahi2025Automating,Pu2025piflow,Ghafarollahi2025autonomous,Wang2025S1MatAgent,Jia2024LLMatDesign,Ding2025SciToolAgent,Chiang2024LLaMP,Zhang2025TopoMAS,Petkovic2025Towards,Zhou2025MAPPS,Xia2025Agentic,Liu2025VASPilot,Wang2025Dreams,Pham2025ChemGraph,Polat2025xChemAgents,Gadde2025Chatbot,Chandrasekhar2025automating,Campbell2025mdcrow,Shi2025A, Masters2025Arcane,Zou2025ElAgente,Guilbert2025dynamate,Soleymanibrojeni2025Genius,Lahouari2026Automated,Li2026Autonomous,Vriza2026Multi,Kim2026materealize,Yang2026QUASAR}.

{\cinzel El Agente} is an LLM-based multi-agentic system featuring hierarchical task decomposition into different sub-problems and dynamical tool selection, with{\cinzel El Agente Q} as its foundational implementation for solving computational quantum chemistry problems \cite{Zou2025ElAgente}. {\cinzel El Agente Q} demonstrates the ability to run quantum chemistry simulations, analyze results, and refine strategies based on feedback from the tools. Moreover, {\cinzel El Agente Q} operates at the level of a university student and demonstrated flexibility while performing a restricted set of quantum chemistry tasks: DFT geometry optimization to equilibrium geometries, vibrational analysis, and single-point excited-state (TD-DFT) calculations. Currently, the architecture of{\cinzel El Agente} is being exploited to cover a wide range of fields, including solid-state chemistry \cite{Govind2026El}, geometry generation and manipulation \cite{choi2026elagenteEstructural}, Quantum computing and dynamics simulations \cite{gustin2026elagentecuantico}, workflow orchestration for GPU-accelerated molecular and materials simulations \cite{bai-aldossary2026grafico}, among others.

In this work, we introduce{\cinzel El Agente Quntur}.{\cinzel Quntur} comprehensively surpasses the capabilities of {\cinzel El Agente Q} by reaching an entirely new level in both breadth and depth. In breadth, its design allows it to support a huge range of quantum chemistry calculations, encompassing advanced structure generation, transition-state search and Intrinsic Reaction Coordinate (IRC) calculations, relativistic and dispersion corrections, accurate energy barriers and kinetic isotope effects, spectroscopy calculations (IR/Raman, UV/Vis, NMR, EPR), excited-state kinetics, etc. {\cinzel Quntur} is a name inspired by the Quechua word {\sl kuntur}, which is the native name for the c\'ondor: the bird that is often used as a symbol of South American culture.

We achieve these capabilities by following three improved design strategies: removing certain hard-coded policy choices, designing tools for generalization and efficiency, and conducting guided deep research during execution. These strategies grant{\cinzel Quntur} complete control over the{\cinzel ORCA}~6.0 quantum chemistry package \cite{Neese2012The, Neese_ORCA6}. In-depth, {\cinzel Quntur} reviews and analyzes software documentation and scientific literature. Therefore, it can plan, justify, execute, and validate diverse multi-stage workflows in accordance with best practices and scientific rigour, while also analyzing the results and comparing them with the literature (see Table \ref{tab:comparison}). To evaluate the practical consequences of our design, we use two complementary tests. First, a benchmark comparison against a Claude Code baseline with access to the same quantum-chemistry software identifies recurring failure modes of general-purpose AI agents in computational chemistry and tests whether {\cinzel Quntur}'s architecture improves reliability on such tasks. Second, research-level case studies stress-test {\cinzel Quntur} on open-ended problems where success depends on method choice, error recovery, and scientific interpretation; these cases evaluate the extent to which the system behaves as a \textit{research collaborator} and expose remaining limitations.

\begin{table*}[ht]
    \centering
    \caption{ Architectural and functional comparison between 
    {\cinzel El Agente Q} and {\cinzel El Agente Quntur}
    }
    \label{tab:comparison}
    \resizebox{\linewidth}{!}{
    \begin{tabular}{@{}ccccccccc@{}}
    \toprule
    System & \makecell{Agents} & \makecell{Tools} & \makecell{Knowledge\\Base} & \makecell{Parallel\\Execution} & \makecell{Image\\Input} & \makecell{Human\\in-the-Loop} & \makecell{Domain\\Breadth} & \makecell{Research\\Depth}\\ 
    \midrule
    \midrule 
    {\cinzel El Agente Q}  & 23 & 18 & Context files & No & No & No & Restricted & Undergraduate \\
    {\cinzel El Agente Quntur} & 58 & 34 & Deep-research & Yes & Yes & Optional & General & Research collaborator \\
    \bottomrule
    \end{tabular}
    }
\end{table*}

The denomination of \textit{research collaborator} implies that the agent operates at the level of a graduate student. It can design multi-step computational workflows that go beyond a single prescribed protocol. These workflows include method selection, intermediate analysis, and conditional branching based on results drawn from user input, the scientific literature, or both. Indeed, {\cinzel Quntur} references published works to motivate methodological choices and to analyze the results during post-analysis. The agent's behaviour is not purely procedural: it exercises judgment when workflows encounter ambiguity, proposes alternative methods, weighs competing convergence strategies, and considers trade-offs between accuracy and cost. It also allows human input for crucial methodological choices or to determine the next steps (human-in-the-loop).

As one research-level example, in our own internal group research, {\cinzel El Agente Quntur} is already being employed to study highly excited electronic states in atoms and to analyze exceptions to Hund's rules at several levels of theory. This task is highly non-trivial, as it requires fine-tuning the self-consistent field settings to preserve spherical symmetry and judiciously designing the model chemistry and the corresponding active space.{\cinzel Quntur} streamlines workflow production in ORCA, which would otherwise have taken much longer to prepare and debug using traditional methods \cite{cga2026hunds}.

\section*{The architecture of {\fontfamily{Cinzel-LF}\selectfont El Agente Quntur}}

{\cinzel Quntur} is built on top of {\cinzel El Agente}'s cognitive architecture and graphical user interface (GUI). {\cinzel El Agente}'s multi-agent approach serves as a long-term memory module, with procedural memory emerging from a team hierarchy, and agent-specific semantic and episodic memory attached to each agent node \cite{Zou2025ElAgente}. Such a design allows the agentic system to dynamically synthesize and execute reasoning-driven workflows, handle errors flexibly, and form a hierarchical memory that reduces each agent's cognitive load and supports long-term behaviour. %

Although{\cinzel Quntur} did not change{\cinzel El Agente Q}'s core cognitive architecture,  it significantly expanded its capabilities thanks to the newly employed three design principles (see \hyperref[sec:methods]{Methods} for a detailed explanation): 
\begin{enumerate}
    \item \textit{Removal of hard-coded policies} to enable flexible scientific reasoning across diverse tasks.
    \item \textit{Tool design for generalization and efficiency}, allowing tools to be reused and composed across domains.
    \item \textit{Guided deep research during execution}, enabling on-demand retrieval of task-relevant context at decision time.
\end{enumerate}

Thanks to these three design principles, we can significantly expand the domain covered with far less design overhead. Compared to{\cinzel El Agente Q}, we increase the number of agents from 23 to 58 and the number of tools from 18 to 34. This enables all the agents to reliably reason and execute across different quantum chemistry domains, as well as in the post-analysis stage.

\subsection*{The procedural memory}

\begin{figure}[ht]
    \centering
    \includegraphics[width=1\textwidth]{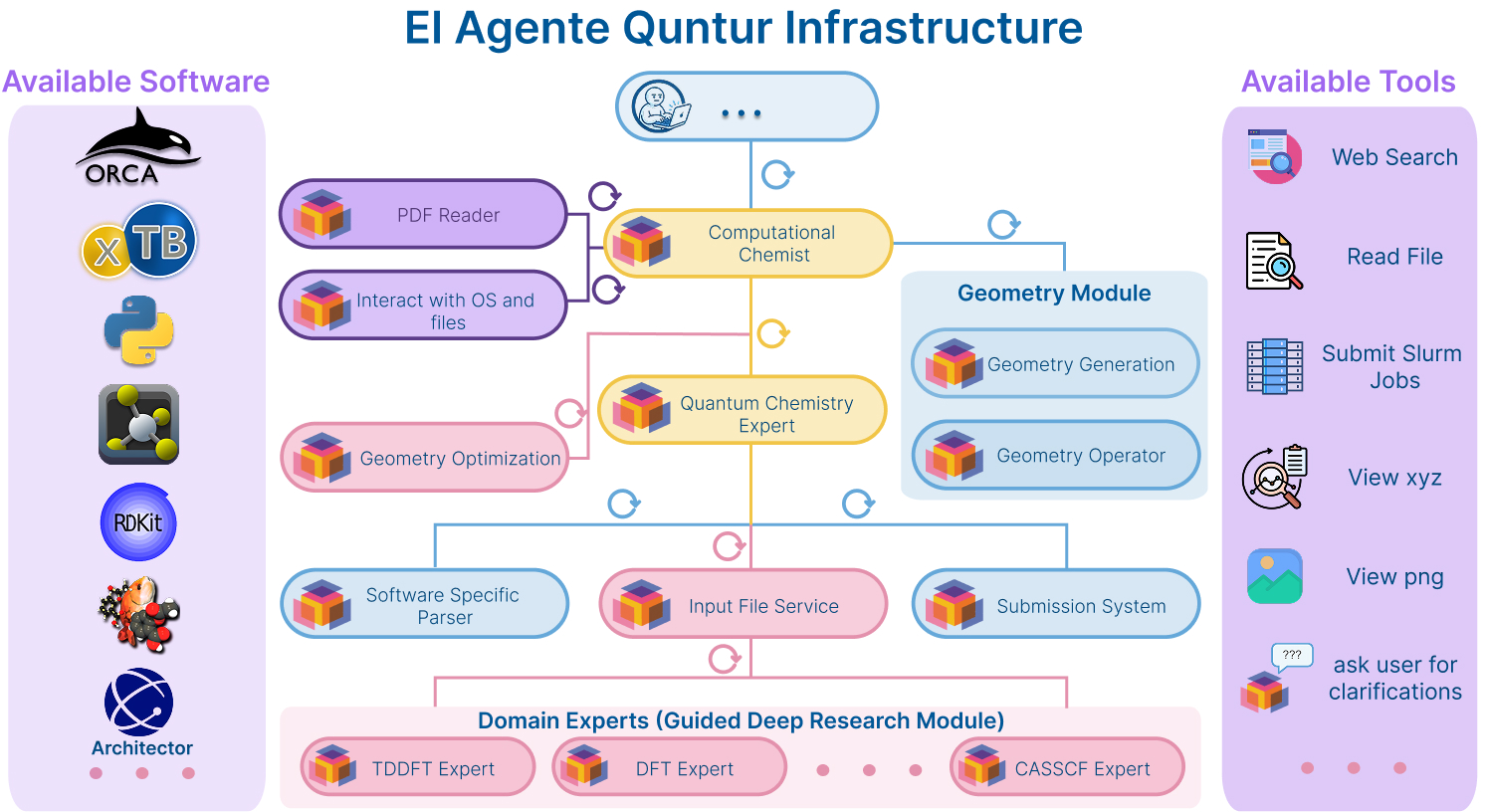}
    \caption{High-level description of the multi-agentic hierarchical architecture of{\cinzel El Agente Quntur}. Each module comprises one or more LLMs, each with the tools necessary to fulfill its role.}
    \label{fig:network}
\end{figure}

Following {\cinzel El Agente}'s original template, {\cinzel Quntur} presents a multi-agent hierarchical architecture that models task decomposition, well-defined job scopes, and coordination mechanisms that characterize human organizations such as companies, governments, and research groups (see Fig. \ref{fig:network}). At the top level, we have the \texttt{computational chemist} (i.e., a principal investigator), a strategic sub-agent responsible for goal formulation, high-level planning, and result analysis. Instead of interacting with the environment directly, it translates the user's question into well-defined steps and delegates them to downstream experts.{\cinzel Quntur} also uses web search and other tools to reason freely over scientific literature, PDFs, and software documentation, rather than relying on human-crafted instructions that limit the agent's reasoning and the range of problems it can tackle.

In the second layer, we have sub-agents that embody domain expertise---akin to postdocs, engineers, or analysts---each operating under a constrained mandate and with a specific vocabulary and toolset. For{\cinzel Quntur}, these are sub-agents specialized in geometry generation, quantum chemistry calculations, reading PDF files, and interactions with file management and post-processing. In addition to preventing cognitive overload, this architectural feature enables parallel processing, task decomposition, and the use of optimal models for each task (e.g., the best-performing LLM for geometry generation in visual recognition). Crucially, hierarchy is complemented by feedback loops: lower-level sub-agents report results or failures to higher levels, allowing the system to revise plans or reassign work. 

The PDF reader is a specialized sub-agent that reads, summarizes, and answers questions about PDF documents. It uses the MinerU tool to transform PDFs to Markdown files \cite{Wang2024mineru}. The geometry generation module showcases tools and other sub-agents for task and tool management. The \texttt{interact with OS and files} sub-agent consists of a single LLM with specialized tools, such as an output-section indexer, Python, and Bash, to parse output files. The quantum chemistry expert sub-agent comprises an input-file service, a guided, in-depth research module, a submission system, and a post-processing system. The \texttt{input file service} sub-agent, the most complex of our sub-agents, has access to our \hyperref[sec:deep]{guided deep research module}, a team of several quantum chemistry sub-agents that act as domain experts for quantum chemistry reasoning and software adaptation.

\section*{Results}

\subsection*{Evaluation of {\fontfamily{Cinzel-LF}\selectfont El Agente Quntur} in computational quantum chemistry tasks}

To rigorously evaluate the reliability and breadth of {\cinzel Quntur}, we introduce a benchmark that extends beyond traditional workflows (see Table \ref{tab:benchmark}) to encompass a wider range of computational chemistry tasks than what we covered in {\cinzel El Agente Q}. The benchmark spans a broad range of domains, including relativistic and correlated wavefunction properties, thermodynamic cycles, potential-energy-surface (PES) mapping, reaction kinetics, and diverse spectroscopic observables. Furthermore, it covers a range of difficulty levels, spanning from undergraduate exercises to research-level problems, and it systematically probes the agent’s ability to operate across multiple levels of theory.

We asked{\cinzel Quntur} to \emph{autonomously} execute all questions five times independently to evaluate its robustness and reliability. To contextualize these results against a frontier general coding-agent baseline, we also evaluated Claude Code with Claude Opus 4.5, the same base model used for all subagents of {\cinzel Quntur} and the most advanced Claude model at the time. Claude Code was run with ORCA skills enabled, so it had basic support for ORCA file preparation, execution, and artifact handling. Claude Code was also allowed to use its web-search capabilities, but we did not write benchmark-specific or additional computational chemistry-specialized skills for Claude Code. 

The baseline allows us to expose failure modes of LLM-based agents in computational chemistry and the extent to which{\cinzel Quntur}'s architecture mitigates them. We did not use human computational chemists as the baseline for this autonomous-run benchmark, because such a comparison would require controlling for experience level, prior familiarity with quantum-chemistry software, time allowed, access to literature and external tools, supervision, and criteria for evaluating both learning and scientific output, and hence it is outside the scope of the present work.

Evaluating scientific agents presents a fundamental challenge: many tasks admit multiple valid methodological choices and interpretation pathways, even when the underlying physical question is well-defined. Consequently, the flexibility inherent to each benchmark question cannot be encoded into a deterministic, fully automated grading scheme without penalizing legitimate alternatives. For this reason, we adopt a benchmarking strategy that mirrors how computational chemists are evaluated in practice, emphasizing expert judgment over strict procedural conformity. The rubrics assess both agents' performance by decomposing each task into planning, geometry generation, input file generation and execution, and post-analysis, thereby allowing expert graders to account for multiple scientifically valid solutions. 

\begin{figure}[ht]
    \centering
    \includegraphics[width=1\linewidth]{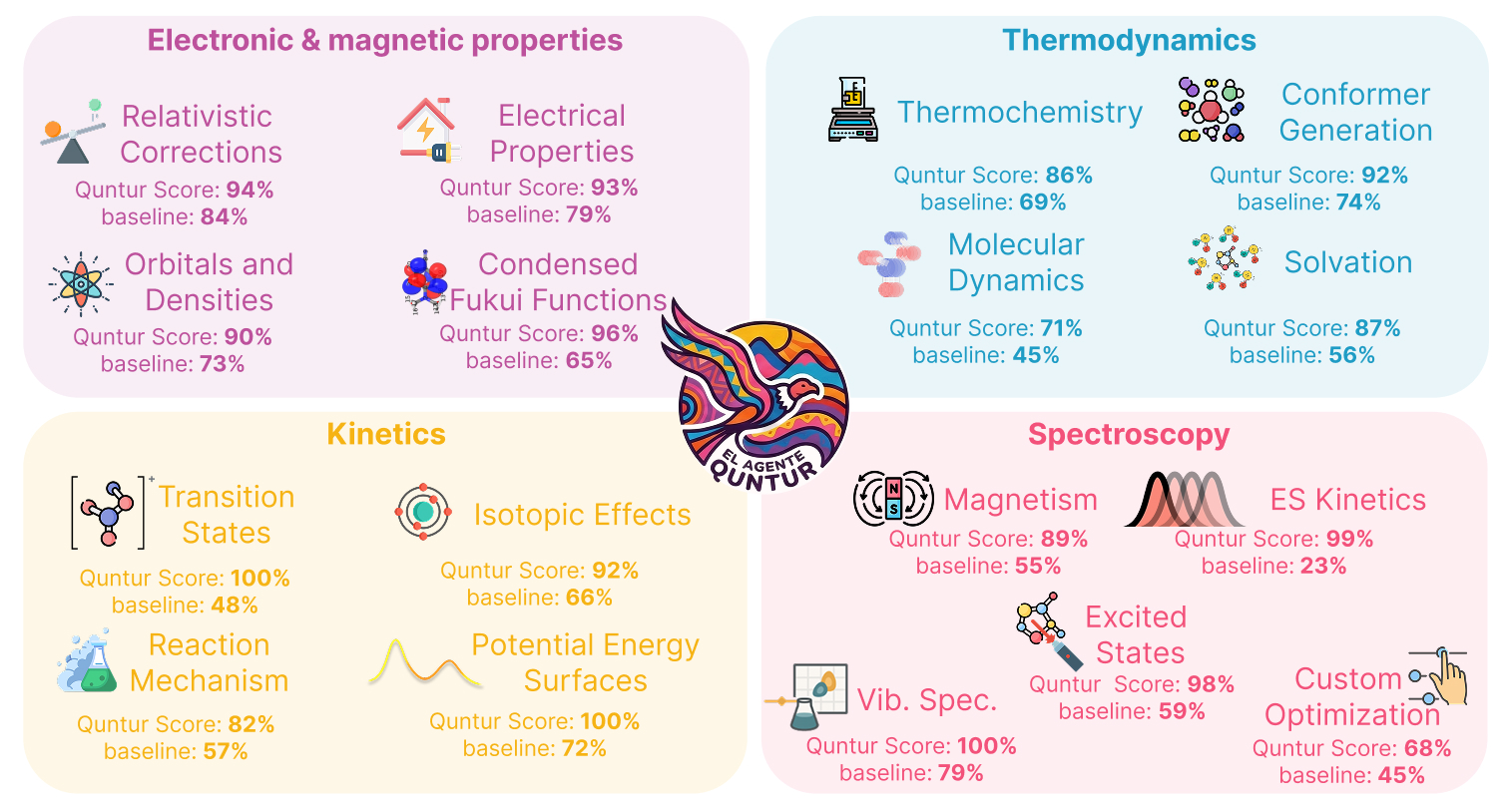}
    \caption{\textbf{Capabilities of {\cinzel El Agente Quntur} and its average benchmark score relative to the Claude Code baseline.} The benchmark comprises 17 computational quantum chemistry exercises spanning electronic and magnetic properties, thermodynamics, kinetics, spectroscopy, and  varying levels of theory and difficulty. The benchmark questions can be found in the \hyperref[sec:benchmark]{Supplemental Information Section \ref{sec:benchmark}}. The rubric for each question is in the \hyperref[SI:QunturEval]{Supplemental Information Section} \ref{SI:QunturEval}. Here we show the average score over 5 repetitions per question, but detailed scores for planning, geometry generation,  input-file generation and execution, and post-analysis are provided in Figs.~\ref{fig:benchmark} and \ref{fig:benchmark2}, for{\cinzel Quntur} and Claude Code, respectively. All benchmark executions used Claude Opus 4.5 as the base language model. For Quntur, all subagents used \texttt{high} reasoning effort. For Claude Code, runs 1--3 used \texttt{high} effort and runs 4--5 used \texttt{xhigh} effort. No performance increase was observed with the higher Claude Code reasoning setting. Token usage is reported in Supplemental Information Section \ref{sec:token_cost_analysis}.}
    \label{fig:breadth}
\end{figure}

The numerical scores in Fig.~\ref{fig:breadth} provide compact summaries of an artifact-based evaluation rather than the full benchmark record. Across the 17 tasks, the mean score was 90.5/100 for {\cinzel Quntur} and 61.7/100 for the Claude Code baseline, both computed from five independent repetitions per task. Importantly, this performance difference was not obtained by using more tokens: {\cinzel Quntur} used an average of 28.3 million tokens, whereas the Claude Code baseline used an average of 110.4 million tokens (Fig.~\ref{fig:tokenanalysis1}). The underlying benchmark questions, task-specific rubrics, point-deduction rationales, detailed failure modes analysis, and token cost analysis are provided in Supplemental Information Sections\hyperref[sec:benchmark]{~\ref{sec:benchmark}},\hyperref[SI:QunturEval]{~\ref{SI:QunturEval}}, and\hyperref[sec:token_cost_analysis]{~\ref{sec:token_cost_analysis}}. Detailed{\cinzel Quntur} and baseline scores and task-level comparisons are provided in Supplementary Figs.~\ref{fig:benchmark} and~\ref{fig:benchmark2}, respectively. The files generated by {\cinzel Quntur} and Claude Code while answering the questions are available in our data repository~\cite{Perez-Zou2026QunturFiles}.

\subsection*{{\fontfamily{Cinzel-LF}\selectfont El Agente Quntur} in research-level scenarios}

\begin{figure}[ht]
    \centering
    \includegraphics[width=1\linewidth]{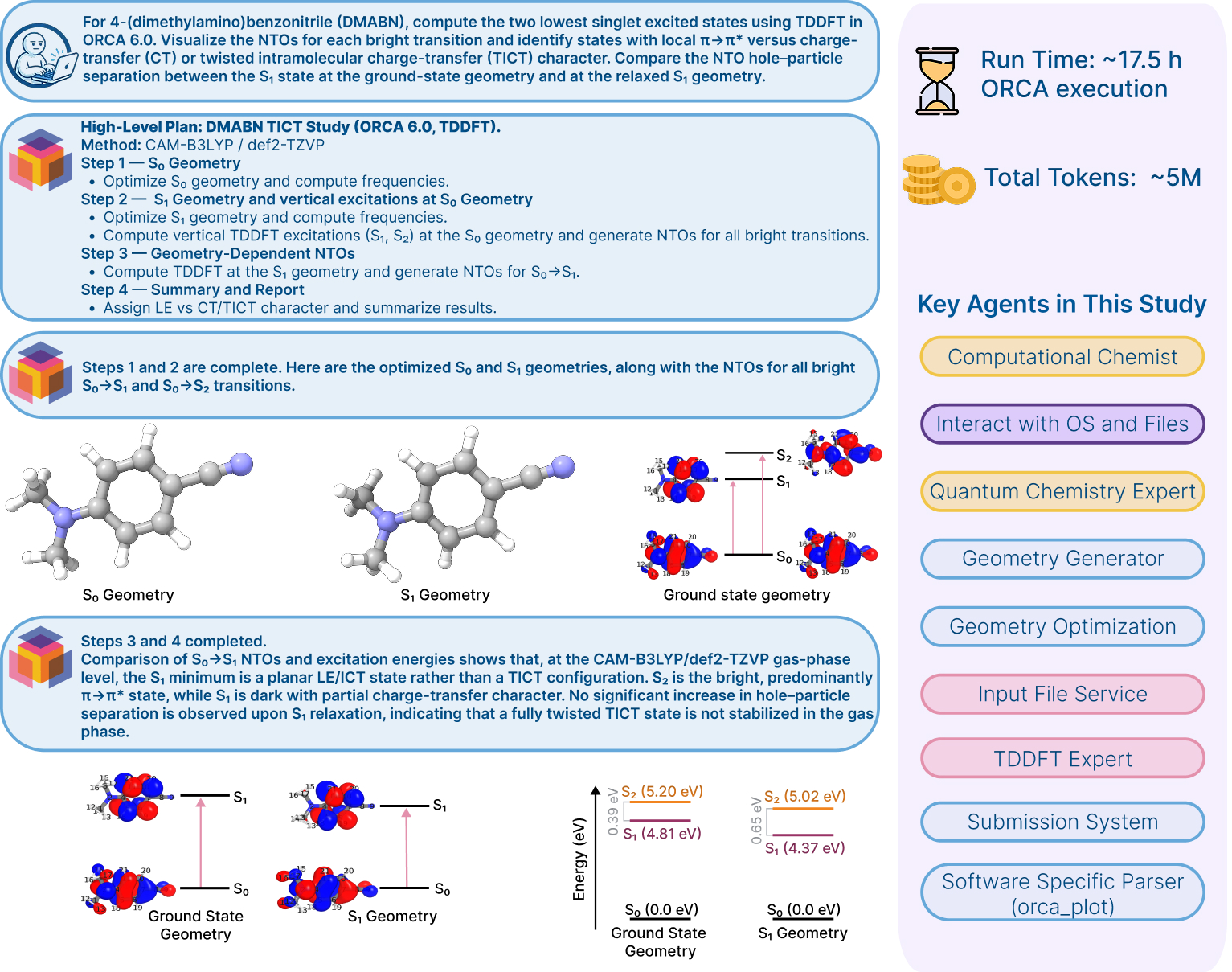}
    \caption{Simplified sketch of the case study \textit{Natural Transition Orbitals (NTOs) visualization and excited states characterization} as solved by {\cinzel El Agente Quntur}. Decisions are based on LLMs' reasoning rather than being hardcoded. The case study required 5 million tokens with Claude Opus 4.5 as the base model for all agents, and an approximate cost of 25 USD at the time it was run. The cost refers to LLM/API usage and does not include quantum-chemistry compute costs. Molecular structures were visualized using ChimeraX 1.11.1 \cite{pettersen2021ucsf,mengUCSFChimeraXTools2023}.
    }
    \label{fig:Juan_workflow}
\end{figure}

To assess{\cinzel Quntur}'s performance as a research collaborator, we use a small set of research-oriented case studies that reflect the types of open-ended questions encountered in computational chemistry research. These case studies also span a broad range of topics (electronic and magnetic properties, thermodynamics, kinetics, and spectroscopy). The prompts for case studies include method selection guided by the literature, comparative analysis across models or levels of theory, options for human-in-the-loop at various levels, and the requirement for post-analysis to obtain chemically meaningful information and interpretations from raw computational data. We present the case studies and the summary of each trajectory in the \hyperref[sec:case_studies]{Supplemental Information Section} \ref{sec:case_studies}, while the generated files can be found in a data repository \cite{Perez-Zou2026QunturFiles}. Yet, to illustrate what{\cinzel Quntur} can do, we present a simplified sketch of the computational workflow designed to address the \hyperref[sec:NTOs]{\textit{Natural Transition Orbitals visualization and excited states characterization} case study} in Fig. \ref{fig:Juan_workflow}. We also base our analysis of the capabilities and current bottlenecks of{\cinzel Quntur} on the case studies. 

\subsection*{Application of Natural Transition Orbitals to excited-state analysis}

4-(Dimethylamino)benzonitrile (DMABN) is a prototypical donor–acceptor system that exhibits dual fluorescence arising from the coexistence of locally excited (LE) and twisted intramolecular charge-transfer (TICT) states \cite{Lippert1962Dual}. We prompt{\cinzel Quntur} to calculate the lowest two singlet excited states of DMABN and to visualize the associated Natural Transition Orbitals (NTOs), to classify the transitions as either local or charge-transfer-like. From the results, it inferred the nature of each excited state from the final optimized geometries and the qualitative features of the NTOs.{\cinzel Quntur} also realized that the calculations contradicted the known result that a shift in character between vertical excitation and the relaxed excited-state geometry occurs. A literature search correctly identified that the inconsistency arises because the calculation is performed in the gas phase rather than in solution \cite{Georgieva2015Intramolecular}. {\cinzel Quntur} solved this case study fully autonomously (without human intervention).

\section*{Discussion}

\subsection*{{\fontfamily{Cinzel-LF}\selectfont El Agente Quntur} vs Claude Code baseline on the benchmark set: the importance of a domain-specific reasoning architecture}

Geometry generation and validation remained a recurring weakness of the Claude Code baseline across the five benchmark repetitions. Structures used for property calculations must be correctly posed, optimized, and checked because errors at this stage propagate into every subsequent result. In the Excited States task, run 1 performed an excited-state rather than ground-state optimization and retained imaginary modes, whereas runs 2-5 used static force-field or hand-written structures without DFT optimization and frequency validation. In the Excited-State Kinetics task, all five runs retained a planar S$_1$ formaldehyde structure and omitted the S$_1$ Hessian needed to identify the pyramidalized minimum. The double-proton-transfer task showed a related pathway-definition problem: runs 1 and 5 followed invalid pathways, while runs 2-4 recovered a plausible intended saddle but did not obtain it from a completed valid NEB dependency; consequently, none completed the intended sequence of validated endpoints, pathway search, and thermochemical analysis. Fixed or insufficiently relaxed structures also affected the Condensed Fukui Functions and Relativistic Corrections tasks in runs 1-5, while Molecular Dynamics runs 1-4 omitted relaxation and run 5 carried forward an unconverged optimization. {\cinzel Quntur} was not error-free, but it more consistently preserved the dependency between structure preparation, validation, and the requested observable. Representative geometry failures are illustrated in Supplemental Information Section \ref{SI:QunturEval}.

Method selection showed the same general pattern. Claude Code often generated technically valid quantum-chemistry calculations without fully determining which method was required for the scientific question. In Excited-State Kinetics runs 1-5, it calculated standard electronic transitions rather than the vibrationally enabled fluorescence rate requested by the task. In Thermochemistry runs 1-5, it used default Conductor-like Polarizable Continuum Model (CPCM)~\cite{Barone1998Quantum} without literature-based validation. In NMR and Magnetic Properties, runs 2-5 omitted one or both reference calculations needed to convert computed responses into chemical shifts, whereas run 1 completed the reference workflow. Similarly, in Custom Optimizations, all five runs ultimately relied on scans or external penalty/numerical optimizers rather than delivering a successful specialized crossing-optimization workflow; run 3 additionally attempted native \texttt{CI-OPT}, but with incomplete state specification and failure at the initial gradient. These recurring examples reflect insufficient literature-based and property-specific method selection, rather than an inability to write syntactically valid input files.

The five-repetition evaluation also gives a more nuanced picture of error recovery. Claude Code sometimes recovered useful calculations after an earlier branch failed, including plausible transition structures and potential-energy surfaces. The recurring weakness was deciding whether a recovery actually satisfied the original scientific objective. Incomplete, nonconverged, or incorrectly defined branches were sometimes carried into the final analysis as if they were successful, particularly in the solvation, molecular-dynamics, and electronic-state-crossing tasks.{\cinzel Quntur} also produced imperfect and unconverged branches, but more often detected the problem, returned to the intended workflow, and kept the limitation from silently becoming the basis of the final answer.

Postprocessing introduced an additional class of errors after calculations had finished. These failures were heterogeneous, but they repeatedly changed the requested quantity or its interpretation. Examples included treating unreferenced magnetic responses as chemical shifts, computing reaction barriers from an inconsistent reference state, using incorrect units or thermochemical references in isotope-effect calculations, and drawing conclusions from incomplete calculation sets. Such errors are not presentation details: they can reverse a trend, change a numerical result, or replace the requested observable with a different quantity.

The evidence for these recurring differences is provided in Supplemental Information Section \ref{SI:QunturEval}, which reports the scores and point-deduction rationales for every repetition and presents direct comparisons of representative geometries and input files. The complete reports, inputs, outputs, structures, and analysis files for both agents are available in the Supplemental Information repository dataset \cite{Perez-Zou2026QunturFiles}. These records show that the Claude Code failure modes persisted across independent repetitions, including two at higher reasoning effort. They also document the corresponding{\cinzel Quntur} failures and deductions. Taken together, the evidence supports greater workflow reliability for Quntur—not error-free performance—and is consistent with its domain-specific architecture improving preservation of scientific dependencies, literature-guided method selection, validation, recovery, and numerical analysis.

The benchmark comparison identifies where these capabilities affect the reliability of standardized computational tasks. We next examine how the same architecture performs in longer, open-ended research workflows, where the relevant capabilities interact and where success cannot be assessed from task completion alone.

\subsection*{Performance of {\fontfamily{Cinzel-LF}\selectfont El Agente Quntur} in research-level scenarios: strengths and limitations}

We organize the research-level case studies around five capabilities that recur across the benchmark analysis: \textit{i)} planning and task decomposition, partially handled by the \texttt{computational\_chemist} sub-agent but distributed across the system; \textit{ii)} geometric reasoning, primarily handled by the geometry generation module; \textit{iii)} error recovery and adaptability, distributed across multiple modules; \textit{iv)} quantum chemistry and software understanding, handled by the \texttt{input\_file\_service} sub-agent and supported by guided deep research; and \textit{v)} post-analysis, mainly carried out by the \texttt{interact\_with\_os\_and\_files} and \texttt{computational\_chemist} sub-agents. Although discussed separately below, these capabilities are interdependent: a case-study outcome may therefore reveal strengths or limitations in more than one category.

\subsubsection*{Planning and task decomposition}

In the research-level case studies, planning was evaluated through{\cinzel Quntur}'s ability to decompose an open-ended question into a coherent computational workflow and to revise that workflow when its initial hypothesis proved inadequate. In the study of \hyperref[sec:esd]{\textit{Fluorescence and intersystem crossing rates of anthracene}},{\cinzel Quntur} correctly identified the need for a multi-stage workflow including ground state optimization, vertical excitation analysis to probe relevant excited states, excited state optimizations, and explicit rate calculations using ORCA's excited-state dynamics module \cite{deSouza2018On,deSouza2019Predicting}. The general workflow for computing these properties is clearly explained in the software's documentation, but {\cinzel Quntur} adapts it to handle unforeseen calculation details or the properties of the molecule of interest.

The importance of a good initial plan cannot be overstated. An error at this stage can silently propagate throughout execution, wasting resources and producing incorrect results. A common example occurs when comparing different DFT functionals to calculate some property. As observed in the \hyperref[sec:NMR]{\textit{31P NMR absolute shielding and chemical shift tensor of triphenylphosphine oxide}} case study,{\cinzel Quntur} decides to run NMR calculations with B3LYP \cite{beckeDensityfunctionalThermochemistryIII1993,leeDevelopmentColleSalvettiCorrelationenergy1988,stephensInitioCalculationVibrational1994} and $\omega$B97X-D3 methods \cite{linLongrangeCorrectedHybrid2013}, using the same geometry optimized using $\omega$B97X-D3. 
We also observe{\cinzel Quntur}'s ability to modify its original plan. For studying \hyperref[sec:isomerization]{\textit{Competing mechanisms of Isocyanide isomerization}},{\cinzel Quntur} mistakenly proposed a reaction mechanism that the calculations revealed to be unfeasible. Subsequently, it removed that mechanism from the workflow and proposed a new triplet carbene pathway, which was explored upon human confirmation. After the newly proposed pathway led to a thermally inaccessible state, it focused solely on the concerted mechanism.

\subsubsection*{Geometric reasoning}

In addition to generating geometries, {\cinzel Quntur} must be able to map between natural-language chemical descriptions and three-dimensional molecular representations (e.g., \texttt{xyz} files) and use this mapping to plan, evaluate, and modify geometries as part of a computational workflow.
{\cinzel Quntur} reads a set of coordinates (text-based analysis) or directly views a molecular structure in \texttt{png} format (image-based analysis), either to handle local parameters such as atomic indices, connectivity, and atomic identities, or to recognize global concepts such as symmetry. For example, in the \hyperref[sec:Fukui]{\textit{Global softness and chemoselectivity}} case study, ORCA reports atomic charges for each atomic index.
{\cinzel Quntur} must be able to assign atomic indices to each atom correctly to report their reactivity. Similarly, during the \hyperref[sec:NTOs]{\textit{Natural Transition Orbitals visualization and excited states characterization}} case study, the \texttt{computational\_chemist} must realize that the equilibrium geometry of S$_1$ of DMABN is not twisted but relatively planar.

\begin{figure}[ht]
    \centering
    \includegraphics[width=1\textwidth]{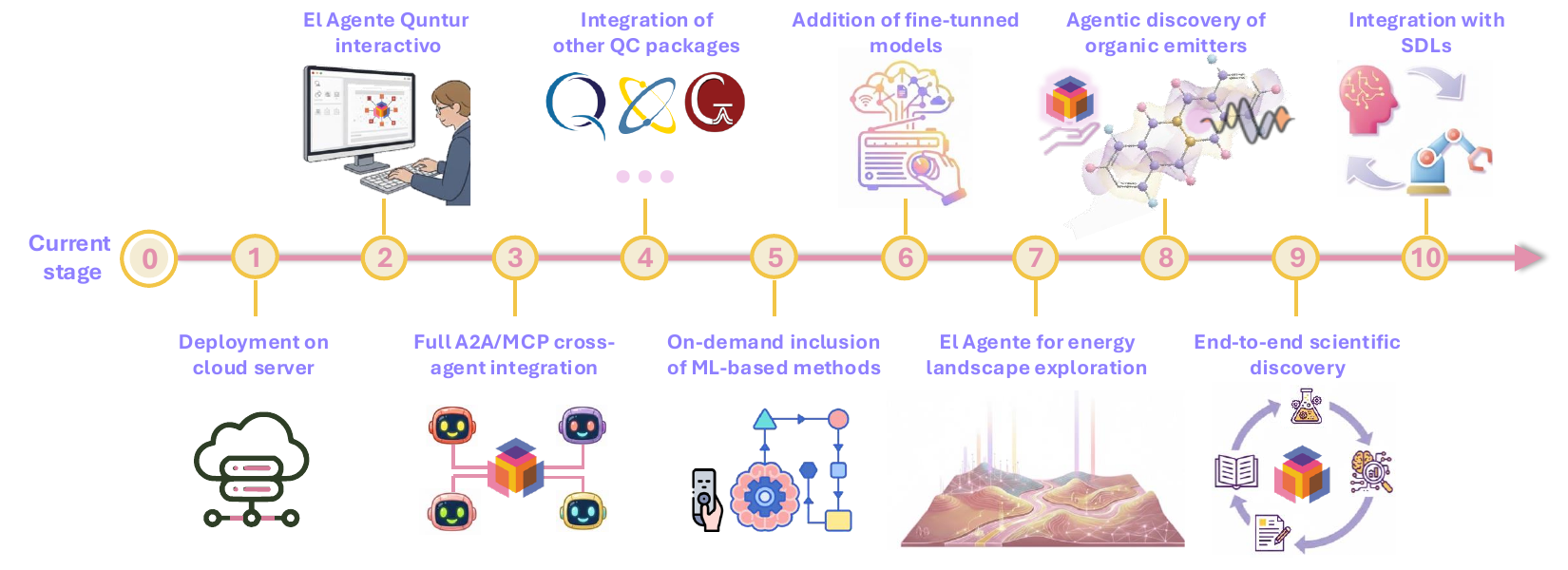}
    \caption{\textbf{Development roadmap.} From computational quantum chemistry research collaborator to autonomous research agent.}
    \label{fig:roadmap}
\end{figure}

Although these results are promising, we observe that LLMs struggle with more complex geometry-related tasks. For example, in the \hyperref[sec:isomerization]{\textit{Competing mechanisms of Isocyanide isomerization}} case study,{\cinzel Quntur} mistakenly misrepresented a triplet carbene isomerization pathway due to its inability to recognize that the ``\ce{CH3NC} (T$_1$)'' and ``Carbene (T$_1$)'' optimized structures were actually the same structure (see Fig. \ref{fig:Han_case_study} in the Supplemental Information). Yet, the most significant limitation arises during the geometry-generation stage of transition-state searches. Performing Nudged Elastic Band (NEB) calculations to find transition states and reaction paths \cite{Mills1995Reversible,Berne1998Classical,Graeme2000Improved} is a particularly challenging problem for{\cinzel Quntur}. For simple problems (see benchmark questions), a TS guess can be directly generated, or a relaxed PES scan can be performed through the changing internal coordinate to find a guess for the TS structure. However, many transformations (e.g., those found in catalysis) involve changes in multiple internal coordinates through the TS, and a multidimensional PES scan over several internal coordinates quickly becomes prohibitively expensive (for context, ORCA allows one to vary up to three internal coordinates in a PES scan).{\cinzel Quntur} was particularly limited in generating geometries when cases required more advanced spatial and geometric understanding and reasoning. We performed two case studies involving NEB: \hyperref[sec:migratory]{\textit{Migratory insertion of \ce{CH3} into \ce{C2H4} on Ni}} (based on References \cite{Musaev1997A}), and \hyperref[sec:transeffect]{\textit{Trans effect of ligands on a square planar Pt complex}} (based on References \cite{Shoubhik2025Benchmark, Pinter2013trans, Zumdahl1968A}). Attempts to allow{\cinzel Quntur} to generate the appropriate geometries for the TS search failed. For more complex TS structures, as in the case studies, interpolating from the optimized structures of the endpoints of a reaction path is often a useful way to generate an approximate reaction path with a reasonable TS guess.
However, these interpolation methods require the reactant and product to be pre-aligned so that there is a one-to-one mapping between their structures. In other words, not only do the atomic indices of the reactant and product have to match exactly, but the positions of the atoms---except for the ones directly involved in the changing reaction coordinate, have to be at a similar location in three-dimensional space (i.e., the Cartesian coordinates of one molecule cannot be at a completely different location even if the internal coordinates of the two molecules are the same).
We were able to complete the case studies only after providing the geometries for the fixed endpoints of the interpolation. Based on our observations, we conclude that geometric reasoning is currently a major bottleneck for AI agents in chemistry. This observation is consistent with recent works showing that current LLMs are notoriously limited in their ability to understand and manipulate 3D structure information \cite{Lv2025atomworld, Yamada2024evaluating, Jang2025structural, choi2026elagenteEstructural}. As discussed in our roadmap (see Fig. \ref{fig:roadmap}), we will re-address these types of computational quantum chemistry problems upon the imminent integration of{\cinzel El Agente Quntur} and{\cinzel El Agente Estructural} \cite{choi2026elagenteEstructural}, a specialized agent for geometry generation, manipulation, and reasoning that was released as a preprint the same day this work was reported, in the same repository \cite{choi2026elagenteEstructural}.

\subsubsection*{Quantum chemistry and software understanding}

A computational quantum chemistry agent must translate each planned step into executable input files. For any quantum chemistry package, this requires a precisely formatted input file specifying the calculation type, molecular structure, chosen level of theory, the properties to be computed, and other settings. In other words, the agent must specify the calculation method in the geometry section and provide additional settings/keywords for special tasks. Indeed, the diversity of input files reflects the field's vast complexity.
In the ORCA~6.0 example, the input file contains a keyword line with dozens of ``run types'' (e.g., \texttt{DFT}, \texttt{HF}, \texttt{OPT}, \texttt{FREQ}, \texttt{IRC}, etc.).
It also comprises dozens of specialized blocks, each with dozens of adjustable settings that govern the calculation of nuanced aspects \cite{Neese_ORCA6}.
Thanks to our guided deep research strategy, described in the \hyperref[sec:methods]{Methods} Section, El Agente{\cinzel Quntur} excels at this task, as evidenced by the breadth and sophistication of the benchmark questions and case studies it can address. A particular example is the \hyperref[sec:N2]{\textit{Potential energy surface of \ce{N2}}} case study, where the challenge lies in assembling input files that correctly combine geometry scans, multiple electronic-structure formalisms, and consistent reference states within a single workflow. Successful assembly of such an input depends on information distributed across different sections of the software manual and distinct conceptual domains of electronic structure theory.{\cinzel Quntur}'s guided deep research enables efficient setup of these calculations, as the input files are generated in parallel by the corresponding domain experts.

\subsubsection*{Error recovery and adaptability}

Self-debugging is arguably the most crucial sub-skill for achieving robust autonomy. Like its predecessor, {\cinzel Quntur} is resilient: when a job fails, produces a warning, or the computed properties (e.g., energies, spin, and vibrational frequencies) indicate that a change in methodology is needed, it does not stop. Instead, it diagnoses the issue and tries a solution. This behaviour is particularly useful during geometry optimizations, where the converged geometry can correspond to a minimum, a first-order saddle point, a second-order saddle point, or other types of extrema. Once again, during the case study \hyperref[sec:isomerization]{\textit{Competing mechanisms of Isocyanide isomerization}}, the transition state of the concerted mechanism converges to a second-order saddle point instead of a first-order saddle point (TS), and{\cinzel Quntur} successfully eliminates the imaginary frequency along the mode orthogonal to the reaction coordinate. In addition, in the same case study, the agent begins with a workflow hypothesis that encompasses two reaction pathways. Upon computational validation that one proposed pathway is unfeasible, the agent autonomously reformulates the workflow to explore an alternative mechanism involving the triplet PES. This instance demonstrates that{\cinzel Quntur}'s design allows it to revise its high-level plan, indicating that adaptability is not limited to low-level error handling or numerical diagnostics. The same happens for the case study \hyperref[sec:solvation]{\textit{solvation effects on the keto–enol equilibrium of acetylacetone in water}}. The original user prompt specifies the use of both an implicit CPCM water model \cite{Barone1998Quantum}, and an explicit water-cluster–CPCM hybrid continuum model to compare solvation free energies. However, both methods fail qualitatively to reproduce the experimentally observed equilibrium. When users subsequently request a comparison with previous computational studies, {\cinzel Quntur} recommends using the Solvation Model based on Density (SMD) \cite{Marenich2009Universal}, re-runs the calculations with it, and confirms that it provides the best agreement with experimental results within chemical accuracy ($< 1$~\kcal{}).
These examples demonstrate the agent's ability to reconsider its initial hypothesis and formulate alternative ones.

\subsubsection*{Post-analysis}

A research collaborator would not simply run calculations; they must also assess the meaningfulness of the results, verify the success of the computations, and present the findings coherently. This set of tasks is particularly challenging for LLMs, which tend to be overconfident in their results rather than rigorously checking for methodological errors. To mitigate this effect, {\cinzel Quntur} analyzes the results by cross-checking the methodology and contextualizing the findings through targeted web searches over relevant literature. Yet we still observe remnants of the problematic behaviour. For example, in the \hyperref[sec:transeffect]{\textit{Trans effect of ligands on a square planar Pt complex}} case study, most of the transition-state results are likely incorrect due to limitations in what the agent can achieve with NEB at the time of this manuscript. The trend is clearly different from that reported in the literature; yet the agent rationalized the differences as methodological, even though different methods should still yield the same trends.

On the other hand, the raw outputs often require additional post-processing, as chemically relevant observables are rarely provided directly by the electronic structure code, e.g., thermal populations from energies, reaction rates from energy barriers, reactivity from electronic properties, spectra plots, or reaction pathway diagrams. At present,{\cinzel Quntur} performs mathematical operations using Python; however, additional safeguards are still required to ensure correct theoretical and mathematical processing, thereby guaranteeing a more reliable transformation of raw outputs into chemically meaningful quantities and visualizations. Finally, we once again showcase {\cinzel Quntur}'s multimodality across all case studies, as it directly inspects the images it generates during post-analysis. This is central in the case study \hyperref[sec:NTOs]{\textit{Natural Transition Orbitals visualization and excited states characterization}}, where{\cinzel Quntur} correctly assigns qualitative features of the NTOs by inspecting \texttt{.png} images generated with the computed .cube files. Visualization of wavefunctions and densities has the potential to provide not only intuitive interpretations of electronic-state characters, but also autonomous active-space selection for multi-reference methods and autonomous diagnostics for SCF-convergence issues \cite{Hughes2025Fundamental}.

\subsection*{Future directions}\label{sec:roadmap}

The alpha version of the {\cinzel El Agente} cloud platform will be released alongside the {\cinzel Quntur} version. The exact release date and practical details for accessing and using the system will be posted on the website \href{https://elagente.ca/}{elagente.ca} soon (Stage 1).
Following its release, we plan to leverage{\cinzel Quntur}'s interactive (human-in-the-loop) capabilities to enhance its value as a classroom tool for teaching computational quantum chemistry at the undergraduate and graduate levels (Stage 2). The remainder of the roadmap focuses on transitioning from a research collaborator to a fully autonomous research agent capable of conducting computational chemistry research independently and integrating smoothly into broader end-to-end chemistry research and self-driving laboratory pipelines. The first step in this direction is the increase in breadth: we will enable communication between{\cinzel Quntur} and other agents within the {\cinzel El Agente} network \cite{Zou2025ElAgente,mueller2026elagenteFarmaco,bai-aldossary2026grafico, gustin2026elagentecuantico, Govind2026El, Zhang2026El} (Stage 3).
We will also add support for other quantum chemistry packages \cite{Epifanovsky2021Qchem,g16, Galvan2019OpenMolcas, Qiming2018PySCF, Alexeev2012GAMESS, Turney2012Psi4}, as well as state-of-the-art machine-learning-based methods (Stages 4 and 5).
These capabilities will allow {\cinzel El Agente} to address interdisciplinary questions that require computational tools and abilities across all scientific fields. Second, we will work to further advance our agent's capabilities in depth. We will address current bottlenecks in computational chemistry research, such as structure reasoning, by training our own specialized sub-agents (Stage 6). For example, we will develop an agent to explore potential energy landscapes and identify chemically relevant critical points, such as minima and saddle points. This task is highly hands-on and central to the study of chemical reaction mechanisms \cite{Tian2025Bulky} (Stage 7). With these developments in hand, we will begin performing autonomous end-to-end chemical research in collaboration with self-driving laboratories \cite{Tom2024Self} (Stages 8-10).

\section*{Methods}\label{sec:methods}

{\cinzel El Agente Quntur} is an API-based multi-agent orchestration framework. Each sub-agent is a separately prompted LLM procedure node with a defined role, context, memory access, tool access, and graph connections to other agents. The \texttt{computational\_chemist} agent performs high-level planning and delegates specialized tasks, including geometry generation, input-file generation, quantum chemistry execution, validation analysis and parsing, and post-processing. The foundation model is accessed through hosted API backends rather than run locally, and was not retrained or fine-tuned for this work; in the benchmark experiments, Claude Opus 4.5 was used as the base model.
{\cinzel Quntur} therefore does not alter the intrinsic weights or chemistry knowledge of the underlying model. Instead, it changes the effective problem-solving process around the model by decomposing tasks across specialized agents, constraining context and tool access, preserving intermediate artifacts for transparent audit and validation, using semantic memories and curated reference material, and grounding decisions in chemistry software outputs. Quantum chemistry calculations are executed by external programs on the available computational infrastructure, such as ORCA jobs submitted through HPC/Slurm, while the LLM is used for interfacing with humans, reasoning, task decomposition, tool selection, retrieving information from literature, and output interpretation.

\subsection*{Removing hard-coded policies to facilitate scientific reasoning}
As foundation models improve over time, agent design is shifting away from hard-coded policies (e.g., fixed step-by-step instructions) toward higher-level prompts that encourage the model to reason and adaptively select tools. This shift scales better: it reduces the amount of expert-crafted context required and leverages the model's internal knowledge more effectively, as human-authored policies can, in practice, become a bottleneck.

As an example, consider a hard-coded rule to always run a GFN\textit{n}-xTB \cite{grimmeRobustAccurateTightBinding2017,bannwarthGFN2xTBAccurateBroadly2019,bannwarthExtendedTightbindingQuantum2021} semi-empirical relaxation (cheap) before an \textit{ab initio} calculation (e.g., DFT, expensive). This approach is often beneficial because pre-relaxation can accelerate subsequent \textit{ab initio} geometry convergence by providing a reasonable initial guess above direct geometry generation. However, such a rigid rule overlooks important edge cases. For instance, when studying transition or excited states, semi-empirical ground state relaxation may distort the geometry and erase features that can only be captured with the \textit{ab initio} method targeting the correct state.

When we remove these hard-coded rules and instead let the agent decide, it can select methods and workflows based on scientific reasoning and the specifics of the input. For example, the agent may consider that a geometry optimization should proceed directly to the \textit{ab initio} stage rather than pre-relaxing the structure with a semi-empirical method. Rigid policies require continuous patching to address exceptions and edge cases, which is time-consuming and still not general. In contrast, reasoning-first approaches tend to generalize better while requiring lower design and maintenance effort.

\subsection*{Design tools that facilitate generalization and efficiency}

Tools are the core interface between an agent's reasoning and its environment. An agent decides which actions to take using its tools, observes the results, and iterates until the goal is achieved. Designing tools that are both general and efficient is therefore fundamental. If the action space is too narrow, the agent may reason correctly about what to do, yet still be unable to complete the task. Worse, overly specialized tools can mislead the agent by hiding relevant out-of-domain context. In other words, general tools enable the agent to tackle a far broader problem space by allowing it to read, discover, execute, and process, and by facilitating the flexible handling of constraints such as a limited LLM context window.

A concrete example of the points discussed above is the task of parsing ORCA outputs. Some quantum chemistry calculations can produce files exceeding 10{,}000 lines, far beyond what most LLMs can read within a context window. These outputs contain many properties, and their structures vary by calculation type in ways that are neither consistent nor thoroughly documented. Traditional solutions rely on case-specific parsers (often regex-based), which are not scalable: it is impractical to manually implement robust, manageable parsers for every property and output variant.
In{\cinzel El Agente Q}, we initially addressed this challenge by predefining a set of output sections and allowing the model to read only the relevant ones \cite{Zou2025ElAgente}. While this strategy has achieved strong results, it remained brittle: new calculation modes introduce additional sections, and even a single section can remain extremely long, particularly those containing tables or charts. Early scaling revealed frequent parsing errors driven by these limitations.

{\cinzel Quntur}'s solution is expanding the action space with two tools: (1) a dynamic section indexer that discovers section names and their line ranges directly from the output, and (2) a bash terminal interface that enables targeted inspection (e.g., reading specific line ranges or searching keywords) without loading entire sections into the LLM context. For large tables and chart-like blocks, the agent can sample the header and a few rows to infer the format, then programmatically extract the table and convert it to CSV for downstream analysis in Python. This pattern allows the agent to generalize during parsing, as tools do not prevent it from accessing any context in long documents, and the Bash shell interface is flexible enough for the agent to handle context smartly rather than forcing it to read a long context.

\subsection*{Guided Deep Research for ORCA input file synthesis}\label{sec:deep}

Deep research on public sources or custom documentation expands the agent's capabilities by enabling it to retrieve task-relevant context on demand through iterative reasoning. A \emph{guided} deep research variant further constrains this process by steering retrieval toward specific sources, scopes, or investigative directions, improving efficiency and reducing spurious exploration. In{\cinzel Quntur}, this skill is instantiated during \emph{input file synthesis}. Instead of encoding quantum chemistry and software-syntactic knowledge for every task in advance, we rely on execution-time reasoning and retrieval.

The focus on input file synthesis is deliberate. Synthesizing computational quantum chemistry input files couples three forms of expertise. First, the agent must perform \emph{quantum chemistry reasoning}: selecting physically meaningful electronic-structure methods, approximations, and observables that are consistent with the research question and the molecular system. Second, it must perform \emph{physical chemistry reasoning}: In some computational chemistry packages like ORCA, a substantial portion of what is often treated as post-processing is specified directly in the main input file. Population analysis, excited-state analysis, calculation of vibrational and electronic spectra, thermochemistry, kinetics, and wavefunction transformations are encoded as input directives rather than post-run scripts \cite{Neese2012The, Neese_ORCA6}. Finally, it must perform \emph{software realization}: mapping those conceptual decisions onto software-specific keywords, blocks, dependencies, and execution pathways. These three aspects are inseparable in practice: a scientifically sensible choice or objective that is incorrectly encoded yields an invalid calculation, while a syntactically correct input that encodes the wrong conceptual assumptions produces meaningless results. As a result, generating the correct ORCA input files is a particularly demanding process in which LLMs are known to perform poorly \cite{Pollice2025Developing}.

To support this level of complexity,{\cinzel Quntur} organizes input file generation around a central \emph{orchestrator} sub-agent (\texttt{input file service}) and a set of \emph{domain specialists} (guided deep research module). The orchestrator operates at a high level. It determines \textit{i)} whether additional research is required at a given decision point, \textit{ii)} which sources are admissible for that research (e.g., ORCA documentation, scientific literature), and \textit{iii)} which domain specialists should be consulted. Its primary role is not to generate detailed input sections directly, but to reason about the overall structure of the calculation: which conceptual components are relevant, how they interact, and how they should be assembled into a coherent whole. Domain specialists are also sub-agents. Each is biased toward a particular class of scientific subproblems (e.g., multi-reference methods, geometry handling, NEB and scanning parameters, excited-state analysis blocks, etc.). They conduct in-depth research across relevant sections of the software manual and related literature, removing hallucinations, and enabling appropriate method selection in line with best practices. The transition from the ORCA block experts of{\cinzel El Agente Q} to quantum chemistry domain experts (~47 domains) in {\cinzel Quntur} is the first step towards integration of other quantum chemistry packages such as{\cinzel Q-Chem} \cite{Epifanovsky2021Qchem},{\cinzel Gaussian} \cite{g16},{\cinzel OpenMolcas} \cite{Galvan2019OpenMolcas},{\cinzel
PySCF} \cite{Qiming2018PySCF},{\cinzel GAMESS} \cite{Alexeev2012GAMESS}, and{\cinzel PSI4} \cite{Turney2012Psi4}.

The hierarchical organization of the input file generation emerged from practical experience rather than theoretical arguments. In particular, we observed that single-agent systems exhibited a recurring failure mode: over-confidence driven by shallow retrieval. A monolithic agent would often read a single relevant documentation page, conclude that it had sufficient context, and proceed to plan and execute based solely on that limited view. For example, we observe that LLMs usually do not adapt parallelization or memory settings to the molecule of interest during geometry optimization. 
By contrast, the hierarchical system first reasons at a higher level about the structure of the calculation, identifying that a \texttt{\%pal} section may be required. Thus, guided deep research at the orchestrator level expands the search space conceptually (``\textit{what sections might be relevant?}''), while deeper investigation by domain specialists refines the details (``\textit{how many cores should I use for this molecule and in this cluster?}''). We do not claim that hierarchy is strictly necessary in all cases, nor that a sufficiently capable single agent could not be instructed on similar behaviours. Rather, this architecture provides evidence that combining hierarchical planning with guided deep research during execution mitigates common failure modes of monolithic agents, enabling more reliable and higher-quality input synthesis.

Guided deep research allows the system to scale across any software's feature space without relying on human-crafted surrogate documentation or static semantic memories. Instead of encoding procedural knowledge for each calculation type, the system design focuses on providing correct sources, role biases, and orchestration logic to guide agents' search and reasoning. Overall, this system demonstrates that execution-time guided deep research, combined with hierarchical orchestration and domain-focused investigation, can support complex scientific workflows through direct input-file synthesis.

\section*{Resource Availability}

\subsection*{Data and code availability}

All the data required to evaluate the presented conclusions are available via \url{https://doi.org/10.5683/SP3/RDSOEA}.

\subsection*{Lead Contact}

Requests for further information and resources should be directed to and will be fulfilled by the lead contacts, Varinia Bernales and Al\'an Aspuru-Guzik.

\section*{Acknowledgments}
We gratefully acknowledge the longstanding contributions of the Matter Lab’s current
and past group members (\url{matter.toronto.edu}), and in particular from El Agente's team. We also thank the reviewers for their insightful comments and constructive suggestions, particularly regarding the Claude Code baseline comparison, which substantially improved the rigor, clarity, and overall quality of this work. This work was supported by the Defense Advanced Research Projects Agency (DARPA) under Agreement No. HR0011262E022. This work was also supported by the AI2050 program of Schmidt Sciences. I.G., J.A.C.G.A., and J.B.P.S. acknowledge funding of this project by the National Sciences and Engineering Research Council of Canada (NSERC) Alliance Grant \#ALLRP587593-23 (Quantamole).
T. W. K. acknowledges the support of the Vector Distinguished Postdoctoral Fellowship.
J.A.C.G.A. acknowledges support from the Council for Science, Technology and Innovation (CSTI), Cross-ministerial Strategic Innovation Promotion Program (SIP), ``Promoting the application of advanced quantum technology platforms to social issues'' (Funding agency: QST).
M.M. was partially supported through a collaborative partnership with Merck KGaA, Darmstadt, Germany.
A. W. acknowledges support from NSERC through their Canada Graduate Scholarships - Doctoral (CGS-D) program as well as support from the Lawson Climate Institute. C.C. acknowledges support from the DOE grant with the University of Minnesota award \#A006801504 and the Basic Science Research Program through the National Research Foundation of Korea (NRF) funded by the Ministry of Education (RS-2025-02634334).
 A.A.-G. thanks Anders G. Fr{\o}seth for his generous support. A.A.-G. and V.B. acknowledge the generous support of Natural Resources Canada and the Canada 150 Research Chairs program and the University of Toronto’s Acceleration Consortium, which receives funding from the CFREF-2022-00042 Canada First Research Excellence Fund.

\section*{Author contributions}
J.B.P.S. and Y.Z. conceived the research, developed the software, designed and performed the benchmark and case-study evaluations, analyzed the results, and prepared the manuscript. J.A.C.G.A. contributed to the conceptual development of the research, tested the system, performed and evaluated benchmark and case-study executions, analyzed the results, and contributed to manuscript preparation. M.M., I.G., and A.W. tested the system, performed and evaluated benchmark and case-study executions, and contributed to manuscript preparation. H.H., T.W.K., Changhyeok Choi, and E.S.I. performed and evaluated benchmark and case-study executions. M.G.V., H.X., and Chris Crebolder maintained the system, contributed to software development and deployment, and supported the system's availability during review. V.B. and A.A.-G. supervised the research. All authors reviewed and approved the final manuscript.

\section*{Declaration of interests}
The authors declare no competing interests. At the time of submission, no author holds equity, receives royalties or licensing income, or is party to a commercial agreement related to the system. Although possible future dissemination models may be considered, no current financial interest exists. The source code is presently non-public for reasons separate from commercialization.

\section*{Declaration of generative AI and AI-assisted technologies in the writing process}
During the preparation of this work, the authors used chatGPT to improve the readability of parts of the manuscript. After using this service, the authors reviewed and edited the content as needed and take full responsibility for the content of the article.
\clearpage
{
\small
\bibliography{references}
\bibliographystyle{numbered}
}

\clearpage

\clearpage

\appendix

\begin{bibunit}[numbered]
\makeatletter
\def\hyper@natlinkstart#1{%
  \Hy@backout{#1}%
  \hyper@linkstart{cite}{cite.SI.#1}%
  \def\hyper@nat@current{#1}%
}
\def\hyper@natlinkbreak#1#2{%
  \hyper@linkend#1\hyper@linkstart{cite}{cite.SI.#2}%
}
\def\hyper@natanchorstart#1{%
  \Hy@raisedlink{\hyper@anchorstart{cite.SI.#1}}%
}
\let\main@citenum\citenum
\renewcommand{\citenum}[1]{%
  \@startbibunitorrelax
  \begingroup
  \let\@auxout\@bibunitaux
  \main@citenum{#1}%
  \endgroup
}
\makeatother
\newpage
\section*{Supplemental Information}
\setcounter{figure}{0}
\renewcommand{\thefigure}{S\arabic{figure}}

\setcounter{table}{0}
\renewcommand{\thetable}{S\arabic{table}}

\renewcommand{\theHfigure}{S\arabic{figure}}
\renewcommand{\theHtable}{S\arabic{table}}

\begingroup
\setcounter{tocdepth}{2}
\tableofcontents
\endgroup
\clearpage

\setcounter{figure}{0}
\renewcommand{\thefigure}{S\arabic{figure}}
\renewcommand{\theHfigure}{S\arabic{figure}}
\setcounter{table}{0}
\renewcommand{\thetable}{S\arabic{table}}
\renewcommand{\theHtable}{S\arabic{table}}
\section*{Supplemental Methods}
\addcontentsline{toc}{section}{Supplemental Methods}
\section{List of Agents}

In this section, we summarize the agents that constitute{\cinzel El Agente Quntur}.

\renewcommand{\arraystretch}{0.95}
\begin{table}[H]
\centering
\caption{\textbf{Agent roster}}
\small
\begin{tabularx}{\linewidth}{@{}>{\centering\arraybackslash}p{0.34\linewidth}|Y@{}}
\toprule
Label & Description \\
\midrule
autoci\_expert & Configures internally contracted multi-reference (FIC-MR) methods using canonical orbitals\\
basis\_expert & Defines the primary orbital basis and auxiliary fitting bases, segmented vs general contraction, decontracting, adding polarization/diffuse functions, element-specific basis overrides, and other basis-related settings\\
casresp\_expert & Provides keywords or blocks that configure CASSCF response property calculations \\
casscf\_expert & Provides casscf settings for multiconfigurational SCF calculations \\
cim\_expert & Configures Cluster-in-Molecule (CIM) settings\\
cis\_tddft\_expert & Provides computational settings for excited-state calculations using Configuration Interaction Singles (CIS), Time-Dependent Density Functional Theory (TDDFT), and Spin-Flip TDA (SF-TDA) \\
compound\_expert & Expert on ORCA's embedded scripting/driver language compound \\
computational\_chemist & Designs high-level plans, delegates all engine-specific details to specialist agents, and perform post-analysis\\
configuration\_recommender & Recommend a suitable wavefunction method, DFT functional, semi-empirical method, as well as suggests when dispersion corrections are needed, that fits the user request\\
conical\_expert & Controls how ORCA searches conical intersections (CIs) between two electronic states \\
cosmors\_expert & cosmors configures the OpenCOSMO-RS solvation interface in ORCA\\
cpcm\_expert & Controls settings for implicit solvation modeling using the Conductor-like Polarizable Continuum Model or other solvation models \\
docker\_expert & Performs automated docking of one or more GUEST structures onto a HOST \\
el\_prop\_expert & The elprop expert expert provides settings for calculating electrical properties: dipole, quadrupole, static/dynamic polarizabilities, mixed dipole-quadrupole and quadrupole-quadrupole polarizabilities\\
eprnmr\_expert & The eprnmr expert configures EPR and NMR property calculations\\
esd\_expert & The esd expert configures excited-state kinetics and spectroscopic calculations including vibronic effects \\
freq\_expert & Controls frequency analysis, sets numerical differentiation controls, applies frequency scaling, controls projection/invariance, thermochemistry (T, P). and other related tasks \\
geom\_expert & Set up geometry constraints, relaxed (not rigid) scans (1D or simultaneous), TS search, non-default step/trust/tolerances, Hessian control, fragment connectivity/constraints, external optimizer hooks, or wall potentials \\
geometry\_generation & Handles the initial geometry creation process with diverse toolkits\\
geometry\_operator & Perform arbitrary geometry editing \\
geometry\_optimization & Expert that optimizes molecular geometries at different levels of theory\\
geometry\_section & Generates the ORCA geometry section, referencing external xyz files, inline coordinates, or internal coordinates\\
goat\_expert & Configures ORCA's GOAT global optimizer, controlling workers, iterations, conformer filtering, entropy-based stopping, and optional free-topology/reactive settings\\
ice\_expert & Configures Iterative Configuration Expansion CI \\
imaginary\_frequency\_expert & Handles imaginary-frequency analysis and correction, including selective mode removal and verification \\
\bottomrule
\end{tabularx}
\end{table}

\renewcommand{\arraystretch}{0.95}
\begin{table}[H]
\centering
\caption{\textbf{Agent roster (continuation)}}
\small
\begin{tabularx}{\linewidth}{@{}>{\centering\arraybackslash}p{0.34\linewidth}|Y@{}}
\toprule
Label & Description \\
\midrule
input\_file\_service & Creates and debugs ORCA 6.0 input files using the ORCA manual and scientific literature \\
input\_keywords\_expert & Controls ORCA input keywords and legacy ORCA blocks (e.g., base, maxcore, moinp, pointcharges) \\
interact\_with\_os\_and\_files & Performs file-system operations such as reading, parsing, moving, copying, and deleting files \\
irc\_expert & Configures Intrinsic Reaction Coordinate (IRC) calculations from transition states \\
loc\_expert & loc expert provides loc settings to control post-SCF molecular orbital localization (e.g., Pipek-Mezey or Foster-Boys) \\
mcrpa\_expert & Provides settings to computes excitation energies and transition properties from the linear response of a converged CASSCF reference \\
md\_expert & Configures and runs ORCA's ab initio molecular dynamics (AIMD) and related MD tools \\
mdci\_expert & Configures Matrix Driven Configuration Interaction engine for correlated wavefunction methods and excited-state solvers \\
mecp\_expert & Configures Minimum Energy Crossing Point (MECP) optimizations to find the lowest-energy geometry where two potential energy surfaces (PES1 and PES2) intersect \\
metal\_complex\_geometry\_generation & Generates metal complex geometry files \\
method\_expert & Configures global, method-level options shared across ORCA modules \\
mp2\_expert & Configures second-order Møller-Plesset correlation \\
mrcc\_expert & Links ORCA to Mihael Kállay's MRCC program. It selects high-level coupled-cluster methods (e.g., CCSDT, CC3, CCSD(T) variants) to run on an SCF reference produced by ORCA \\
mrci\_expert & Configures ORCA's traditional uncontracted multi-reference correlation module (orca\_mrci) \\
mtr\_expert & Performs mode tracking scans along selected vibrational normal modes using a precomputed Hessian \\
nbo\_expert & Forwards literal NBO keylists (\$NBO, \$DEL, etc.) from ORCA to an external NBO6/7 executable \\
ndoparas\_expert & Customizes semiempirical NDO model parameters (CNDO/INDO/NDDO) by setting element- and index-specific constants, INDO/S interaction factors, and the INDO/S Coulomb-integral parameter \\
neb\_expert & Configures the Nudged Elastic Band (NEB) method to locate a minimum energy path (MEP) between given reactant and product structures \\
numgrad\_expert & Configures numerical differentiation for energy gradients \\
orca\_plotting\_agent & Generating visualizations using orca utility programs and other tools \\
output\_expert & Controls properties to be printed in the output \\
pal\_expert & Controls settings for parallel execution across multiple processors or nodes \\
paras\_expert & Defines named numeric parameters (scalars, ranges, or explicit lists) for later substitution elsewhere in the input \\
plots\_expert & Defines settings for the input file to generate 2D contour plots and 3D surface plots for atomic/molecular/natural/corresponding orbitals and (spin) densities \\
quantum\_chemistry\_calculation\_expert & Expert in performing wavefunction-based quantum chemistry calculations, semi-empirical calculations, DFT calculations, conformer search, molecular dynamics calculations, DOCKER for posing pairs of molecules, explicit solvation, and all other calculations supported by ORCA 6.0 \\
rel\_expert & Controls relativistic Hamiltonians and spin-orbit coupling (SOC) operators. Provides settings for scalar-relativistic methods (DKH/ZORA/IORA/X2C), picture-change effects for properties, finite nucleus models, and detailed construction of spin orbit couplings \\
\bottomrule
\end{tabularx}
\end{table}

\renewcommand{\arraystretch}{0.95}
\begin{table}[H]
\centering
\caption{\textbf{Agent roster (continuation)}}
\small
\begin{tabularx}{\linewidth}{@{}>{\centering\arraybackslash}p{0.34\linewidth}|Y@{}}
\toprule
Label & Description \\
\midrule
rocis\_expert & Configures calculations of excited states via ROCIS (CIS on ROHF/UHF/RHF references) and spectra\\
rr\_expert & The rr expert configures resonance Raman-related tasks that rely on a precomputed nuclear Hessian \\
run\_orca & Slurm Job Submitter. Submit Slurm job/jobs to HPC to run ORCA calculations \\
runtypes\_keyword\_expert & This expert provides runtype keywords for basic calculation settings (SP, EnergyGrad, Opt, FREQ, PrintThermoChem, PropertiesOnly, EDA, HF, DFT, FREQ, etc) \\
scf\_expert & Controls the target precision of the energy and the wavefunction \\
solvator\_expert & Configures automatic explicit solvation of molecules \\
symmetry\_expert & The symmetry expert controls symmetry handling for a calculation \\
\bottomrule
\end{tabularx}
\end{table}

\newpage
\section{The benchmark set}\label{sec:benchmark}

The benchmark was constructed to cover representative calculation families commonly exposed in ORCA 6.0 documentation, including single-reference and correlated electronic-structure methods, geometry optimization and frequency analysis, spectroscopy, excited states, reaction-path methods, conformer and solvation workflows, and property calculations~\cite{Neese_ORCA6}.

The benchmark should be understood as an expert-designed, artifact-based evaluation framework. At present, there is no generally accepted benchmark for autonomous agents performing computational quantum chemistry workflows. We therefore designed the rubrics around the dependency structure of practical calculations: a defensible plan, appropriate molecular or state geometries, valid input files and execution settings, and postprocessing/reporting that extracts the requested physical quantity. Their relative weights are task-specific because different observables depend on different bottlenecks; for example, transition-state searches depend strongly on geometry and stationary-point validation, whereas NMR and fluorescence calculations depend more heavily on reference calculations or property-specific input blocks. Scores were assigned by inspecting the generated geometries, input files, outputs, scripts, plots, parsed values, and final reports. Thus, the numerical scores are compact summaries; the reusable benchmark objects are the questions, rubrics, generated artifacts, and point-deduction rationales.

\newpage

{
\renewcommand{\arraystretch}{0.90}
\begin{table}[!htbp]
\centering
\caption{
\textbf{The benchmark.}
A set of 17 computational quantum chemistry exercises covering several topics from electronic and magnetic properties, thermodynamics, kinetics, and spectroscopy, as well as various levels of theory.
}
\label{tab:benchmark}

\footnotesize %

\begin{tabularx}{\linewidth}{@{}c| >{\hsize=0.22\hsize}M| >{\hsize=0.22\hsize}M |Y@{}}
\toprule
Field & Topic & Level of theory & Question \\
\midrule

\multirow{4}{*}{%
  \raisebox{+45pt}[\dimexpr\height+\depth\relax]{%
    \rotatebox[origin=c]{90}{Electronic and magnetic properties}%
  }%
}

& Relativistic Corrections & DFT/ZORA ZORA-def2-TZVP \cite{weigendBalancedBasisSets2005,lentheRelativisticRegularTwocomponent1993} & Calculate the \ce{^{195}Pt} NMR shielding constant of the square-planar Pt(II) complex \ce{[PtCl4]^{2-}} using ORCA. Compare two methods: a non-relativistic DFT/PBE0-D4 NMR calculation with a def2-TZVP basis, and a scalar-relativistic DFT/PBE0-D4/ZORA calculation using the ZORA-def2-TZVP basis. Generate a final Markdown report.\\

& Electrical Properties & MP2/aug-cc-pVDZ \cite{dunning1989,kendall1992} & Compute the permanent dipole moment and the static polarizability tensor of nitrobenzene. Then compare these values to those of benzene and explain how the \ce{-NO2} substituent alters the molecular charge distribution and the anisotropy of the polarizability. Use MP2/aug-cc-pVDZ for all calculations. Generate a final markdown report.\\

& Orbitals and Densities & UHF/cc-pVDZ \cite{dunning1989} &
Visualize and analyze the SOMO and spin density of the Piperidine N-oxide radical and the pyridinium N-oxide cation radical; compare their radical stabilities. Use UHF/cc-pVDZ. Generate a final markdown report. \\

& Condensed Fukui Functions & ROHF/cc-pVTZ \cite{dunning1989} &
Using ROHF/cc-pVTZ, compute condensed Fukui functions $f^{+}$ (for nucleophilic attack) and $f^{-}$ (for electrophilic attack) for the carbon atoms in toluene using ORCA by finite-difference of electron number (N, N+1, N-1). Compare the condensed Fukui indices for different population schemes (Mulliken, Hirshfeld, and Loewdin). Generate a final markdown report.\\

\midrule

\multirow{4}{*}{%
  \raisebox{-15pt}[\dimexpr\height+\depth\relax]{%
    \rotatebox[origin=c]{90}{Thermodynamics}%
  }%
}
& Thermochemistry & DLPNO-CCSD(T)/def2-TZVPP \cite{guoCommunicationImprovedLinear2018,weigendBalancedBasisSets2005} & Using DLPNO-CCSD(T)/def2-TZVPP, use a thermodynamic cycle to compute $\Delta G_{\text{deprot}}$ for acetic acid in the gas phase and in water (implicit solvent). Compute the predicted pKa. Generate a final markdown report.\\

& Conformer Generation & GFN2-xTB and B3LYP/def2-SVP \cite{bannwarthGFN2xTBAccurateBroadly2019,beckeDensityfunctionalThermochemistryIII1993,leeDevelopmentColleSalvettiCorrelationenergy1988,stephensInitioCalculationVibrational1994} & Using ORCA GOAT \cite{desouzaGOATGlobalOptimization2025}, generate a diverse set of conformers for serine at the GFN2-xTB level. Re-optimize at a B3LYP/def2-SVP level and identify the lowest-energy structure.
Compute Boltzmann populations at 298 K in the gas phase. Generate a final markdown report.\\

& Molecular Dynamics & GFN2-xTB \cite{bannwarthGFN2xTBAccurateBroadly2019,grimmeR2SCAN3cSwissArmy2021} &
Simulate a hydrated chloride ion cluster,\ce{Cl^{-}(H2O)n} ($n \ge 12$) at 300 K using AIMD. From the trajectory, compute the \ce{Cl-O} radial distribution function g(\ce{Cl-O}) and coordination number.
Use GFN2-xTB.
Simulate 0.1 ps with a 1 fs timestep after equilibration. Generate a final markdown report. \\

& Explicit solvation & r$^{2}$SCAN-3c \cite{grimmeR2SCAN3cSwissArmy2021} & Build explicit microsolvated clusters of acetic acid with 3, 5, and 7 water molecules [\ce{AcOH.(H2O)n}] and optimize with r$^{2}$SCAN-3c in the gas phase and with CPCM water. Compute deprotonation free energies for each cluster size and compare the convergence of the predicted pKa toward experiment as n increases. Generate a final markdown report. \\

\midrule

\multirow{4}{*}{%
  \raisebox{-85pt}[\dimexpr\height+\depth\relax]{%
    \rotatebox[origin=c]{90}{Kinetics}%
  }%
}

& Potential Energy Surfaces & CASSCF(4,4)/cc-pVDZ \cite{dunning1989} & Using CASSCF(4,4)/cc-pVDZ, calculate the potential energy surface for rotation between s-cis and s-trans butadiene conformers. Identify all stationary points along the dihedral: both minima and the maximum energy structure. Compute the energy barrier and generate a final markdown report. \\

& Transition State Methods & PBE0-D3(BJ)/def2-TZVP \cite{Grimme2010AConsistent,Grimme2011Effect,Adamo1999Toward,weigendBalancedBasisSets2005} & Set up a nudged elastic band (NEB) calculation in ORCA for the double proton transfer in the formic acid dimer. Use the optimized symmetric dimer as the reactant and the fully transferred dimer as the product, generate a reasonable band of images, and refine the NEB path to obtain an approximate transition structure. Then perform an OPTTS calculation starting from the highest-energy NEB image and follow it by an IRC to obtain $\Delta G^{\ddagger}$, $\Delta H^{\ddagger}$, and $\Delta S^{\ddagger}$ for the process. Use PBE0-D3(BJ)/def2-TZVP in all calculations. Generate a final markdown report.\\

& Reaction Mechanisms & MP2/cc-pVTZ \cite{dunning1989} & Using MP2/cc-pVTZ, locate and verify the transition state for the chloride attack on methyl chloride. Validate connectivity using IRC. Extract $\Delta G^{\ddagger}$, $\Delta H^{\ddagger}$, and $\Delta S^{\ddagger}$. Generate a final markdown report.\\

& Isotopic Effects & HF/6-31G(d) \cite{hehre1972} &
Consider the solvolysis of para-substituted benzyl chlorides as a model S$_\text{N}$1 reaction in implicit water. Use ORCA to compute secondary kinetic isotope effects by substituting ring hydrogen atoms with deuterium and evaluating $\Delta G^{\ddagger}$ for H and D cases. Analyze how changes in hybridization and resonance stabilization between the ground state and carbocation intermediate influence the magnitude and sign of the secondary KIE. Use HF/6-31G(d) for all calculations. Generate a final markdown report. \\

\bottomrule
\end{tabularx}

\end{table}
}

{
\renewcommand{\arraystretch}{0.90}
\begin{table}[!htbp]
\centering
\caption{
\textbf{The benchmark (continuation).}
A set of 17 computational quantum chemistry exercises covering several topics from electronic and magnetic properties, thermodynamics, kinetics, and spectroscopy, as well as various levels of theory.
}
\label{tab:benchmark2}

\footnotesize %

\begin{tabularx}{\linewidth}{@{}c| >{\hsize=0.22\hsize}M| >{\hsize=0.22\hsize}M |Y@{}}
\toprule
Field & Topic & Level of theory & Question \\

\midrule

\multirow{5}{*}{%
  \raisebox{-40pt}[\dimexpr\height+\depth\relax]{%
    \rotatebox[origin=c]{90}{Spectroscopy}%
  }%
}
& Excited States & TD-DFT (PBE0 / $\omega$B97X-D)/def2-TZVP \cite{Adamo1999Toward,chaiLongrangeCorrectedHybrid2008,weigendBalancedBasisSets2005} & Compare TD-DFT excitation energies of \textit{p}-nitroaniline using PBE0 and $\omega$B97X-D, both using the def2-TZVP basis set. Generate a final markdown report. \\

& Excited-State Kinetics & TD-DFT CAM-B3LYP/def2-TZVP \cite{yanaiNewHybridExchange2004,weigendBalancedBasisSets2005} & Compute the fluorescence rate of formaldehyde from the S$_1$ state using TDDFT CAM-B3LYP def2-TZVP. Assume gas phase. Generate a final markdown report.\\

& NMR and Magnetic Properties & HF/aug-cc-pVTZ \cite{dunning1989,kendall1992} & Using HF/aug-cc-pVTZ, compute $^{19}$F and $^1$H chemical shifts for fluorobenzene and pentafluorobenzene. Compare shielding changes with respect to substituent effects, ring current anisotropy, and substituent inductive/mesomeric effects. Generate a final markdown report.\\

& Vibrational Spectroscopy & CCSD/cc-pVDZ \cite{dunning1989} & Using CCSD/cc-pVDZ, compute vibrational spectra for \ce{H2O} and \ce{D2O} and compare theoretical isotopic shifts against  $\sqrt{\mu_D/\mu_H}$ scaling predictions. Generate a final markdown report. \\

& Custom Optimizations & TD-DFT CAM-B3LYP/6-31G \cite{yanaiNewHybridExchange2004,hehre1972} & Locate (i) the S$_0$/T$_1$ MECP and (ii) the S$_0$/S$_1$ CI for twisted ethylene (\ce{C2H4}). Use TDDFT/TDA with CAM-B3LYP/6-31G; employ spin-flip TDDFT where necessary to ensure a stable description in the biradical region near the CI or MECP. Start from a ~90° twisted geometry. Report optimized geometries and final residual gaps at convergence. Produce a Markdown report.\\

\bottomrule
\end{tabularx}

\end{table}
}

\newpage

\section{Evaluation of {\fontfamily{Cinzel-LF}\selectfont El Agente Quntur} and the Claude Code baseline on the benchmark set}\label{SI:QunturEval}

\subsection{General evaluation rubric for benchmark questions}

The following is a general rubric to evaluate the performance of{\cinzel El Agente Quntur} on the benchmark set (see Table 2 in the manuscript). The percentages shown are default and are subsequently adjusted to adapt to each problem (e.g., geometry-generation percentages are increased for geometry-heavy tasks).

\subsubsection{Planning (\texorpdfstring{$\sim$}{TEXT}15\%)}

\begin{itemize}
    \item \textbf{Efficiency:} Calculations that are independent of each other are executed in parallel, while calculations that require outputs from previous steps are executed sequentially.
    \item \textbf{Flexibility:} The computational plan is modified by the agent when needed (e.g., upon failure, convergence issues, or new information). These methodological deviations must be properly justified.
    \item \textbf{Scientific validity:} The final computational plan answers the prompt.
\end{itemize}

\subsubsection{Geometry Generation (\texorpdfstring{$\sim$}{TEXT}20\%)}

\begin{itemize}
    \item \textbf{Geometries generated correctly:} All required geometries for the task are generated.
    \item \textbf{Additional geometry-related tasks:} Correct handling of geometry-related operations such as distortions, trajectory files, in-line geometry sections, use of internal coordinates, or inclusion of isotopic effects, etc.
\end{itemize}

\subsubsection{Input File Generation and Execution (\texorpdfstring{$\sim$}{TEXT}35\%)}

Multiple computational strategies are acceptable as long as the input is internally consistent, executable, and appropriate for the task. The input file is validated during execution. Therefore, if there is nothing obviously incorrect, the calculation completes successfully, and the final numbers are correct, the input file is considered correct.

\subsubsection*{Keyword Line}
\begin{itemize}
    \item Correct run type for the task (e.g., \texttt{OPT}, \texttt{FREQ}, \texttt{SP}, \texttt{IRC}).
    \item Correct method choice for the problem.
    \item Correct basis set.
\end{itemize}
\textit{Extra keywords are acceptable if optional and non-conflicting.}

\subsubsection*{Blocks}
\begin{itemize}
    \item All necessary blocks for the task are present.
    \item Commands inside blocks are syntactically correct and consistent with the keyword line.
    \item Blocks ensure that the requested outputs are produced.
\end{itemize}
\textit{Extra blocks or options are acceptable if optional and non-conflicting.}

\subsubsection*{Geometry Section}
\begin{itemize}
    \item Correct charge and multiplicity.
    \item Valid ORCA geometry syntax.
\end{itemize}

\subsubsection{Postprocessing and reporting (\texorpdfstring{$\sim$}{TEXT}30\%)}

\begin{itemize}
    \item \textbf{Extraction of data from output files:} Relevant quantities are correctly extracted from ORCA output files.
    \item \textbf{Postprocessing into chemical information:} Extracted data are correctly transformed into chemically meaningful results (e.g., energies, spectra, thermodynamic quantities).
    \item \textbf{Final report:} Generation of report, comparison with literature, inclusion of references.
\end{itemize}

\subsection{Rubric and evaluation of each benchmark question:}

\subsubsection{Evaluation of the benchmark question representative of the topic ``Relativistic Corrections''.}

\renewcommand{\arraystretch}{0.90}
\begin{table}[H]
\centering
\caption{\textbf{Evaluation Rubric.}
The rubric used to grade the benchmark question representative of the topic ``Relativistic Corrections''.}
\small
\begin{tabularx}{\linewidth}{@{}
p{0.04\linewidth}|
>{\centering\arraybackslash}p{0.14\linewidth}|
c|
>{\raggedright\arraybackslash}X
@{}}
\toprule
Topic & Criterion & Weight & Task\\
\midrule

\multirow{4}{*}{ \raisebox{-10pt}[\dimexpr\height+\depth\relax]{\rotatebox[origin=c]{90}{Relativistic Corrections}}}

& Planning & 15\% & Single geometry used for all calculations.

Parallel execution of independent NMR jobs.

Same XC functional in both calculations.

Correct handling of relativistic vs non-relativistic workflow.

\\

& Geometry generation & 15\% & Square-planar \ce{[PtCl4]^{2-}} structure.

Correct connectivity and symmetry.

Charge = –2, multiplicity = 1.

Geometry reused consistently. \\

& Input file generation and execution & 40\% &  

\textbf{Non-relativistic NMR:} 

DFT + def2-TZVP (ECP on Pt).

NMR shielding enabled.

\textbf{Scalar-relativistic NMR:} 

ZORA Hamiltonian.

ZORA-def2-TZVP / SARC-ZORA-TZVP.

Correct auxiliary bases.

Valid ORCA syntax and block structure.

\\

& Postprocessing and reporting & 30\% & Extraction of \ce{^{195}Pt} isotropic shielding.

Side-by-side comparison of methods.

Identification of relativistic effect.

Quality of final report.

\\

\bottomrule
\end{tabularx}
\end{table}

\renewcommand{\arraystretch}{0.95}
\begin{table}[H]
\centering
\caption{\textbf{Per-iteration evaluation score for the benchmark question representative of the topic ``Relativistic Corrections''.}
Scores assigned to each independent execution of the benchmark task, using the rubric defined above.}
\small
\begin{tabularx}{\linewidth}{@{}
>{\centering\arraybackslash}p{0.14\linewidth}|
>{\centering\arraybackslash}p{0.13\linewidth}|
>{\centering\arraybackslash}p{0.13\linewidth}|
>{\centering\arraybackslash}p{0.13\linewidth}|
>{\centering\arraybackslash}p{0.13\linewidth}|
>{\centering\arraybackslash}p{0.14\linewidth}
@{}}
\toprule
Iteration & Planning (\%) & Geometry (\%) & Input (\%) & Postproc. (\%) & Total (\%) \\
\midrule
Run 1 & 13 & 13 & 38 & 26 & 90 \\
Run 2 & 15 & 15 & 40 & 30 & 100 \\
Run 3 & 12 & 15 & 30 & 25 & 82 \\
Run 4 & 15 & 15 & 40 & 30 & 100 \\
Run 5 & 15 & 15 & 40 & 30 & 100 \\
\midrule
\textbf{Average} & 14.0 & 14.6 & 37.6 & 28.2 & 94.4 \\
\bottomrule
\end{tabularx}
\end{table}

\begin{itemize}
    \item Run 1: The ECP and ZORA shieldings used different optimized geometries, so the reported difference is not an electronic-method-only comparison.\\
    \item Run 2: The requested branches used one validated geometry and were correctly extracted.\\
    \item Run 3: The requested def2-TZVP/ECP comparator was replaced by SARC-DKH-TZVP without a DKH Hamiltonian, making the near-cancellation result methodologically inappropriate.\\
    \item Run 4: The complete workflow used one validated geometry, both requested branches, and a Pt-specific grid.\\
    \item Run 5: The workflow recovered from an initial script error and completed both requested branches on one validated geometry.
\end{itemize}

\newpage

\subsubsection{Evaluation of the benchmark question representative of the topic ``Electrical Properties''.}

\renewcommand{\arraystretch}{0.90}
\begin{table}[H]
\centering
\caption{\textbf{Evaluation Rubric.}
The rubric used to grade the benchmark question representative of the topic ``Electrical Properties''.}
\small
\begin{tabularx}{\linewidth}{@{}
p{0.04\linewidth}|
>{\centering\arraybackslash}p{0.14\linewidth}|
c|
>{\raggedright\arraybackslash}X
@{}}
\toprule
Topic & Criterion & Weight & Task\\
\midrule

\multirow{4}{*}{ \raisebox{-20pt}[\dimexpr\height+\depth\relax]{\rotatebox[origin=c]{90}{Electrical Properties}}}

& Planning & 10\% & Generation of the initial nitrobenzene and benzene structures.

Optimization of the structure with a reasonable level (same level of theory, semiempirical quantum mechanical or efficient DFT method).

Electronic structure calculation at the MP2/aug-cc-pVDZ level of theory of all compounds to obtain the static dipole moments and polarizability tensors. \\

& Geometry generation & 10\% & Successful generation of the benzene and nitrobenzene structures.\\

& Input file generation and execution & 40\% & 
Optimization at a consistent and appropriate level of theory (same level of theory as property calculation, semiempirical quantum mechanical or efficient DFT method).

keyword line (property calculation): \texttt{! RI-MP2 aug-cc-pVDZ}  

blocks: \texttt{
\%elprop
        Dipole       true
        Polar        true
end
}.
Dipole is set by default, thus leaving it out is acceptable. \\

& Postprocessing and reporting & 40\% & Read out the dipole moment (single value as the Euclidean norm) and the dipole polarizability tensor (six values) from the outputs for both structures. \textit{Optional}: Read the isotropic polarizability.

Comparison of the absolute dipole moments of benzene and nitrobenzene and analogously, comparison of anisotropy of the static polarizability tensor.

Quality of final report, particularly regarding the analysis of how and why the electron-withdrawing nitro group changes both properties. \\

\bottomrule
\end{tabularx}
\end{table}

\renewcommand{\arraystretch}{0.95}
\begin{table}[H]
\centering
\caption{\textbf{Per-iteration evaluation score for the benchmark question representative of the topic ``Electrical Properties''.}
Scores assigned to each independent execution of the benchmark task, using the rubric defined above.}
\small
\begin{tabularx}{\linewidth}{@{}
>{\centering\arraybackslash}p{0.14\linewidth}|
>{\centering\arraybackslash}p{0.13\linewidth}|
>{\centering\arraybackslash}p{0.13\linewidth}|
>{\centering\arraybackslash}p{0.13\linewidth}|
>{\centering\arraybackslash}p{0.13\linewidth}|
>{\centering\arraybackslash}p{0.14\linewidth}
@{}}
\toprule
Iteration & Planning (\%) & Geometry (\%) & Input (\%) & Postproc. (\%) & Total (\%) \\
\midrule
Run 1 & 10 & 10 & 40 & 24 & 84 \\
Run 2 & 10 & 10 & 40 & 40 & 100 \\
Run 3 & 10 & 10 & 40 & 30 & 90 \\
Run 4 & 10 & 10 & 40 & 40 & 100 \\
Run 5 & 10 & 10 & 40 & 30 & 90 \\
\midrule
\textbf{Average} & 10.0 & 10.0 & 40.0 & 32.8 & 92.8 \\
\bottomrule
\end{tabularx}
\end{table}

\begin{itemize}
    \item Run 1: The report omitted sizable off-diagonal tensor elements and consequently computed the wrong invariant anisotropy for nitrobenzene.\\
    \item Run 2: The full tensors and invariants were correctly reported.\\
    \item Run 3: All requested scientific stages completed and receive full execution credit; the deduction is solely for omitting off-diagonal Cartesian tensor elements from the report.\\
    \item Run 4: The complete optimized-geometry workflow and full tensors were correct.\\
    \item Run 5: The report omitted the six Cartesian tensor elements; report phrasing about the response level is not separately penalized.
\end{itemize}

\newpage

\subsubsection{Evaluation of the benchmark question representative of the topic ``Orbitals and densities''.}

\renewcommand{\arraystretch}{0.90}
\begin{table}[H]
\centering
\caption{\textbf{Evaluation Rubric.}
The rubric used to grade the benchmark question representative of the topic ``Orbitals and densities ''.}
\small
\begin{tabularx}{\linewidth}{@{}
p{0.04\linewidth}|
>{\centering\arraybackslash}p{0.14\linewidth}|
c|
>{\raggedright\arraybackslash}X
@{}}
\toprule
Topic & Criterion & Weight & Task\\
\midrule

\multirow{4}{*}{ \raisebox{-80pt}[\dimexpr\height+\depth\relax]{\rotatebox[origin=c]{90}{Orbitals and densities}}}

& Planning & 10{\%} & Generation of both structures.

Geometry optimizations and plots for molecular orbitals and densities for both structures.

Repeat the previous step with a different method to correct for spin contamination (optional).

Generation of cube files from .gbw files using orca\_plot (optional). \\

& Geometry generation & 20\% & Successful generation of piperidine noxide radical geometry and pyridinium N-oxide cation radical. \\

& Input file generation and execution & 20\% & \textbf{1. Piperidine N-oxide radical optimization:} 

keyword line: ! HF cc-pVDZ Opt UNO UCO, UHF, and convergence or grid settings allowed.  
blocks (optional): \%plots Format Gaussian\_Cube dim1 80 dim2 80 dim3 80 min1 0 max1 0 min2 0  max2 0 min3 0 max3 0 SpinDens(``piperidine\_noxide\_spindens.cube''); MO(``piperidine\_noxide\_somo.cube'', 27, 0); end.

geometry section: * xyzfile 0 2 piperidine\_noxide\_radical\_initial.xyz. 

\textbf{2. pyridinium N-oxide radical cation optimization:} 

keyword line: ! HF cc-pVDZ Opt UNO UCO, UHF, and convergence or grid settings allowed. Repeating optimizations with a new method to correct for spin contamination is optional.

blocks (optional): \%plots Format Gaussian\_Cube dim1 80 dim2 80 dim3 80 min1 0 max1 0 min2 0  max2 0 min3 0 max3 0 SpinDens(``pyridinium\_noxide\_spindens.cube''); MO(``pyridinium\_noxide\_somo.cube'', 24, 0); end.

geometry section: * xyzfile 1 2 pyridinium\_noxide\_radical\_initial.xyz.\\

& Postprocessing and reporting & 50{\%} & Tracking spin contamination after optimization of both structures.

Suggesting or performing a change of method.

Orbitals cube files generation.

Densities cube files generation.

plots visualization and interpretation: spin density in N-O fragment vs delocalized over the ring.

Quality of final report. No quantitative results should be claimed for the pyridinium due to significant spin contamination.\\

\bottomrule
\end{tabularx}
\end{table}

\renewcommand{\arraystretch}{0.95}
\begin{table}[H]
\centering
\caption{\textbf{Per-iteration evaluation score for the benchmark question representative of the topic ``Orbitals and densities''.}
Scores assigned to each independent execution of the benchmark task, using the rubric defined above.}
\small
\begin{tabularx}{\linewidth}{@{}
>{\centering\arraybackslash}p{0.14\linewidth}|
>{\centering\arraybackslash}p{0.13\linewidth}|
>{\centering\arraybackslash}p{0.13\linewidth}|
>{\centering\arraybackslash}p{0.13\linewidth}|
>{\centering\arraybackslash}p{0.13\linewidth}|
>{\centering\arraybackslash}p{0.14\linewidth}
@{}}
\toprule
Iteration & Planning (\%) & Geometry (\%) & Input (\%) & Postproc. (\%) & Total (\%) \\
\midrule
Run 1 & 10 & 20 & 20 & 50 & 100 \\
Run 2 & 10 & 20 & 20 & 50 & 100 \\
Run 3 & 10 & 20 & 20 & 50 & 100 \\
Run 4 & 5 & 9 & 10 & 24 & 48 \\
Run 5 & 10 & 20 & 20 & 50 & 100 \\
\midrule
\textbf{Average} & 9.0 & 17.8 & 18.0 & 44.8 & 89.6 \\
\bottomrule
\end{tabularx}
\end{table}

\begin{itemize}
    \item Run 1: The minima, requested cubes, and visualization artifacts were correct.\\
    \item Run 2: The optimized minima, orbital selection, cube generation, and visualizations were complete and correct.\\
    \item Run 3: Both requested species were valid local minima with the correct cubes; a higher local conformer and report hypotheses were not penalized.\\
    \item Run 4: The pyridinium branch was a constitutional isomer; this identity error invalidated its optimization, properties, and the corresponding comparison.\\
    \item Run 5: The frequency-validated minima, requested cubes, and working images were complete; literature-backed hypotheses were not penalized.
\end{itemize}

\subsubsection{Evaluation of the benchmark question representative of the topic ``Condensed Fukui functions''.}

\renewcommand{\arraystretch}{0.90}
\begin{table}[H]
\centering
\caption{\textbf{Evaluation Rubric.}
The rubric used to grade the benchmark question representative of the topic ``Condensed Fukui functions''.}
\small
\begin{tabularx}{\linewidth}{@{}
p{0.04\linewidth}|
>{\centering\arraybackslash}p{0.14\linewidth}|
c|
>{\raggedright\arraybackslash}X
@{}}
\toprule
Topic & Criterion & Weight & Task\\
\midrule

\multirow{4}{*}{ \raisebox{-50pt}[\dimexpr\height+\depth\relax]{\rotatebox[origin=c]{90}{Condensed Fukui functions}}}

& Planning & 10\% & Optimization only for neutral species. 

Single point calculations for neutral, anion, and cation, at the optimized geometry of neutral toluene.\\

& Geometry generation & 10\% & Generate reference geometry. \\

& Input file generation and execution & 40\% &  \textbf{1. Neutral toluene optimization:}

keyword line: ! HF cc-pVTZ Opt Freq, other convergence or grid settings allowed.

blocks: none required. 

geometry section: * xyzfile 0 1 toluene\_initial.xyz. 

\textbf{2. Neutral toluene single point calculation} 

keyword line: ! HF cc-pVTZ Hirshfeld, other convergence or grid settings allowed. 

blocks: \%output Print[ P\_Mulliken ]  1 Print[ P\_Loewdin  ]  1 Print[ P\_Hirshfeld]  1 end. ROHF can also be specified in the main keyword line.

geometry section: * xyzfile 0 1 neutral\_toluene\_opt.xyz.  

\textbf{3. Anion toluene single point calculation} 

keyword line: ! HF cc-pVTZ Hirshfeld, other convergence or grid settings allowed. 

blocks: \%scf HFTyp ROHF end. \%output Print[ P\_Mulliken ]  1 Print[ P\_Loewdin  ]  1  Print[ P\_Hirshfeld]  1 end. ROHF can also be specified in the main keyword line.

geometry section: * xyzfile -1 2 neutral\_toluene\_opt.xyz    

\textbf{4. Cation toluene single point calculation} 

keyword line: ! HF cc-pVTZ Hirshfeld, other convergence or grid settings allowed. 

blocks: \%scf HFTyp ROHF end. \%output Print[ P\_Mulliken ]  1 Print[ P\_Loewdin  ]  1  Print[ P\_Hirshfeld]  1 end. ROHF can also be specified in the main keyword line.

geometry section: * xyzfile 1 2 neutral\_toluene\_opt.xyz  \\

& Postprocessing and reporting & 40\% & Angles and energies for minima and transition state. 

Extraction of thermodynamic properties ($\Delta G$, $\Delta H$, $\Delta S$). 

Quality of final report.\\

\bottomrule
\end{tabularx}
\end{table}

\renewcommand{\arraystretch}{0.95}
\begin{table}[H]
\centering
\caption{\textbf{Per-iteration evaluation score for the benchmark question representative of the topic ``Fukui functions''.}
Scores assigned to each independent execution of the benchmark task, using the rubric defined above.}
\small
\begin{tabularx}{\linewidth}{@{}
>{\centering\arraybackslash}p{0.14\linewidth}|
>{\centering\arraybackslash}p{0.13\linewidth}|
>{\centering\arraybackslash}p{0.13\linewidth}|
>{\centering\arraybackslash}p{0.13\linewidth}|
>{\centering\arraybackslash}p{0.13\linewidth}|
>{\centering\arraybackslash}p{0.14\linewidth}
@{}}
\toprule
Iteration & Planning (\%) & Geometry (\%) & Input (\%) & Postproc. (\%) & Total (\%) \\
\midrule
Run 1 & 10 & 8 & 37 & 27 & 82 \\
Run 2 & 10 & 10 & 40 & 40 & 100 \\
Run 3 & 10 & 10 & 40 & 40 & 100 \\
Run 4 & 10 & 10 & 40 & 40 & 100 \\
Run 5 & 10 & 10 & 40 & 40 & 100 \\
\midrule
\textbf{Average} & 10.0 & 9.6 & 39.4 & 37.4 & 96.4 \\
\bottomrule
\end{tabularx}
\end{table}

\begin{itemize}
    \item Run 1: The neutral remained a small-imaginary-mode structure; the report reversed both Fukui signs and reported a positive electron affinity where its energies give approximately $-2.91$ eV.\\
    \item Run 2: The optimized-neutral dependency, three charge states, population analysis, and final indices were correct.\\
    \item Run 3: The numerical workflow and all reported Fukui tables were complete and traceable.\\
    \item Run 4: The optimized-neutral and RHF/ROHF population workflow was complete.\\
    \item Run 5: The required states, method, optimized geometry, and final Fukui analysis were correct.
\end{itemize}

\subsubsection{Evaluation of the benchmark question representative of the topic ``Thermochemistry''.}

\renewcommand{\arraystretch}{0.90}
\begin{table}[H]
\centering
\caption{\textbf{Evaluation Rubric.}
The rubric used to grade the benchmark question representative of the topic ``Thermochemistry''.}
\small
\begin{tabularx}{\linewidth}{@{}
p{0.04\linewidth}|
>{\centering\arraybackslash}p{0.14\linewidth}|
c|
>{\raggedright\arraybackslash}X
@{}}
\toprule
Topic & Criterion & Weight & Task\\
\midrule
\multirow{4}{*}{ \raisebox{-50pt}[\dimexpr\height+\depth\relax]{\rotatebox[origin=c]{90}{Thermochemistry}}}

& Planning & 30\% & Literature search of thermal chemistry cycle and thermal dynamic data of \ce{AcOH} and proton

Geometry generation of all species.

Parallel optimization and frequency calculation + Single point calculations with implicit solvation.
\\

& Geometry generation & 10\% & Generate geometries of \ce{AcOH} and \ce{AcO-} \\

& Input file generation and execution & 30\% &  
\textbf{1. \ce{AcOH} and \ce{AcO-} optimization and frequency in gas phase:}
keyword line: B3LYP Def2-SVP OptFreq TightOpt, other reasonable functionals, equal or higher levels of basis set, other convergence or grid settings allowed.

geometry section: * xyzfile 0 1 acetic\_acid\_xtb.xyz for \ce{AcOH}, * xyzfile -1 1 acetate\_xtb.xyz for \ce{AcO-}

\textbf{2. \ce{AcOH} and \ce{AcO-} single point in gas phase:} 

keyword line: DLPNO-CCSD(T) def2-TZVPP
 TightSCF. 

geometry section: * xyzfile 0 1 acetic\_acid\_opt.xyz for \ce{AcOH}, * xyzfile -1 1 acetate\_opt.xyz for \ce{AcO-}

\textbf{3. \ce{AcOH} and \ce{AcO-} single point in gas phase:} 

keyword line: DLPNO-CCSD(T) def2-TZVPP TightSCF

blocks: \%cpcm smd true SMDsolvent ``water'' end

geometry section: * xyzfile 0 1 acetic\_acid\_opt.xyz for \ce{AcOH}, * xyzfile -1 1 acetate\_opt.xyz for \ce{AcO-}

\\

& Postprocessing and reporting & 30\% & 
Thermaldynamic cycle equation for pKa calculation.

Calculated pKa of \ce{AcOH} in gas phase and in implicit water.

Comparison of calculated pKa with literature value and rationalization of potential errors.

Quality of final report.
\\

\bottomrule
\end{tabularx}
\end{table}

\renewcommand{\arraystretch}{0.95}
\begin{table}[H]
\centering
\caption{\textbf{Per-iteration evaluation score for the benchmark question representative of the topic ``Thermochemistry''.}
Scores assigned to each independent execution of the benchmark task, using the rubric defined above.}
\small
\begin{tabularx}{\linewidth}{@{}
>{\centering\arraybackslash}p{0.14\linewidth}|
>{\centering\arraybackslash}p{0.13\linewidth}|
>{\centering\arraybackslash}p{0.13\linewidth}|
>{\centering\arraybackslash}p{0.13\linewidth}|
>{\centering\arraybackslash}p{0.13\linewidth}|
>{\centering\arraybackslash}p{0.14\linewidth}
@{}}
\toprule
Iteration & Planning (\%) & Geometry (\%) & Input (\%) & Postproc. (\%) & Total (\%) \\
\midrule
Run 1 & 25 & 10 & 30 & 25 & 90 \\
Run 2 & 25 & 10 & 30 & 21 & 86 \\
Run 3 & 25 & 10 & 30 & 22 & 87 \\
Run 4 & 30 & 10 & 30 & 18 & 88 \\
Run 5 & 30 & 10 & 23 & 18 & 81 \\
\midrule
\textbf{Average} & 27.0 & 10.0 & 28.6 & 20.8 & 86.4 \\
\bottomrule
\end{tabularx}
\end{table}

\begin{itemize}
    \item Run 1: SMD was incorporated through a defensible DFT solvation correction, but the resulting pKa of 8.48 remained too high.\\
    \item Run 2: SMD was incorporated through a defensible B3LYP-D4 solvation correction, but the reported pKa was 11.66.\\
    \item Run 3: SMD was applied directly in the DLPNO-CCSD(T) solvent calculations, but the pKa of 9.95 remained substantially inaccurate.\\
    \item Run 4: SMD was incorporated through a defensible DFT correction, but the direct pKa was 10.54 and the semidirect route gave an internally inconsistent 21.93.\\
    \item Run 5: The solvent workflow used CPCM rather than the literature-supported SMD treatment and produced a pKa of 13.02.
\end{itemize}

\subsubsection{Evaluation of the benchmark question representative of the topic ``Conformer Generation''.}

\renewcommand{\arraystretch}{0.90}
\begin{table}[H]
\centering
\caption{\textbf{Evaluation Rubric.}
The rubric used to grade the benchmark question representative of the topic ``Conformer Generation''.}
\small
\begin{tabularx}{\linewidth}{@{}
p{0.04\linewidth}|
>{\centering\arraybackslash}p{0.14\linewidth}|
c|
>{\raggedright\arraybackslash}X
@{}}
\toprule
Topic & Criterion & Weight & Task\\
\midrule

\multirow{4}{*}{ \raisebox{-60pt}[\dimexpr\height+\depth\relax]{\rotatebox[origin=c]{90}{Conformer Generation}}}

& Planning & 20\% & Generation of the initial Serine structure.

Conformer search using an appropriate tool such as CREST or GOAT.

Parsing of the conformers as single three-dimensional structures.

DFT reoptimization of each of the structures.

\textit{Optional}: Check for conformers that fell into the same local minimum (similar total energy and virtually zero structural RMSD).

Re-calculate the Boltzmann weights for the conformer ensemble and report the lowest-energy structure together with the total ensemble. \\

& Geometry generation & 20\% & Successful generation of the serine geometry. L-serine should be chosen, as it's the biologically active form and is usually used. The protonation state can be either, but in the gas phase, it's the neutral form. \\

& Input file generation and execution & 30\% & \textbf{1. Serine conformer generation:} 

keyword line: \texttt{! XTB GOAT}  

blocks (optional): Specifications about GOAT, which might be reasonable. Not critical for the run.

\textbf{2. Re-optimization of the structures on the DFT level:} 

keyword line: \texttt{! B3LYP D4} (or \texttt{D3(BJ)}), other convergence or grid settings allowed. 
{\cinzel Quntur} shall either include the dispersion correction as suggested or inform the user that it's missing.
The basis set can be explicitly stated or not, since it is the default anyway.\\

& Postprocessing and reporting & 30\% & \textit{Between step 1 and 2:} Parsing of the conformers as single three-dimensional structures and preparing distinct calculations for them.
\textit{Optional}: Set an upper threshold for high-energy conformers.

\textit{After step 2:} Reading out DFT energies from ORCA. 

\textit{Optional}: Check for conformers that fell into the same local minimum (similar total energy and virtually zero structural RMSD).

Re-calculate Boltzmann weights for the conformer ensemble. 

Quality of the final report, particularly regarding the Boltzmann ensemble and the lowest-energy structure. \\

\bottomrule
\end{tabularx}
\end{table}

\renewcommand{\arraystretch}{0.95}
\begin{table}[H]
\centering
\caption{\textbf{Per-iteration evaluation score for the benchmark question representative of the topic ``Conformer Generation''.}
Scores assigned to each independent execution of the benchmark task, using the rubric defined above.}
\small
\begin{tabularx}{\linewidth}{@{}
>{\centering\arraybackslash}p{0.14\linewidth}|
>{\centering\arraybackslash}p{0.13\linewidth}|
>{\centering\arraybackslash}p{0.13\linewidth}|
>{\centering\arraybackslash}p{0.13\linewidth}|
>{\centering\arraybackslash}p{0.13\linewidth}|
>{\centering\arraybackslash}p{0.14\linewidth}
@{}}
\toprule
Iteration & Planning (\%) & Geometry (\%) & Input (\%) & Postproc. (\%) & Total (\%) \\
\midrule
Run 1 & 20 & 20 & 30 & 22 & 92 \\
Run 2 & 20 & 20 & 30 & 19 & 89 \\
Run 3 & 20 & 20 & 27 & 19 & 86 \\
Run 4 & 20 & 20 & 30 & 23 & 93 \\
Run 5 & 20 & 20 & 30 & 30 & 100 \\
\midrule
\textbf{Average} & 20.0 & 20.0 & 29.4 & 22.6 & 92.0 \\
\bottomrule
\end{tabularx}
\end{table}

\begin{itemize}
    \item Run 1: Exact duplicate DFT minima materially biased the reported populations; the defensible no-dispersion choice was not penalized.\\
    \item Run 2: The global minimum was correct, but exact-copy pairs double-counted each of the two leading conformer families and invalidated their reported populations.\\
    \item Run 3: Duplicate leaders were counted separately and the search missed the repeated same-level minimum by approximately 1.62 kcal/mol.\\
    \item Run 4: The minimum and imaginary-mode recovery were correct, but one optimized conformer family was counted three times.\\
    \item Run 5: The completed cutoff, frequency, Gibbs-population, and duplicate-free refinement workflow received full credit.
\end{itemize}

\subsubsection{Evaluation of the benchmark question representative of the topic: ``Molecular Dynamics''.}

\renewcommand{\arraystretch}{0.90}
\begin{table}[H]
\centering
\caption{\textbf{Evaluation Rubric.}
The rubric used to grade the benchmark question representative of the topic ``Molecular Dynamics''.}
\small
\begin{tabularx}{\linewidth}{@{}
p{0.04\linewidth}|
>{\centering\arraybackslash}p{0.14\linewidth}|
c|
>{\raggedright\arraybackslash}X
@{}}
\toprule
Topic & Criterion & Weight & Task\\
\midrule

\multirow{4}{*}{ \raisebox{-40pt}[\dimexpr\height+\depth\relax]{\rotatebox[origin=c]{90}{Molecular Dynamics}}}

& Planning & 10\% & Generating an initial structure for the Chloride ion, along with 12 water molecules

Performing the geometry optimization using GFN2-xTB

Performing AIMD simulations with short equilibration and then production 

Extracting the AIMD trajectory and then analysing the radial distribution function of Cl and O
 \\

& Geometry generation & 20\% & Generate physically reasonable initial geometry for hydrated chloride ion cluster, with correct charge, and spin multiplicity. 

\\

& Input file generation and execution & 30\% &

\textbf{1.  Geometry relaxation \ce{Cl^{-}(H2O)12}:} 

Check if the agent used the xTB tool to perform geometry relaxation.

geometry section: *{xyz file -1 1 cl\_12water\_initial.xyz}

\textbf{2. AIMD \ce{Cl^{-}(H2O)12}:} 

keyword line: \texttt{! XTB2 MD}

block: \% md Timestep = 0.5 for step size 

          Initvel 300\_K  for velocity initialization  
          
          Thermostat settings for equilibration

          Run the equilibration command with the specified number of steps. 

          Thermostat settings for production 
          
          Dump settings for storing production trajectory ``prod\_traj.xyz'' 
          
          Run command for 0.1 ps production.  

geometry section: * xyz file -1 1 cl\_12water\_optimized.xyz

\\

& Postprocessing and reporting & 40\% &

Correct extraction of the production MD trajectories. Computing the radial distribution function of Cl and O.

Results should be presented in a final markdown report that is comprehensive and easy to interpret with primary literature sources. 

\\

\bottomrule
\end{tabularx}
\end{table}

\renewcommand{\arraystretch}{0.95}
\begin{table}[H]
\centering
\caption{\textbf{Per-iteration evaluation score for the benchmark question representative of the topic ``Molecular Dynamics''.}
Scores assigned to each independent execution of the benchmark task, using the rubric defined above.}
\small
\begin{tabularx}{\linewidth}{@{}
>{\centering\arraybackslash}p{0.14\linewidth}|
>{\centering\arraybackslash}p{0.13\linewidth}|
>{\centering\arraybackslash}p{0.13\linewidth}|
>{\centering\arraybackslash}p{0.13\linewidth}|
>{\centering\arraybackslash}p{0.13\linewidth}|
>{\centering\arraybackslash}p{0.14\linewidth}
@{}}
\toprule
Iteration & Planning (\%) & Geometry (\%) & Input (\%) & Postproc. (\%) & Total (\%) \\
\midrule
Run 1 & 7 & 17 & 17 & 12 & 53 \\
Run 2 & 5 & 5 & 15 & 31 & 56 \\
Run 3 & 7 & 13 & 22 & 32 & 74 \\
Run 4 & 8 & 18 & 22 & 27 & 75 \\
Run 5 & 10 & 18 & 29 & 40 & 97 \\
\midrule
\textbf{Average} & 7.4 & 14.2 & 21.0 & 28.4 & 71.0 \\
\bottomrule
\end{tabularx}
\end{table}

\begin{itemize}
    \item Run 1: The cluster dissociated before production, averaged approximately 391 K, and the report substituted early equilibration frames for the requested production RDF and coordination number.\\
    \item Run 2: No optimization was performed and the initial shell was crowded and O-in; equilibration recovered a useful shell, but production averaged approximately 345 K.\\
    \item Run 3: The retained initial and optimized geometries were identical with no optimization output, MD referenced the initial structure, and the reported first RDF minimum was too distant.\\
    \item Run 4: The optimization converged, but equilibration was only 100 fs, production averaged approximately 403 K, and the report did not disclose the thermal failure.\\
    \item Run 5: This was the strongest two-stage workflow near 305 K; the limited deduction reflects retained optimization provenance, not finite-cluster limitations intrinsic to the prompt.
\end{itemize}

\subsubsection{Evaluation of the benchmark question representative of the topic ``Explicit Solvation''.}

\renewcommand{\arraystretch}{0.90}
\begin{table}[H]
\centering
\caption{\textbf{Evaluation Rubric.}
The rubric used to grade the benchmark question representative of the topic ``Explicit Solvation''.}
\small
\begin{tabularx}{\linewidth}{@{}
p{0.04\linewidth}|
>{\centering\arraybackslash}p{0.14\linewidth}|
c|
>{\raggedright\arraybackslash}X
@{}}
\toprule
Topic & Criterion & Weight & Task\\
\midrule

\multirow{4}{*}{ \raisebox{-40pt}[\dimexpr\height+\depth\relax]{\rotatebox[origin=c]{90}{Explicit Solvation}}}

& Planning & 10\% & Generating an initial structure for AcOH(H\textsubscript{2}O)\textsubscript{3}, AcOH(H\textsubscript{2}O)\textsubscript{5}, AcOH(H\textsubscript{2}O)\textsubscript{7}, AcO\textsuperscript{-}(H\textsubscript{2}O)\textsubscript{3}, AcO\textsuperscript{-}(H\textsubscript{2}O)\textsubscript{5}, AcO\textsuperscript{-}(H\textsubscript{2}O)\textsubscript{7}

Perform geometry optimizations using r2SCAN-3c, with and without an implicit CPCM solvent model.

Extract Gibbs free energies from all calculations and compute pKa values for each cluster size in both gas-phase and CPCM environments, analyzing convergence toward experimental pKa values.

Generate a summary report that includes pKa convergence plots across different cluster sizes, with direct comparisons between gas-phase and CPCM results as well as against literature values.

\\

& Geometry generation & 30\% & Generate physically reasonable initial geometry for the solvation of Acetic Acid by Water Molecules and implicit solvent, with correct charge, and spin multiplicity. 

\\

& Input file generation and execution & 30\% &

\textbf{1.  Geometry relaxation AcOH/AcO\textsuperscript{-}(H\textsubscript{2}O)\textsubscript{n}:} 

keyword line: ! r2SCAN-3c TightOpt FREQ

geometry section: *{xyz file -1 1 AcO\_3water\_initial.xyz}, *{xyz file -1 1 AcO\_5water\_initial.xyz}, *{xyz file -1 1 AcO\_7water\_initial.xyz}, *{xyz file 0 1 AcOH\_3water\_initial.xyz} *{xyz file 0 1 AcOH\_5water\_initial.xyz} *{xyz file 0 1 AcOH\_7water\_initial.xyz}

\\

& Postprocessing and reporting & 30\% &

Electronic energies and thermal corrections obtained from frequency calculations were combined to yield Gibbs free energies, which were then used to compute deprotonation free energies and pKa values.

Results should be presented in a final markdown report that is comprehensive and easy to interpret with primary literature sources. 

\\

\bottomrule
\end{tabularx}
\end{table}

\renewcommand{\arraystretch}{0.95}
\begin{table}[H]
\centering
\caption{\textbf{Per-iteration evaluation score for the benchmark question representative of the topic ``Explicit Solvation''.}
Scores assigned to each independent execution of the benchmark task, using the rubric defined above.}
\small
\begin{tabularx}{\linewidth}{@{}
>{\centering\arraybackslash}p{0.14\linewidth}|
>{\centering\arraybackslash}p{0.13\linewidth}|
>{\centering\arraybackslash}p{0.13\linewidth}|
>{\centering\arraybackslash}p{0.13\linewidth}|
>{\centering\arraybackslash}p{0.13\linewidth}|
>{\centering\arraybackslash}p{0.14\linewidth}
@{}}
\toprule
Iteration & Planning (\%) & Geometry (\%) & Input (\%) & Postproc. (\%) & Total (\%) \\
\midrule
Run 1 & 10 & 20 & 30 & 30 & 90 \\
Run 2 & 10 & 15 & 30 & 30 & 85  \\
Run 3 & 10 & 20 & 30 & 30 & 90  \\
Run 4 & 10 & 15 & 30 & 30 & 85 \\
Run 5 & 10 & 15 & 30 & 30 & 85 \\
\midrule
\textbf{Average} & 10 & 17 & 30 & 30 & 87 \\
\bottomrule
\end{tabularx}
\end{table}

\begin{itemize}
    \item Run 1: \\ The agent first performed a geometry optimization using GFN2-xTB, followed by refinement with r2SCAN-3c. When the r2SCAN-3c optimization reached the maximum number of iterations, the agent successfully restarted the calculation. Gibbs free energies were extracted from the output files and used to compute deprotonation free energies and the corresponding pKa values. However, for the initial configuration, the agent usually failed to generate the lowest-energy structure of AcOH, which negatively affected the accuracy of the free energy calculations.
    \item Run 2: \\ The agent first performed a geometry optimization using GFN2-xTB, followed by refinement with r2SCAN-3c. When the r2SCAN-3c optimization reached the maximum number of iterations, the agent successfully restarted the calculation. Gibbs free energies were extracted from the output files and used to compute deprotonation free energies and the corresponding pKa values. However, for the initial configuration, the agent always failed to generate the lowest-energy structure of AcOH, which negatively affected the accuracy of the free energy calculations.
    \item Run 3: \\The agent first performed a geometry optimization using GFN2-xTB, followed by refinement with r2SCAN-3c. When the r2SCAN-3c optimization reached the maximum number of iterations, the agent successfully restarted the calculation. Gibbs free energies were extracted from the output files and used to compute deprotonation free energies and the corresponding pKa values. However, for the initial configuration, the agent usually failed to generate the lowest-energy structure of AcOH, which negatively affected the accuracy of the free energy calculations.
    \item Run 4: \\ The agent first performed a geometry optimization using GFN2-xTB, followed by refinement with r2SCAN-3c. When the r2SCAN-3c optimization reached the maximum number of iterations, the agent successfully restarted the calculation. Gibbs free energies were extracted from the output files and used to compute deprotonation free energies and the corresponding pKa values. However, for the initial configuration, the agent always failed to generate the lowest-energy structure of AcOH, which negatively affected the accuracy of the free energy calculations.
    \item Run 5:  \\ The agent first performed a geometry optimization using GFN2-xTB, followed by refinement with r2SCAN-3c. When the r2SCAN-3c optimization reached the maximum number of iterations, the agent successfully restarted the calculation. Gibbs free energies were extracted from the output files and used to compute deprotonation free energies and the corresponding pKa values. However, for the initial configuration, the agent always failed to generate the lowest-energy structure of AcOH, which negatively affected the accuracy of the free energy calculations.
\end{itemize}

\subsubsection{Evaluation of the benchmark question representative of the topic ``Potential Energy Surfaces''.}

\renewcommand{\arraystretch}{0.90}
\begin{table}[H]
\centering
\caption{\textbf{Evaluation Rubric.}
The rubric used to grade the benchmark question representative of the topic ``Potential Energy Surfaces''.}
\small
\begin{tabularx}{\linewidth}{@{}
p{0.04\linewidth}|
>{\centering\arraybackslash}p{0.14\linewidth}|
c|
>{\raggedright\arraybackslash}X
@{}}
\toprule
Topic & Criterion & Weight & Task\\
\midrule

\multirow{4}{*}{ \raisebox{-30pt}[\dimexpr\height+\depth\relax]{\rotatebox[origin=c]{90}{Potential Energy Surfaces}}}

& Planning & 10\% & Complete dihedral relaxed scan between 0 and 180$^{0}$, no optimization needed.

Optimizations for each stationary point are optional.

Rigid scan instead of a relaxed scan (half credit, no credit if no optimization was run before rigid scan). \\

& Geometry generation & 20\% & Generate reference geometry (trans or cis)

Correct choice of minima and top-of-barrier geometries. \\

& Input file generation and execution & 40\% &  \textbf{1. Butadiene trans optimization (optional):} 

keyword line: CASSCF cc-pVDZ TightOpt NumFreq, other convergence or grid settings allowed. 

blocks: \%casscf  nel 4 norb 4 mult 1 nroots 1 end. 

geometry section: * xyzfile 0 1 butadiene\_initial.xyz (it can use trans or cis initial geometry; it can also use inline or internal coordinates). 

This step is not necessary when geom scan is used.

\textbf{2. Butadiene PES scan:}

keyword line: ! CASSCF cc-pVDZ Opt, other convergence or grid settings allowed. 

blocks: \%casscf  nel 4 norb 4 mult 1 nroots 1 end. \%geom Scan D 0 1 2 3 = 180.0, 0.0, 19 end end. The number of grid points must be reasonable. Scan directionality must be consistent with the initial geometry.

geometry section: * xyzfile 0 1 butadiene\_initial.xyz (again, cis or trans). 

\textbf{3. Butadiene single-point calculations (optional):} 

keyword line: ! CASSCF cc-pVDZ, other convergence or grid settings allowed. 

blocks: \%casscf  nel 4 norb 4 mult 1 nroots 1 end. 

geometry section: * xyzfile 0 1 some\_geometry.xyz (to replace scan).\\

& Postprocessing and reporting & 30\% & Angles and energies for minima (cis is not planar but gauche). 

Energy barrier along dihedral. 

Quality of final report.\\

\bottomrule
\end{tabularx}
\end{table}

\renewcommand{\arraystretch}{0.95}
\begin{table}[H]
\centering
\caption{\textbf{Per-iteration evaluation score for the benchmark question representative of the topic ``Potential Energy surfaces''.}
Scores assigned to each independent execution of the benchmark task, using the rubric defined above.}
\small
\begin{tabularx}{\linewidth}{@{}
>{\centering\arraybackslash}p{0.14\linewidth}|
>{\centering\arraybackslash}p{0.13\linewidth}|
>{\centering\arraybackslash}p{0.13\linewidth}|
>{\centering\arraybackslash}p{0.13\linewidth}|
>{\centering\arraybackslash}p{0.13\linewidth}|
>{\centering\arraybackslash}p{0.14\linewidth}
@{}}
\toprule
Iteration & Planning (\%) & Geometry (\%) & Input (\%) & Postproc. (\%) & Total (\%) \\
\midrule
Run 1 & 10 & 20 & 40 & 30 & 100 \\
Run 2 & 10 & 20 & 40 & 30 & 100  \\
Run 3 & 10 & 20 & 40 & 30 & 100  \\
Run 4 & 10 & 20 & 40 & 30 & 100 \\
Run 5 & 10 & 20 & 40 & 30 & 100 \\
\midrule
\textbf{Average} & 10 & 20 & 40 & 30 & 100 \\
\bottomrule
\end{tabularx}
\end{table}

\begin{itemize}
    \item Run 1: The scan failed initially because the agent started from an s-trans (180$^{0}$) geometry, but tried to scan from 0$^{0}$ to 180$^{0}$ (instead of from 180$^{0}$ to 0$^{0}$). The agent found the error and fixed it, so no points were deducted.\\
    \item Run 2: The scan was set up properly. The agent also performed optimizations for each stationary point, finding the cis, trans, and main transition states. These calculations are optional and are not graded.\\
    \item Run 3: Same as Run 1.\\
    \item Run 4: Same as Run 1\\
    \item Run 5: Same issue as Run 1 with scan directionality. During debugging, the agent avoided including the trans configuration, but obtained it later via geometry optimization.
\end{itemize}

\subsubsection{Evaluation of the benchmark question representative of the topic ``Transition State Methods''.}

\renewcommand{\arraystretch}{0.90}
\begin{table}[H]
\centering
\caption{\textbf{Evaluation Rubric.}
The rubric used to grade the benchmark question representative of the topic ``Transition State Methods''.}
\small
\begin{tabularx}{\linewidth}{@{}
p{0.04\linewidth}|
>{\centering\arraybackslash}p{0.14\linewidth}|
c|
>{\raggedright\arraybackslash}X
@{}}
\toprule
Topic & Criterion & Weight & Task\\
\midrule

\multirow{4}{*}{ \raisebox{-80pt}[\dimexpr\height+\depth\relax]{\rotatebox[origin=c]{90}{Transition State Methods}}}

& Planning & 20\% & Optimization of reaction substrate and product.

Transition state search using NEB.

Transition state localization using OptTS.

Transition state verification using IRC.\\

& Geometry generation & 10\% & Generate reference geometry with correct atom order and symmetry. \\

& Input file generation and execution & 40\% &  \textbf{1. Substrate and product geometry optimization:}

keyword line: ! PBE0 D3BJ Def2-TZVP Opt Freq,  other convergence or grid settings allowed.

blocks: none required. 

geometry section: * xyzfile 0 1 formic\_acid\_dimer\_reactant.xyz for reactant, * xyzfile 0 1 formic\_acid\_dimer\_product.xyz for product.

\textbf{2. NEB calculation for transition state localization} 

keyword line: ! PBE0 D3BJ Def2-TZVP NEB, other convergence or grid settings allowed. Must be consistent with the previous step. NEB-CI is valid.

blocks: \%neb Product ``product.xyz''  NImages 8 Interpolation IDPP MaxIter 500 ConvType all Tol\_Scale 10.0 PrintLevel 1 end. Other settings on NImages, tolerances, and MaxIter are valid if reasonable. If NEB-CI is chosen, it's better to set Tol\_Turn\_On\_CI.
  
geometry section: * xyzfile 0 1 reactant.xyz  

\textbf{3. Transition state geometry optimization calculation} 

keyword line: ! PBE0 D3BJ Def2-TZVP OptTS, other convergence or grid settings allowed. Must be consistent with the previous step. 

blocks: \%geom Calc\_Hess true end

geometry section: * xyzfile 0 1 TS\_guess.xyz   

\textbf{4. Transition state verification via IRC} 

keyword line: ! PBE0 D3BJ Def2-TZVP IRC,  other convergence or grid settings allowed. Must be consistent with the previous step.  

blocks: \%irc Direction both InitHess read Hess\_Filename   ``TS.hess'' MaxIter 50 end. Other settings, such as Init\_Displ, DE\_Init\_Displ, TolMaxG, TolRMSG, etc., are allowed if reasonable. Calc\_Hess True, valid, but lower marks.

geometry section: * xyzfile 0 1 TS.xyz \\

& Postprocessing and reporting & 30\% & NEB-CI calculation converged.

Imaginary freq verified. 

Backward and forward IRC paths verified.

Extraction of thermodynamic properties ($\Delta G$, $\Delta H$, $\Delta S$). 

Quality of final report.\\

\bottomrule
\end{tabularx}
\end{table}

\renewcommand{\arraystretch}{0.95}
\begin{table}[H]
\centering
\caption{\textbf{Per-iteration evaluation score for the benchmark question representative of the topic ``Transition State Methods''.}
Scores assigned to each independent execution of the benchmark task, using the rubric defined above.}
\small
\begin{tabularx}{\linewidth}{@{}
>{\centering\arraybackslash}p{0.14\linewidth}|
>{\centering\arraybackslash}p{0.13\linewidth}|
>{\centering\arraybackslash}p{0.13\linewidth}|
>{\centering\arraybackslash}p{0.13\linewidth}|
>{\centering\arraybackslash}p{0.13\linewidth}|
>{\centering\arraybackslash}p{0.14\linewidth}
@{}}
\toprule
Iteration & Planning (\%) & Geometry (\%) & Input (\%) & Postproc. (\%) & Total (\%) \\
\midrule
Run 1 & 20 & 10 & 40 & 30 & 100 \\
Run 2 & 20 & 10 & 40 & 30 & 100  \\
Run 3 & 20 & 10 & 38 & 30 & 98  \\
Run 4 & 20 & 10 & 40 & 30 & 100 \\
Run 5 & 20 & 10 & 40 & 30 & 100 \\
\midrule
\textbf{Average} & 20 & 10 & 39.6 & 30 & 99.6 \\
\bottomrule
\end{tabularx}
\end{table}

\begin{itemize}
    \item Run 1: No issues observed.\\
    \item Run 2: No issues observed.\\
    \item Run 3: Used NEB-CI but did not specify a threshold.\\
    \item Run 4: No issues observed.\\
    \item Run 5: No issues observed.
\end{itemize}

\subsubsection{Evaluation of the benchmark question representative of the topic ``Reaction Mechanisms''.}

\renewcommand{\arraystretch}{0.90}
\begin{table}[H]
\centering
\caption{\textbf{Evaluation Rubric.}
The rubric used to grade the benchmark question representative of the topic ``Reaction Mechanisms''.}
\small
\begin{tabularx}{\linewidth}{@{}
p{0.04\linewidth}|
>{\centering\arraybackslash}p{0.14\linewidth}|
c|
>{\raggedright\arraybackslash}X
@{}}
\toprule
Topic & Criterion & Weight & Task\\
\midrule

\multirow{4}{*}{ \raisebox{-20pt}[\dimexpr\height+\depth\relax]{\rotatebox[origin=c]{90}{Reaction Mechanisms}}}

& Planning & 20\% & Optimizes encounter complex and TS followed by IRC on TS structure 
\bigskip
\bigskip
\\
& Geometry generation & 20\% & Generates correct encounter complex and TS guess 
\bigskip
\bigskip
\\
& Input file generation and execution & 20\% &  Key input file settings across all calculations (i.e. functional/basis set, syntax) and for each type of calculation (i.e. ``opt'' and ``freq'' for encounter complex, ``optts'' and ``freq'' for transition state, ``irc'' for IRC) 
\bigskip
\bigskip
\\

& Postprocessing and reporting & 40\% & 
Confirms presence of 1 imaginary frequency that is sufficiently large (10\%)

\medskip

Report generated and contains $\Delta{G}^{\ddagger}$, $\Delta{H}^{\ddagger}$, $\Delta{H}^{\ddagger}$ (10
\%)

\medskip

Overall quality of report (interpretation of results, comparison with literature, formatting, etc., 10\%)

\medskip

Accuracy of results (values of G, H, S  from the output files for encounter complex and TS, reported $\Delta{G}^{\ddagger}$, $\Delta{H}^{\ddagger}$, $\Delta{H}^{\ddagger}$; each compared to a set of manual calculations and weighed equally to form a total of 10\%)

\bigskip
\bigskip

\\

\bottomrule
\end{tabularx}
\end{table}

\renewcommand{\arraystretch}{0.95}
\begin{table}[H]
\centering
\caption{\textbf{Per-iteration evaluation score for the benchmark question representative of the topic ``Reaction Mechanisms''.}
Scores assigned to each independent execution of the benchmark task, using the rubric defined above.}
\small
\begin{tabularx}{\linewidth}{@{}
>{\centering\arraybackslash}p{0.14\linewidth}|
>{\centering\arraybackslash}p{0.13\linewidth}|
>{\centering\arraybackslash}p{0.13\linewidth}|
>{\centering\arraybackslash}p{0.13\linewidth}|
>{\centering\arraybackslash}p{0.13\linewidth}|
>{\centering\arraybackslash}p{0.14\linewidth}
@{}}
\toprule
Iteration & Planning (\%) & Geometry (\%) & Input (\%) & Postproc. (\%) & Total (\%) \\
\midrule
Run 1 & 20 & 20 & 20 & 40 & 100 \\
Run 2 & 20 & 20 & 20 & 40 & 100 \\
Run 3 & 20 & 20 & 20 & 40 & 100 \\
Run 4 & 10 & 10 & 10 & 23 & 53 \\
Run 5 & 10 & 10 & 14 & 23 & 57 \\
\midrule
\textbf{Average} & 16.0 & 16.0 & 16.8 & 33.2 & 82.0 \\
\bottomrule
\end{tabularx}
\end{table}

\begin{itemize}
    \item Run 1: The encounter-complex, transition-state, IRC, and activation-thermochemistry workflow was complete.\\
    \item Run 2: The complete workflow used the defensible RI-MP2/cc-pVTZ and cc-pVTZ/C implementation.\\
    \item Run 3: The Compound workflow correctly optimized the encounter complex and transition state and supported the reported activation quantities.\\
    \item Run 4: No encounter complex was generated, so separated fragments were used as the activation reference and the central thermochemistry was incorrect.\\
    \item Run 5: The encounter complex was again absent, so the report used incorrect separated-fragment activation values; unnecessary Hessian machinery further reduced methodological clarity. The optional CSV-writing failure is not penalized.
\end{itemize}

\subsubsection{Evaluation of the benchmark question representative of the topic ``Isotopic Effects''.}

\renewcommand{\arraystretch}{0.90}
\begin{table}[H]
\centering
\caption{\textbf{Evaluation Rubric.}
The rubric used to grade the benchmark question representative of the topic ``Isotopic Effects''.}
\small
\begin{tabularx}{\linewidth}{@{}
p{0.04\linewidth}|
>{\centering\arraybackslash}p{0.14\linewidth}|
c|
>{\raggedright\arraybackslash}X
@{}}
\toprule
Topic & Criterion & Weight & Task\\
\midrule

\multirow{4}{*}{ \raisebox{-80pt}[\dimexpr\height+\depth\relax]{\rotatebox[origin=c]{90}{Isotopic Effects}}}

& Planning & 10\% & geometry generation of benzyl chlorides and carbocations. TS optimizations are allowed, but they are significantly more complex.

Optimizations only for non-deuterated species.

recalculation of thermodynamic quantities for deuterated species from non-deuterated Hessians.\\

& Geometry generation & 30\% & Geometry generation of benzyl chlorides.

Geometry generation of carbocations. \\

& Input file generation and execution & 30\% &  \textbf{1. geometry optimizations:}

keyword line: ! HF 6-31G(d) OPT FREQ CPCM(Water), other convergence or grid settings allowed. Other solvent models, such as SMD, are allowed.

blocks: No blocks are required.

geometry section: * xyzfile 0 1 benzyl\_chloride\_initial.xyz. * xyzfile 1 1 benzyl\_cation\_initial.xyz.

\textbf{2. Thermochemistry of deuterated species:}

keyword line: ! PRINTTHERMOCHEM.

blocks: \%geom INHESSNAME ``benzyl\_cation\_opt.hess'' end. 

geometry section: * xyzfile 0 1 benzyl\_chloride\_initial.xyz. * xyzfile 1 1 benzyl\_cation\_initial.xyz. Atomic mass change for deuterated species via ``M 2.00141''.

\textbf{3. Transition state optimization (optional):}

keyword line: ! HF 6-31G(d) OptTS Freq CPCM(water).

blocks: \%geom Calc\_Hess true TS\_Active\_Atoms {0 1} end modify\_internal { B 0 1 A } end. 

geometry section: * xyzfile 0 1 benzyl\_chloride\_initial.xyz.\\

& Postprocessing and reporting & 30\% & Extraction of Final Gibbs free energy. 

Calculation of Final Gibbs free energies for deuterated species using thermal correction from the PRINTTHERMOCHEM calculation.

Calculation of KIE.

Quality of final report.\\

\bottomrule
\end{tabularx}
\end{table}

\renewcommand{\arraystretch}{0.95}
\begin{table}[H]
\centering
\caption{\textbf{Per-iteration evaluation score for the benchmark question representative of the topic ``Isotopic Effects''.}
Scores assigned to each independent execution of the benchmark task, using the rubric defined above.}
\small
\begin{tabularx}{\linewidth}{@{}
>{\centering\arraybackslash}p{0.14\linewidth}|
>{\centering\arraybackslash}p{0.13\linewidth}|
>{\centering\arraybackslash}p{0.13\linewidth}|
>{\centering\arraybackslash}p{0.13\linewidth}|
>{\centering\arraybackslash}p{0.13\linewidth}|
>{\centering\arraybackslash}p{0.14\linewidth}
@{}}
\toprule
Iteration & Planning (\%) & Geometry (\%) & Input (\%) & Postproc. (\%) & Total (\%) \\
\midrule
Run 1 & 10 & 30 & 30 & 30 & 100 \\
Run 2 & 10 & 30 & 30 & 30 & 100 \\
Run 3 & 10 & 30 & 30 & 25 & 95 \\
Run 4 & 10 & 30 & 10 & 18 & 68 \\
Run 5 & 10 & 30 & 30 & 28 & 98 \\
\midrule
\textbf{Average} & 10.0 & 30.0 & 26.0 & 26.2 & 92.2 \\
\bottomrule
\end{tabularx}
\end{table}

\begin{itemize}
    \item Run 1: The complete isotope and Gibbs-thermochemistry workflow received full credit.\\
    \item Run 2: The complete Gibbs workflow and imaginary-mode recovery were correct and received full credit.\\
    \item Run 3: Full Gibbs corrections existed but the report used only ZPE estimates and parsed initial rather than final optimization energies.\\
    \item Run 4: Only the methoxy-substituted transition state yielded a usable KIE, and the report conflated separate frequency blocks.\\
    \item Run 5: The fixed-geometry isotope results supported the KIEs; the deduction is confined to parsing initial electronic energies.
\end{itemize}

\subsubsection{Evaluation of the benchmark question representative of the topic ``Excited States''.}

\renewcommand{\arraystretch}{0.90}
\begin{table}[H]
\centering
\caption{\textbf{Example of Evaluation Rubric.}
The rubric used to grade the benchmark question representative of the topic ``Excited States''.}
\small
\begin{tabularx}{\linewidth}{@{}
p{0.04\linewidth}|
>{\centering\arraybackslash}p{0.14\linewidth}|
c|
>{\raggedright\arraybackslash}X
@{}}
\toprule
Topic & Criterion & Weight  & Task\\
\midrule

\multirow{4}{*}{ \raisebox{-30pt}[\dimexpr\height+\depth\relax]{\rotatebox[origin=c]{90}{Excited States}}}

& Planning & 20\% &
Identify the required workflow: ground-state geometry optimization followed by TDDFT excited-state calculations. Correct selection of the two functionals (PBE0 and $\omega$B97X-D) with the def2-TZVP basis set. Clear definition of the quantities to be compared, namely, vertical excitation energies. \\

& Geometry generation & 10\% &
Generate a physically reasonable equilibrium geometry for p-nitroaniline in its ground electronic state, with correct molecular structure, charge, and spin multiplicity. Ground-state optimization at the DFT level is expected and should be reused consistently for TDDFT calculations. \\

& Input file generation and execution & 30\% &
\textbf{1. Ground-state optimization:} keyword line using either PBE0 or $\omega$B97X-D with the def2-TZVP basis set, including correct charge and spin multiplicity.

\textbf{2. TDDFT excitation energies:} TDDFT calculations performed on the optimized ground-state geometry of the appropriate functional, with the same basis set and a consistent number of requested excited states. \\

& Postprocessing & 40\% &
Correct extraction of excitation energies from both TDDFT calculations. Clear tabulation and comparison of the results, and discussion of trends between the two functionals. Results should be presented in a final markdown report with brief physical interpretation and comments on methodological limitations. \\

\bottomrule
\end{tabularx}
\end{table}

\renewcommand{\arraystretch}{0.95}
\begin{table}[H]
\centering
\caption{\textbf{Per-iteration evaluation score for the benchmark question representative of the topic ``Excited States''.}
Scores assigned to each independent execution of the benchmark task, using the rubric defined above.}
\small
\begin{tabularx}{\linewidth}{@{}
>{\centering\arraybackslash}p{0.14\linewidth}|
>{\centering\arraybackslash}p{0.13\linewidth}|
>{\centering\arraybackslash}p{0.13\linewidth}|
>{\centering\arraybackslash}p{0.13\linewidth}|
>{\centering\arraybackslash}p{0.13\linewidth}|
>{\centering\arraybackslash}p{0.14\linewidth}
@{}}
\toprule
Iteration & Planning (20) & Geometry (10) & Input (30) & Postproc. (40) & Total (100) \\
\midrule
Run 1 & 20 & 10 & 27 & 40 & 97 \\
Run 2 & 20 & 10 & 28 & 40 & 98 \\
Run 3 & 20 & 10 & 30 & 40 & 100 \\
Run 4 & 20 & 10 & 30 & 40 & 100 \\
Run 5 & 20 & 10 & 27 & 40 & 97 \\
\midrule
\textbf{Average} & 20.0 & 10.0 & 28.4 & 40.0 & 98.4 \\
\bottomrule
\end{tabularx}
\end{table}

\begin{itemize}
    \item Run 1: The validated geometry used def2-SVP and TDA rather than the literal def2-TZVP/full-TDDFT specification; the common geometry was accepted.\\
    \item Run 2: The PBE0/def2-TZVP minimum and spectra were validated; the small deduction reflects TDA rather than full TDDFT.\\
    \item Run 3: The validated def2-TZVP minimum and full-TDDFT comparison received full credit.\\
    \item Run 4: The validated minimum and full-TDDFT comparison received full credit; the close functional variant was defensible.\\
    \item Run 5: As in Run 1, the small deduction reflects def2-SVP geometry optimization and TDA, not the common validated geometry.
\end{itemize}

\subsubsection{Evaluation of the benchmark question representative of the topic ``Excited-State Kinetics''.}
\renewcommand{\arraystretch}{0.90}
\begin{table}[H]
\centering
\caption{\textbf{Example of Evaluation Rubric.}
The rubric used to grade the benchmark question representative of the topic ``Excited-State Kinetics''.}
\small
\begin{tabularx}{\linewidth}{@{}
p{0.04\linewidth}|
>{\centering\arraybackslash}p{0.14\linewidth}|
c|
>{\raggedright\arraybackslash}X
@{}}
\toprule
Topic & Criterion & Weight  & Task\\
\midrule

\multirow{4}{*}{ \raisebox{-10pt}[\dimexpr\height+\depth\relax]{\rotatebox[origin=c]{90}{Excited-State Kinetics}}}

& Planning & 10\% &
Identify the required workflow: ground-state geometry optimization and an excited-state (S$_1$) calculation using TDDFT. Evaluation of the fluorescence rate using the Hessians. Accurate selection of CAM-B3LYP/def2-TZVP and gas-phase conditions. \\

& Geometry generation & 20\% &
Generate a physically reasonable equilibrium geometry for formaldehyde in its ground electronic state, with correct molecular structure, charge, and spin multiplicity. Ground-state optimization at the DFT level is expected. \\

& Input file generation and execution & 40\% &
\textbf{1. Ground-state optimization and frequencies:} keyword line: ! CAM-B3LYP def2-TZVP Opt NumFreq (or equivalent). Correct charge and multiplicity.

\textbf{2. Excited-state optimization and frequencies:} keyword line: ! CAM-B3LYP def2-TZVP TDDFT Opt NumFreq TightSCF (or equivalent), requesting an appropriate number of excited states and correctly identifying and following the S$_1$ state.

\textbf{3. Fluorescence rate evaluation:} inclusion of the appropriate ESD(Fluor) settings such that the fluorescence rate is obtained directly from the excited-state Hessian and printed in the output. \\

& Postprocessing & 30\% &
Correct extraction of the fluorescence rate as reported by ESD(Fluor)

Verification of its physical plausibility

Clear presentation in a final markdown report. 

Discussion of assumptions and interpretation of the result. \\

\bottomrule
\end{tabularx}
\end{table}

\renewcommand{\arraystretch}{0.95}
\begin{table}[H]
\centering
\caption{\textbf{Per-iteration evaluation score for the benchmark question representative of the topic ``Excited-State Kinetics''.}
Scores assigned to each independent execution of the benchmark task, using the rubric defined above.}
\small
\begin{tabularx}{\linewidth}{@{}
>{\centering\arraybackslash}p{0.14\linewidth}|
>{\centering\arraybackslash}p{0.13\linewidth}|
>{\centering\arraybackslash}p{0.13\linewidth}|
>{\centering\arraybackslash}p{0.13\linewidth}|
>{\centering\arraybackslash}p{0.13\linewidth}|
>{\centering\arraybackslash}p{0.14\linewidth}
@{}}
\toprule
Iteration & Planning (10) & Geometry (20) & Input (40) & Postproc. (30) & Total (100) \\
\midrule
Run 1 & 10 & 20 & 40 & 28 & 98 \\
Run 2 & 10 & 20 & 40 & 30 & 100 \\
Run 3 & 10 & 20 & 40 & 30 & 100 \\
Run 4 & 10 & 20 & 40 & 28 & 98 \\
Run 5 & 10 & 20 & 40 & 30 & 100 \\
\midrule
\textbf{Average} & 10.0 & 20.0 & 40.0 & 29.2 & 99.2 \\
\bottomrule
\end{tabularx}
\end{table}

\begin{itemize}
    \item Run 1: The complete ESD workflow was correct; the small deduction reflects mislabeled structural and transition-energy quantities.\\
    \item Run 2: The full-TDDFT, Hessian-dependent ESD workflow and extracted rate received full credit.\\
    \item Run 3: The planar saddle was repaired and the complete TDA-ESD workflow received full credit.\\
    \item Run 4: The ESD workflow was correct; the small deduction reflects labeling vertical absorption as emission.\\
    \item Run 5: The complete Hessian-dependent ESD workflow received full credit.
\end{itemize}

\subsubsection{Evaluation of the benchmark question representative of the topic ``NMR and Magnetic Properties''.}

\renewcommand{\arraystretch}{0.90}
\begin{table}[H]
\centering
\caption{\textbf{Evaluation Rubric.}
The rubric used to grade the benchmark question representative of the topic ``NMR and Magnetic Properties''.}
\small
\begin{tabularx}{\linewidth}{@{}
p{0.04\linewidth}|
>{\centering\arraybackslash}p{0.14\linewidth}|
c|
>{\raggedright\arraybackslash}X
@{}}
\toprule
Topic & Criterion & Weight & Task\\
\midrule

\multirow{4}{*}{ \raisebox{-20pt}[\dimexpr\height+\depth\relax]{\rotatebox[origin=c]{90}{NMR and Magnetic Properties}}}

& Planning & 20\% & Geometry generation of fluorobenzene, pentafluorobenzene, along with tetramethylsilane (TMS) and trichlorofluoromethane ( \ce{CFCl3}): standard  \ce{^{1}H}  and  \ce{^{19}F} NMR references

Geometry optimizations were performed for all four molecules at the HF/aug-cc-pVTZ level of theory or an appropriate alternative DFT method or basis set.

Absolute NMR shielding constants were calculated for all molecules for the relevant nuclei (\ce{^{1}H} and/or \ce{^{19}F}), using the GIAO method at the HF/aug-cc-pVTZ level of theory.

Convert shielding values ($\sigma_{\text{iso}}$) to chemical shifts in ppm ($\delta$). 
Analyze trends and discuss anisotropy effects. 
 \\

& Geometry generation & 20\% & Generate physically reasonable equilibrium geometries for fluorobenzene, pentafluorobenzene, TMS, \ce{CFCl3} in their ground electronic state, with correct molecular structure, charge, and spin multiplicity. 

\\

& Input file generation and execution & 30\% &

\textbf{1.  Geometry optimization \ce{CFCl3}, fluorobenzene, pentafluorobenzene, TMS:} 

keyword line: ! RHF aug-cc-pVTZ OPT FREQ TightSCF. 

geometry section: * {xyzfile 0 1 CFCl3\_initial.xyz}, * xyzfile 0 1 fluorobenzene\_initial.xyz, * xyzfile 0 1 pentafluorobenzene\_initial.xyz, * xyzfile 0 1 TMS\_initial.xyz.

\textbf{2. NMR shielding \ce{CFCl3}, fluorobenzene, pentafluorobenzene, TMS:} 

keyword line: ! RHF aug-cc-pVTZ TIGHTSCF

geometry section: * xyzfile 0 1 CFCl3\_opt.xyz, * xyzfile 0 1 pentafluorobenzene\_opt.xyz, * xyzfile 0 1 fluorobenzene\_opt.xyz, * xyzfile 0 1 TMS\_opt.xyz

block: \%eprnmr Ori = GIAO Nuclei = all H { shift } Nuclei = all F { shift } end.

Note: NMR keyword can also be used to call \%eprnmr block.
\\

& Postprocessing and reporting & 30\% &

\ce{^{1}H} and  \ce{^{19}F} NMR properties of fluorobenzene and pentafluorobenzene calculated.

Chemical shielding values ($\sigma_{\text{iso}}$) and chemical shifts ($\delta$) are reported to determine the shielding effect of fluorination on an aromatic ring. 

Results should be presented in a final markdown report that is comprehensive and easy to interpret with primary literature sources. 

\\

\bottomrule
\end{tabularx}
\end{table}

\renewcommand{\arraystretch}{0.95}
\begin{table}[H]
\centering
\caption{\textbf{Per-iteration evaluation score for the benchmark question representative of the topic ``NMR and Magnetic Properties''.}
Scores assigned to each independent execution of the benchmark task, using the rubric defined above.}
\small
\begin{tabularx}{\linewidth}{@{}
>{\centering\arraybackslash}p{0.14\linewidth}|
>{\centering\arraybackslash}p{0.13\linewidth}|
>{\centering\arraybackslash}p{0.13\linewidth}|
>{\centering\arraybackslash}p{0.13\linewidth}|
>{\centering\arraybackslash}p{0.13\linewidth}|
>{\centering\arraybackslash}p{0.14\linewidth}
@{}}
\toprule
Iteration & Planning (\%) & Geometry (\%) & Input (\%) & Postproc. (\%) & Total (\%) \\
\midrule
Run 1 & 20 & 20 & 30 & 30 & 100 \\
Run 2 & 20 & 20 & 30 & 30 & 100 \\
Run 3 & 20 & 20 & 30 & 30 & 100 \\
Run 4 & 0 & 10 & 20 & 15 & 45 \\
Run 5 & 20 & 20 & 30 & 30 & 100 \\
\midrule
\textbf{Average} & 16.0 & 18.0 & 28.0 & 27.0 & 89.0 \\
\bottomrule
\end{tabularx}
\end{table}

\begin{itemize}
    \item Run 1: The complete four-species optimization, shielding, and TMS/CFCl$_3$ reference workflow was correct.\\
    \item Run 2: All four species were optimized and frequency-validated, and the requested referenced shifts were reported correctly.\\
    \item Run 3: The alternative validated geometry level and complete reference workflow were defensible; literature interpretation was not penalized.\\
    \item Run 4: TMS and CFCl$_3$ references were absent, so the reported target shieldings were not converted to the requested referenced chemical shifts.\\
    \item Run 5: The complete reference workflow and shielding-to-shift conversion were correct; the level of theoretical discussion was not penalized.
\end{itemize}

\subsubsection{Evaluation of the benchmark question representative of the topic ``Vibrational Spectroscopy''.}

\renewcommand{\arraystretch}{0.90}
\begin{table}[H]
\centering
\caption{\textbf{Evaluation Rubric.}
The rubric used to grade the benchmark question representative of the topic ``Vibrational
Spectroscopy''.}
\small
\begin{tabularx}{\linewidth}{@{}
p{0.04\linewidth}|
>{\centering\arraybackslash}p{0.14\linewidth}|
c|
>{\raggedright\arraybackslash}X
@{}}
\toprule
Topic & Criterion & Weight & Task\\
\midrule

\multirow{4}{*}{ \raisebox{-15pt}[\dimexpr\height+\depth\relax]{\rotatebox[origin=c]{90}{Isotopic Effects}}}

& Planning & 20\% & Water optimization at the CCSD/cc-PDVZ level 

Modification of the Hessian file.

Recalculation of vibrational frequencies.\\

& Geometry generation & 10\% & Generate initial water geometry. \\

& Input file generation and execution & 30\% &  keyword line: ! CCSD cc-pVDZ Opt NumGrad NumFreq, other convergence or grid settings allowed. AUTOCI-CCSD also allowed.

blocks: no blocks required. \%numgrad, \%freq, and \%autoci blocks for controlled settings are optional. 

geometry section: * xyzfile 0 1 water\_initial.xyz. \\

& Postprocessing and reporting & 40\% & 

Extraction of vibrational frequencies. 

Modification of Hessian files and recalculation of vibrational frequencies.

Calculation of isotopic shifts.

Quality of final report.\\

\bottomrule
\end{tabularx}
\end{table}

\renewcommand{\arraystretch}{0.95}
\begin{table}[H]
\centering
\caption{\textbf{Per-iteration evaluation score for the benchmark question representative of the topic ``Vibrational
Spectroscopy''.}
Scores assigned to each independent execution of the benchmark task, using the rubric defined above.}
\small
\begin{tabularx}{\linewidth}{@{}
>{\centering\arraybackslash}p{0.14\linewidth}|
>{\centering\arraybackslash}p{0.13\linewidth}|
>{\centering\arraybackslash}p{0.13\linewidth}|
>{\centering\arraybackslash}p{0.13\linewidth}|
>{\centering\arraybackslash}p{0.13\linewidth}|
>{\centering\arraybackslash}p{0.14\linewidth}
@{}}
\toprule
Iteration & Planning (\%) & Geometry (\%) & Input (\%) & Postproc. (\%) & Total (\%) \\
\midrule
Run 1 & 20 & 10 & 30 & 40 & 100 \\
Run 2 & 20 & 10 & 30 & 40 & 100 \\
Run 3 & 20 & 10 & 30 & 40 & 100 \\
Run 4 & 20 & 10 & 30 & 40 & 100 \\
Run 5 & 20 & 10 & 30 & 40 & 100 \\
\midrule
\textbf{Average} & 20 & 10 & 30 & 40 & 100 \\
\bottomrule
\end{tabularx}
\end{table}

\begin{itemize}
    \item Run 1: No issues observed.\\
    \item Run 2: No issues observed.\\
    \item Run 3: No issues observed.\\
    \item Run 4: No issues observed.\\
    \item Run 5: No issues observed.
\end{itemize}

\subsubsection{Evaluation of the benchmark question representative of the topic ``Custom Optimizations''.}

\renewcommand{\arraystretch}{0.90}
\begin{table}[H]
\centering
\caption{\textbf{Evaluation Rubric.}
The rubric used to grade the benchmark question representative of the topic ``Custom Optimizations''.}
\small
\begin{tabularx}{\linewidth}{@{}
p{0.04\linewidth}|
>{\centering\arraybackslash}p{0.14\linewidth}|
c|
>{\raggedright\arraybackslash}X
@{}}
\toprule
Topic & Criterion & Weight & Task\\
\midrule

\multirow{4}{*}{ \raisebox{-70pt}[\dimexpr\height+\depth\relax]{\rotatebox[origin=c]{90}{Custom Optimizations}}}

& Planning & 10\% & Generation of ethylene geometry and twisted geometry.

S1 geometry optimization from twisted geometry before MECP/CI search. 

CI search optimization from S1 equilibrium geometry using TDDFT. 

CI search optimization from S1 equilibrium geometry using SF-TDDFT (if bare TDDFT fails). \\

& Geometry generation & 10\% & Successful generation of ethylene geometry and twisted geometry. \\

& Input file generation and execution & 50\% & \textbf{1. MECP optimization:} keyword line: ! CAM-B3LYP/6-31 G Opt SurfCrossOpt; other convergence or grid settings allowed. 

blocks: \%mecp Mult 3 end.

geometry section: * xyzfile 0 1 twisted\_ethylene\_initial.xyz. 

\textbf{2. SF-TDDFT CI optimization:} 

keyword line: ! CAM-B3LYP 6-31G CI-OPT, other convergence or grid settings allowed. 

blocks: \%tddft SF TRUE NROOTS 5 IRoot 2 JRoot 1 end. NROOTS can vary. \%conical METHOD GP ETOL $10^{-4}$ end (other methods for CI search are also valid)

geometry section: * xyzfile 0 3 twisted\_ethylene\_initial.xyz. 

The agent must try SF TDDFT and move to TDDFT only if it cannot converge the optimization.

\textbf{3. TDDFT CI optimization (half credit):} 

keyword line: ! CAM-B3LYP 6-31G CI-OPT, other convergence or grid settings allowed. 

blocks: \%tddft  NROOTS 5 IROOT 2 JROOT 1 end.

\%conical METHOD UBP end. Additional tolerance settings, methods (e.g., GP, GP\_nonacme) are allowed.

geometry section: * xyzfile 0 1 twisted\_ethylene\_initial.xyz. \\

& Postprocessing and reporting & 30\% & Extraction of S0, S1, and T1 energies at the optimized geometries.
 
Extraction of geometries (approx. $90^{0}$ twisted geometry for TDDFT, pyramidalized geometry for SF TDDFT).

The limitations of LR TDDFT must be clearly discussed relative to SF-TDDFT or multi-reference methods.

Quality of final report.\\

\bottomrule
\end{tabularx}
\end{table}

\renewcommand{\arraystretch}{0.95}
\begin{table}[H]
\centering
\caption{\textbf{Per-iteration evaluation score for the benchmark question representative of the topic ``Custom Optimizations''.}
Scores assigned to each independent execution of the benchmark task, using the rubric defined above.}
\small
\begin{tabularx}{\linewidth}{@{}
>{\centering\arraybackslash}p{0.14\linewidth}|
>{\centering\arraybackslash}p{0.13\linewidth}|
>{\centering\arraybackslash}p{0.13\linewidth}|
>{\centering\arraybackslash}p{0.13\linewidth}|
>{\centering\arraybackslash}p{0.13\linewidth}|
>{\centering\arraybackslash}p{0.14\linewidth}
@{}}
\toprule
Iteration & Planning (\%) & Geometry (\%) & Input (\%) & Postproc. (\%) & Total (\%) \\
\midrule
Run 1 & 8 & 9 & 43 & 23 & 83 \\
Run 2 & 6 & 8 & 33 & 13 & 60 \\
Run 3 & 5 & 8 & 33 & 10 & 56 \\
Run 4 & 8 & 10 & 44 & 25 & 87 \\
Run 5 & 6 & 8 & 27 & 13 & 54 \\
\midrule
\textbf{Average} & 6.6 & 8.6 & 36.0 & 16.8 & 68.0 \\
\bottomrule
\end{tabularx}
\end{table}

\begin{itemize}
    \item Run 1: The MECP converged, but corrected spin-flip CI attempts did not converge and the final conventional-LR result was a flat artifact; the required S$_1$ pre-optimization was absent.\\
    \item Run 2: The MECP converged, but the CI used a triplet-like root as S$_1$ and reported the resulting singlet--triplet gap as an S$_0$/S$_1$ intersection.\\
    \item Run 3: The CI again optimized the wrong singlet--triplet root pair, and the report presented the flat structure as a successful S$_0$/S$_1$ result.\\
    \item Run 4: This was the strongest recovery and produced a plausible pyramidalized approximation, but the corrected spin-flip search remained unconverged and used an unexplained 6-31G* substitution.\\
    \item Run 5: The MECP reached its cycle limit and the CI converged the wrong singlet--triplet root pair, so both central success claims were overstated.
\end{itemize}

\subsection{Performance summary of {\fontfamily{Cinzel-LF}\selectfont El Agente Quntur} on the benchmark set}

\renewcommand{\arraystretch}{0.95}
\begin{table}[H]
\centering
\caption{\textbf{Per-task scores of{\cinzel El Agente Quntur} on the benchmark set.}}
\small
\begin{tabularx}{\linewidth}{@{}
>{\centering\arraybackslash}p{0.27\linewidth}|
>{\centering\arraybackslash}p{0.12\linewidth}|
>{\centering\arraybackslash}p{0.12\linewidth}|
>{\centering\arraybackslash}p{0.12\linewidth}|
>{\centering\arraybackslash}p{0.12\linewidth}|
>{\centering\arraybackslash}p{0.14\linewidth}
@{}}
\toprule
Topic & Planning (\%) & Geometry (\%) & Input (\%) & Postproc. (\%) & Overall. (\%)   \\
\midrule
Relativistic Corrections & 93 & 97 & 94 & 94 & 94 \\
Electrical Properties & 100 & 100 & 100 & 82 & 93 \\
Orbitals and Densities & 90 & 89 & 90 & 90 & 90 \\
Condensed Fukui Functions & 100 & 96 & 99 & 94 & 96 \\
Thermochemistry & 90 & 100 & 95 & 69 & 86 \\
Conformer Generation & 100 & 100 & 98 & 75 & 92 \\
Molecular Dynamics & 74 & 71 & 70 & 71 & 71 \\
Explicit Solvation & 100 & 57 & 100 & 100 & 87 \\
Potential Energy Surfaces & 100 & 100 & 100 & 100 & 100 \\
Transition State Methods & 100 & 100 & 99 & 100 & 100 \\
Reaction Mechanisms & 80 & 80 & 84 & 83 & 82 \\
Isotopic Effects & 100 & 100 & 87 & 87 & 92 \\
Excited States & 100 & 100 & 95 & 100 & 98 \\
Excited-State Kinetics & 100 & 100 & 100 & 97 & 99 \\
NMR and Magnetic Properties & 80 & 90 & 93 & 90 & 89 \\
Vibrational Spectroscopy & 100 & 100 & 100 & 100 & 100 \\
Custom Optimizations & 66 & 86 & 72 & 56 & 68 \\
\bottomrule
\end{tabularx}
\end{table}

\begin{figure}[ht]
    \centering
    \includegraphics[width=1\linewidth]{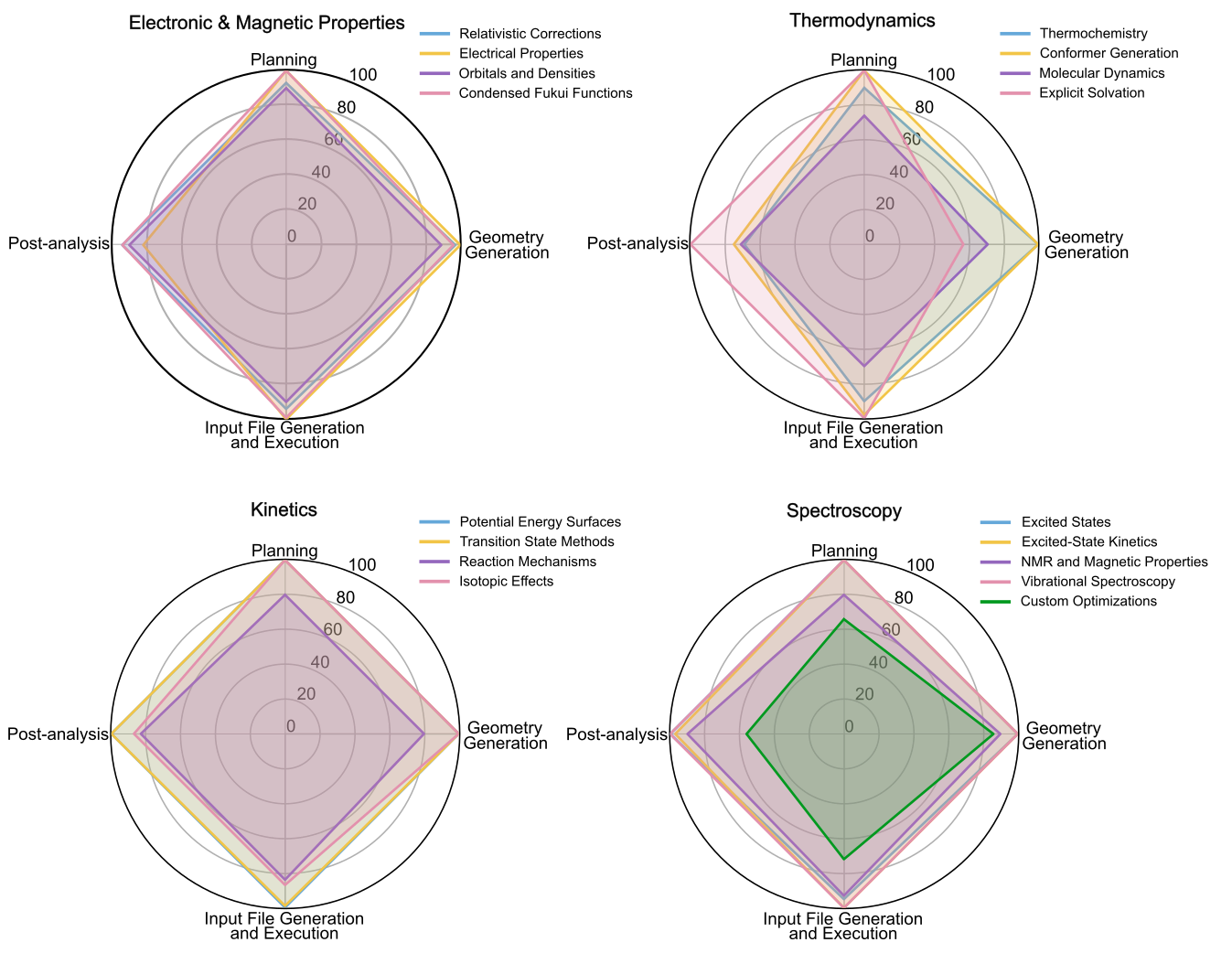}
    \caption{\textbf{Performance of{\cinzel El Agente Quntur} on the benchmark set.} The benchmark consists of a set of 17 computational quantum chemistry exercises covering several topics from electronic and magnetic properties, thermodynamics, kinetics, and spectroscopy, as well as the various levels of theory and difficulty. The benchmark questions are in the \hyperref[sec:benchmark]{Supplemental Information Section \ref{sec:benchmark}}. The rubric for each question is in the \hyperref[SI:QunturEval]{Supplemental Information Section} \ref{SI:QunturEval}. All benchmarks were conducted using Claude Opus 4.5 as the base language model for all agents.}
    \label{fig:benchmark}
\end{figure}

\newpage

\subsection{Performance summary of the Claude Code baseline on the benchmark set}

Claude Code was evaluated as a general coding-agent baseline in five independent executions for each of the 17 benchmark questions. The baseline used Claude Opus 4.5, the frontier Claude model available at the time of the{\cinzel Quntur} benchmark evaluation. Claude Code was run with ORCA skills enabled and with web-search access. No benchmark-specific or computational chemistry-specialized custom skills were written for Claude Code. This choice kept the baseline distinct from Quntur, whose development explicitly includes detailed skills, guided deep-research procedures, and specialized tools.

Claude Code was graded using the same task-specific rubrics used for Quntur. Grading emphasizes whether the generated computational workflow is defensible for the requested observable, not only whether ORCA input files were written or output files terminated normally. Defensible methodological choices, literature-supported interpretation, prompt ambiguity, recovered exploratory failures, and limitations intrinsic to ORCA were not penalized unless they produced a materially incorrect or unsupported requested result.

\subsubsection{Claude Code scores on the benchmark set}

Table \ref{table:CC} presents Claude Code scores averaged over five independent executions of each benchmark task. The four rubric-category columns are normalized to the corresponding category maximum before averaging, and the overall column is the mean total score across the five executions. Values in the table are rounded to the nearest percentage point; the unrounded macro-average across the 17 task means is 61.7/100.

\renewcommand{\arraystretch}{0.95}
\begin{table}[H]
\centering
\caption{\textbf{Per-task scores of Claude Code on the benchmark set.}}\label{table:CC}
\small
\begin{tabularx}{\linewidth}{@{}
>{\centering\arraybackslash}p{0.27\linewidth}|
>{\centering\arraybackslash}p{0.12\linewidth}|
>{\centering\arraybackslash}p{0.12\linewidth}|
>{\centering\arraybackslash}p{0.12\linewidth}|
>{\centering\arraybackslash}p{0.12\linewidth}|
>{\centering\arraybackslash}p{0.14\linewidth}
@{}}
\toprule
Topic & Planning (\%) & Geometry (\%) & Input (\%) & Postproc. (\%)  & Overall (\%)  \\
\midrule
Relativistic Corrections & 80 & 60 & 95 & 82 & 84 \\
Electrical Properties & 80 & 80 & 94 & 63 & 79 \\
Orbitals and Densities & 70 & 80 & 79 & 69 & 73 \\
Condensed Fukui Functions & 40 & 30 & 75 & 70 & 65 \\
Thermochemistry & 76 & 84 & 69 & 58 & 69 \\
Conformer Generation & 91 & 79 & 81 & 53 & 74 \\
Molecular Dynamics & 48 & 30 & 47 & 50 & 45 \\
Explicit Solvation & 78 & 57 & 59 & 45 & 56 \\
Potential Energy Surfaces & 72 & 80 & 79 & 58 & 72 \\
Transition State Methods & 64 & 42 & 55 & 32 & 48 \\
Reaction Mechanisms & 50 & 50 & 70 & 57 & 57 \\
Isotopic Effects & 60 & 73 & 85 & 42 & 66 \\
Excited States & 50 & 34 & 60 & 69 & 59 \\
Excited-State Kinetics & 40 & 40 & 20 & 12 & 23 \\
NMR and Magnetic Properties & 52 & 58 & 57 & 53 & 55 \\
Vibrational Spectroscopy & 80 & 90 & 81 & 75 & 79 \\
Custom Optimizations & 52 & 72 & 40 & 41 & 45 \\
\bottomrule
\end{tabularx}
\end{table}

The five Claude Code executions were not all performed at the same reasoning-effort setting. Runs 1--3 used high effort (recorded as \texttt{default} in the token-cost workbook), whereas Runs 4--5 used \texttt{xhigh}. Table \ref{table:CC_effort} separates the observed performance and resource use of these two groups. The mean benchmark score is the macro-average over the 17 task scores within each execution, subsequently averaged within the effort group. Token counts include direct input, cache creation, cache reads, and output tokens.

\renewcommand{\arraystretch}{1.0}
\begin{table}[H]
\centering
\caption{\textbf{Observed Claude Code performance and resource use by reasoning-effort setting.} Values are averages over three high-effort executions or two \texttt{xhigh}-effort executions.}\label{table:CC_effort}
\small
\begin{tabularx}{\linewidth}{@{}
>{\raggedright\arraybackslash}X|
>{\centering\arraybackslash}p{0.18\linewidth}|
>{\centering\arraybackslash}p{0.18\linewidth}|
>{\centering\arraybackslash}p{0.15\linewidth}
@{}}
\toprule
Metric & \texttt{High}(Runs 1--3) & \texttt{xhigh} (Runs 4--5) & Change \\
\midrule
Mean benchmark score (/100) & 62.02 & 61.29 & $-0.73$ points \\
Mean total tokens per task & $5.98\times10^{6}$ & $7.26\times10^{6}$ & $+21\%$ \\
Mean cost per task (USD) & \$6.17 & \$9.32 & $+51\%$ \\
Mean tokens per complete 17-task pass & $101.7\times10^{6}$ & $123.4\times10^{6}$ & $+21\%$ \\
\bottomrule
\end{tabularx}
\end{table}

Thus, the higher effort setting is clearly reflected in resource use but not in aggregate benchmark performance. The \texttt{xhigh} group improves 7 tasks and worsens 10 relative to the corresponding high-group task means, and the association between the per-task change in token use and score is weakly negative ($r=-0.19$). The cost increase is driven mainly by cache traffic: mean cache-creation tokens per task approximately double, cache-read tokens increase by 15\%, and output tokens increase by only 3\%. This comparison is observational rather than causal because only two \texttt{xhigh} repetitions are available, reasoning effort is confounded with execution order, and token consumption had already increased across the three high runs.

\subsubsection{Point-deduction rationale for the Claude Code baseline}

The notes below summarize the consequential failure modes observed across the five executions of each task. Defensible methodological choices and literature-supported interpretation were not penalized; deductions focus on missing scientific dependencies, incorrect molecular or electronic states, nonconvergence, parsing errors, omitted requested observables, and results materially affected by the selected workflow.

\begin{itemize}
    \item \textbf{Relativistic Corrections:} All five executions completed and parsed the shielding calculations, but used an unoptimized, frequency-unvalidated Pt--Cl geometry despite the strong geometry dependence of the shielding. Run 4 also replaced the requested non-relativistic/ZORA comparison with a mismatched DKH-based comparator. ORCA defaults and intrinsic relativistic-treatment limitations were not penalized.

    \item \textbf{Electrical Properties:} The MP2/aug-cc-pVDZ property calculations generally succeeded. Deductions reflect the missing optimization in Run 2, the lower-level geometry and incorrect anisotropy in Run 3, selection of an unrelaxed dipole block in Run 4, and omission of requested Cartesian tensor elements in Runs 1 and 5.

    \item \textbf{Orbitals and Densities:} Run 1 used the wrong piperidine formula, Run 4 plotted occupied orbitals rather than the requested SOMOs, and Run 5 produced blank SOMO figures because of a cube-parser error. Runs 2 and 3 were substantially correct but lacked frequency validation of the optimized structures. Literature-backed stability interpretations were not penalized.

    \item \textbf{Condensed Fukui Functions:} All five executions completed the neutral, anionic, and cationic population calculations, but omitted the required optimized and frequency-validated neutral HF reference geometry. The fixed geometry materially affected symmetry and the condensed indices; Run 4 additionally reversed the attacking-species interpretation.

    \item \textbf{Thermochemistry:} Most executions generated the correct acid/base species and completed substantial gas-phase thermochemistry, but all selected generic CPCM without literature-based justification rather than recognizing SMD as the better-supported choice for this ionic-solvation problem. The resulting direct-cycle pKa values were near 13--15 rather than the experimental value near 4.8; some runs also retained nonminimum geometries or incomplete solvent-cycle stages.

    \item \textbf{Conformer Generation:} Runs 1 and 3 completed strong workflows. The remaining deductions arise from an incomplete and duplicate-biased ensemble in Run 2, exporting an intermediate rather than final optimized geometry in Run 4, and using a constitutional isomer of serine in Run 5; these errors materially alter populations or molecular identity.

    \item \textbf{Molecular Dynamics:} Four executions omitted relaxation of the initial chloride--water cluster, while the remaining optimization reached its cycle limit. Temperature control, equilibration/production separation, and trajectory analysis were often incomplete, and two coordination-number implementations used incorrect cutoffs. Finite-cluster limitations inherent to the prompt were not penalized.

    \item \textbf{Explicit Solvation:} The requested $n=3$, 5, and 7 cluster branches were frequently incomplete or nonconverged, leaving incomparable electronic-energy and Gibbs-energy data. Several reports consequently inferred pKa convergence without a complete set of validated gas/CPCM minima and thermochemical corrections; single-topology limitations were secondary to these execution failures.

    \item \textbf{Potential Energy Surfaces:} Runs 1 and 5 received full credit for supported relaxed-coordinate workflows. Run 2 reported a complete surface although most scan points lacked completion evidence, Run 3 used a rigid single-point grid without prior optimization, and Run 4 completed the relaxed scan but omitted the required final analysis/report.

    \item \textbf{Transition State Methods:} Run 1 followed the wrong noncyclic pathway. Runs 2--4 independently recovered the intended saddle but did not obtain it from a completed valid NEB dependency, while Run 5 promoted a wrong pathway with five imaginary modes and an unconverged IRC. These are pathway and stationary-point validation failures, not optional-setting deductions.

    \item \textbf{Reaction Mechanisms:} All five executions located a plausible MP2/cc-pVTZ S\textsubscript{N}2 transition state and ran IRC calculations, but none optimized and frequency-validated the required encounter complex. Activation enthalpies, entropies, and free energies were therefore referenced to separated fragments and describe a different thermochemical quantity.

    \item \textbf{Isotopic Effects:} The executions showed useful isotope-Hessian processing, but the final KIE analyses variously contained a unit error, substituted heuristic frequency scaling for isotope thermochemistry, omitted part of the substituent series, or lacked the required final report. Deductions were tied to numerical correctness and task completeness, not isotope-label syntax.

    \item \textbf{Excited States:} Each execution produced TDDFT excitation data, but none supplied the required optimized, frequency-validated ground-state DFT geometry; one optimized an excited state instead and the others used force-field or hand-written structures. The resulting geometry dependence changed state ordering and excitation energies. TDA itself was accepted as defensible.

    \item \textbf{Excited-State Kinetics:} Every execution retained a planar S$_1$ stationary point without the excited-state Hessian needed to identify the out-of-plane instability. None completed ORCA's Hessian-dependent ESD/Herzberg--Teller fluorescence workflow, and the reported oscillator-strength or literature estimates were not the requested vibronic fluorescence rate.

    \item \textbf{NMR and Magnetic Properties:} Run 1 completed the TMS/CFCl$_3$ reference workflow and received high credit. Runs 2--5 calculated target-molecule shieldings but omitted one or both required reference calculations, so absolute shieldings or ad hoc differences were reported instead of referenced $^1$H and $^{19}$F chemical shifts.

    \item \textbf{Vibrational Spectroscopy:} Runs 1 and 4 completed the optimized-water, mass-modified-Hessian, and frequency-recalculation workflow. Run 2 omitted the CCSD optimization and Hessian route, while Runs 3 and 5 did not preserve a traceable D$_2$O recalculation and promoted unsupported or hard-coded frequencies. Recovered exploratory syntax failures were not penalized.

    \item \textbf{Custom Optimizations:} No execution delivered both a validated S$_0$/T$_1$ MECP and S$_0$/S$_1$ conical intersection. The workflows included wrong-root optimization, unconverged or gap-only structures, a faulty spin-flip energy parser, and one chemically malformed geometry; minimum energy gaps alone were not accepted as converged crossings.
\end{itemize}

\begin{figure}[ht]
    \centering
    \includegraphics[width=1\linewidth]{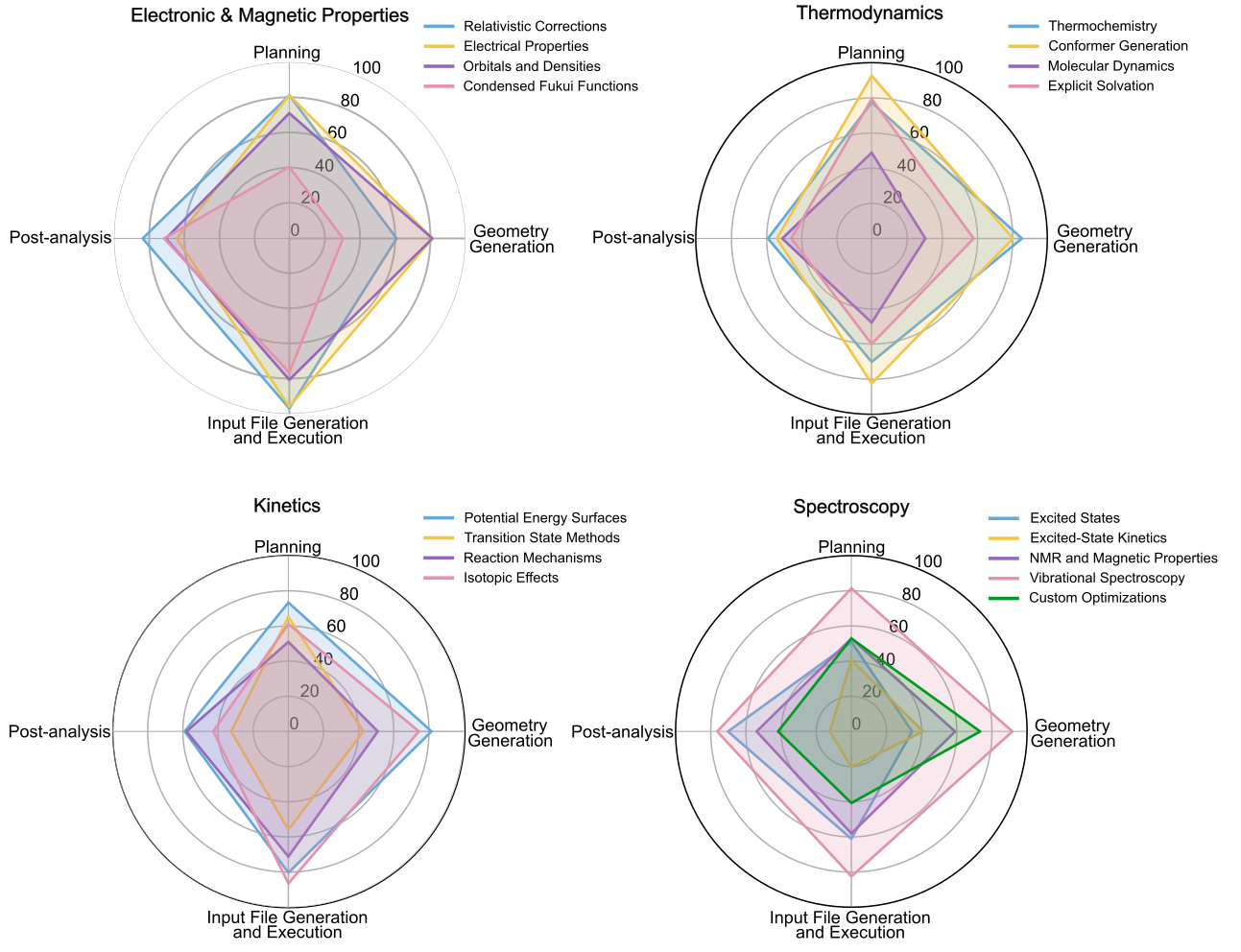}
    \caption{\textbf{Performance of Claude Code on the benchmark set.} The benchmark comprises 17 computational quantum chemistry exercises, each evaluated in five independent Claude Code executions using the same task-specific rubrics applied to {\cinzel El Agente Quntur}. Radar values are five-execution mean category scores normalized to each category maximum. Claude Code used Claude Opus 4.5 with ORCA skills and web-search access but without benchmark-specific computational-chemistry skills.}
    \label{fig:benchmark2}
\end{figure}

\newpage

\subsection{Claude Code failure modes and direct comparison with Quntur}

The results show that the Claude Code baseline struggles greatly in geometry generation and input file generation, error recovery, and post-analysis. Even worse, such errors can propagate silently throughout the workflow, which wastes time and computational resources. Here we focus on a few baseline failure modes related to geometry generation and input file generation, while other baseline failure modes as well as{\cinzel Quntur} failure modes are discussed in detail in the main manuscript, in previous sections of this Supplemental Information, or can be completely accessed from our dataset repository \cite{Perez-Zou2026QunturFiles}.

\subsubsection{Representative Geometry generation Failure Modes}

\begin{figure}[ht]
    \centering
    \includegraphics[width=1\linewidth]{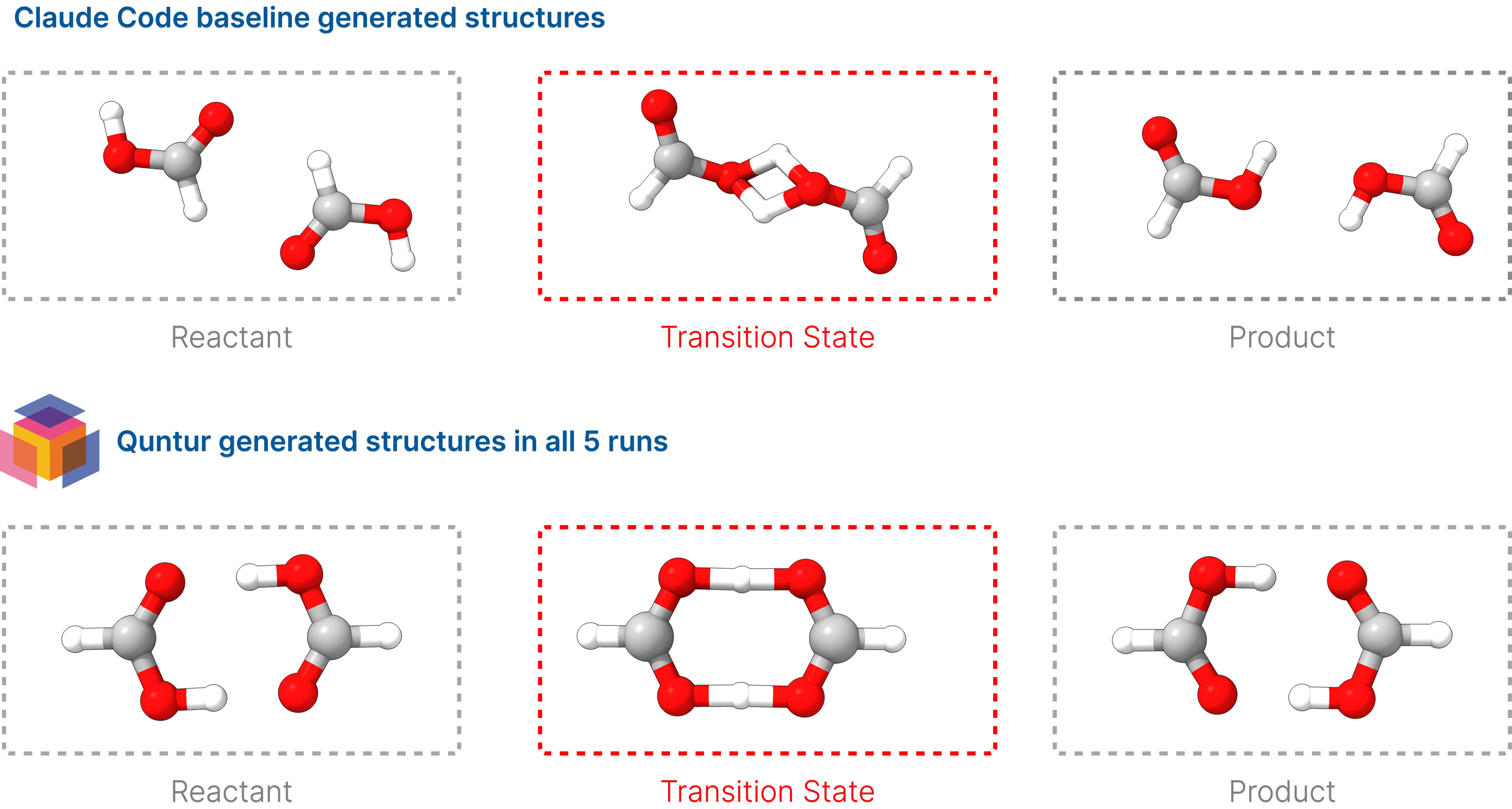}
    \caption{\textbf{Geometry comparison for the formic-acid-dimer double-proton-transfer task.} The upper panels show the optimized reactant, transition structure, and product from Claude Code run 1; the lower panels show the corresponding structures from{\cinzel Quntur} run 1. Claude run 1 generated noncyclic, side-to-side endpoints, so its converged NEB, OptTS, IRC, and activation analysis describe a different proton-transfer pathway.{\cinzel Quntur} run 1 retained the facing cyclic dimer and concerted transfer geometry; the other four{\cinzel Quntur} executions recovered the same pathway.}
    \label{fig:formic_acid_dimer}
\end{figure}

Figure \ref{fig:formic_acid_dimer} compares the geometries from Claude Code run 1 and{\cinzel Quntur} run 1 for the ``Transition State Methods'' benchmark. The{\cinzel Quntur} structures shown are representative of the same cyclic pathway recovered in all five{\cinzel Quntur} executions.

\begin{figure}[ht]
    \centering
    \includegraphics[width=0.5\linewidth]{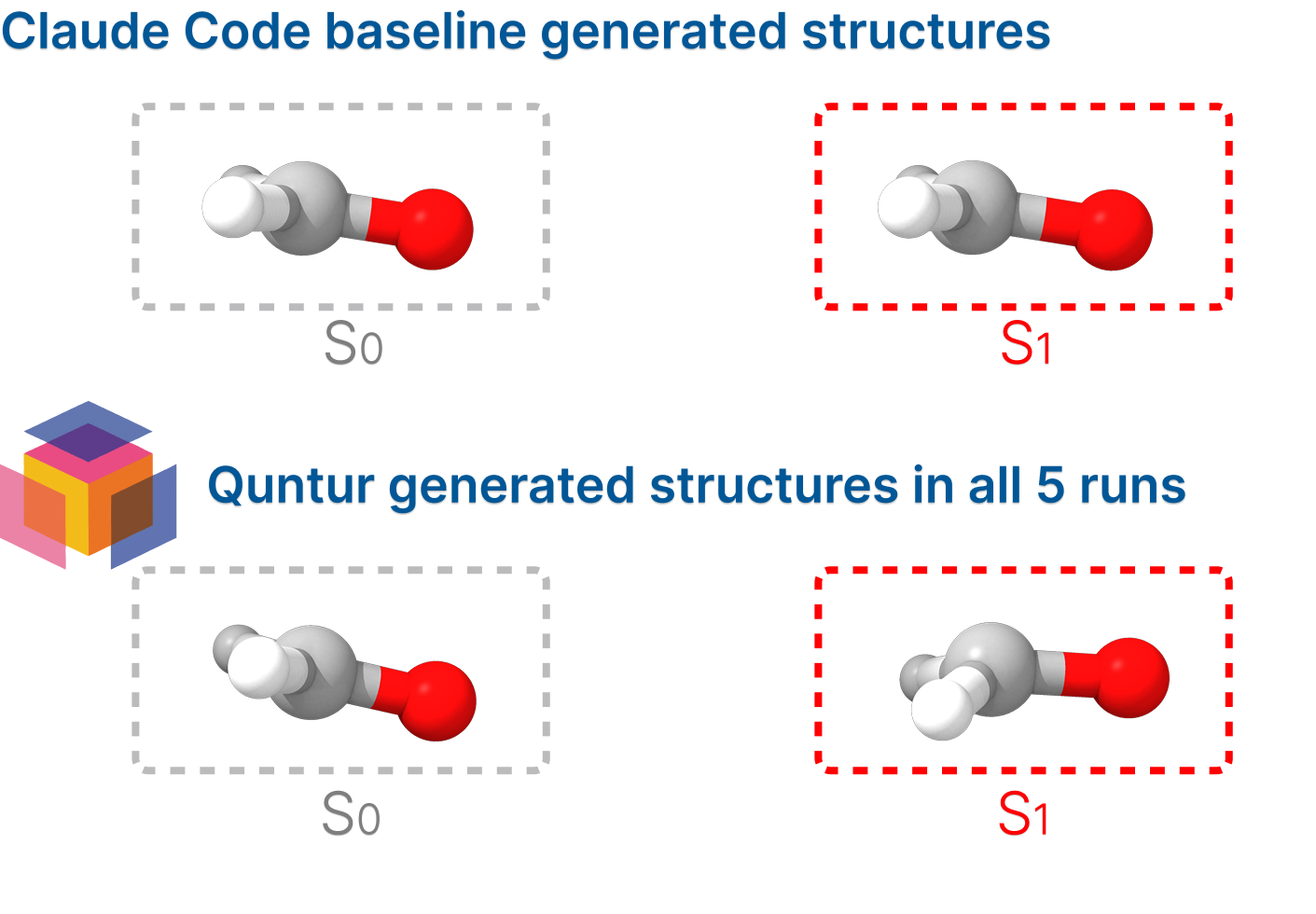}
    \caption{\textbf{Ground- and excited-state geometries for the formaldehyde fluorescence task.} Representative S$_0$ and final S$_1$ structures are shown for Claude Code and{\cinzel Quntur}. Planar S$_0$ formaldehyde is the correct ground-state structure for both agents. In all five Claude executions, optimization from the exactly planar S$_0$ geometry retained a planar S$_1$ stationary point and no S$_1$ Hessian was calculated to detect the out-of-plane instability. All five{\cinzel Quntur} executions obtained an all-real-frequency pyramidal S$_1$ minimum; runs 2--4 explicitly displaced and reoptimized an initially planar saddle.}
    \label{fig:formaldehyde_fluorescence}
\end{figure}

Figure \ref{fig:formaldehyde_fluorescence} shows the systematic geometry difference for the ``Excited-State Kinetics'' benchmark.

\subsubsection{Representative input file generation Failure Modes}

\begin{figure}[ht]
    \centering
    \includegraphics[width=0.9\linewidth]{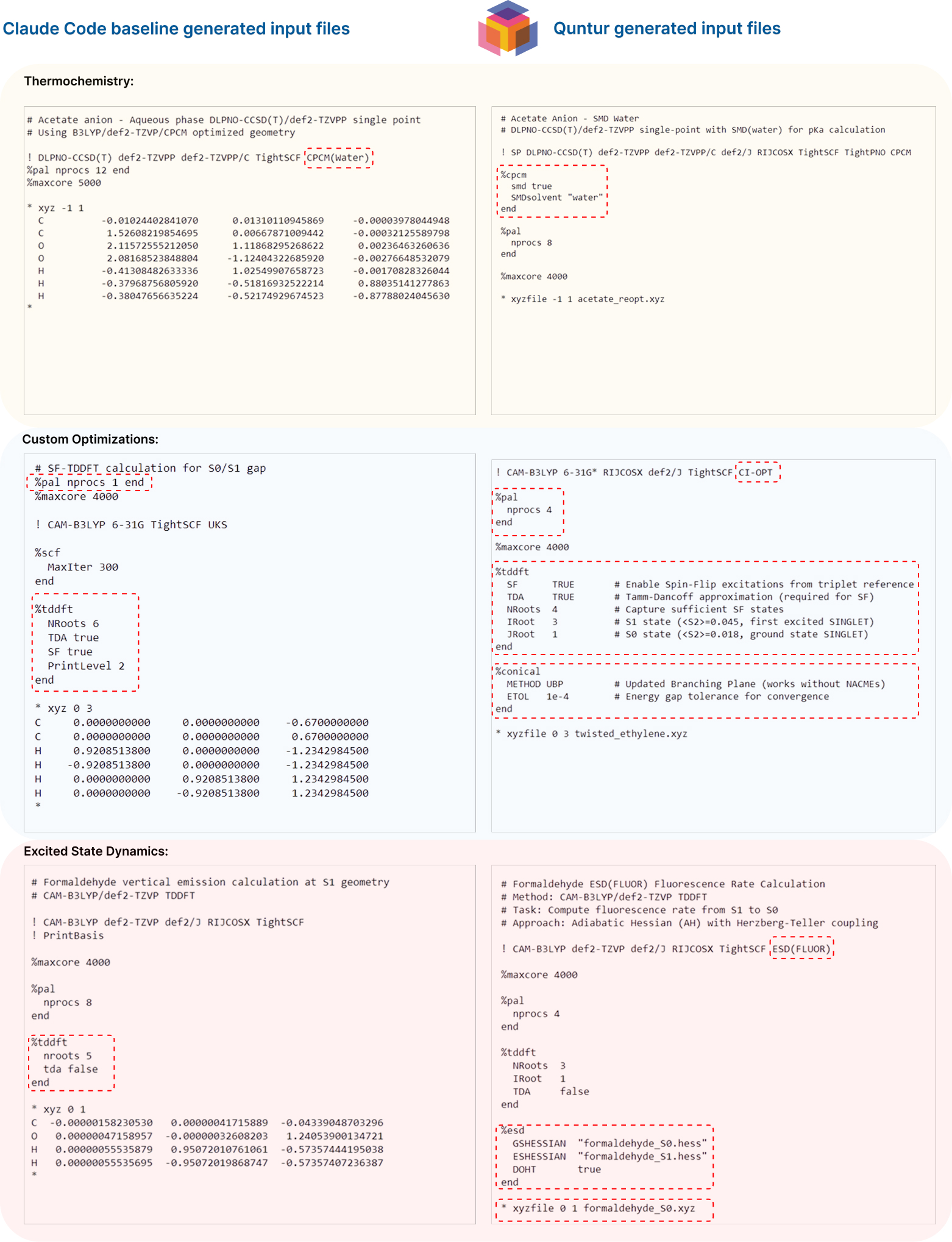}
    \caption{\textbf{Representative input-file comparisons from the five-execution evaluation.} The rebuilt panels compare: (i) Claude Thermochemistry run 1, which uses generic \texttt{CPCM(Water)} for the aqueous acetate single point, with{\cinzel Quntur} run 3, which uses the literature-justified SMD water treatment at DLPNO-CCSD(T)/def2-TZVPP; (ii) Claude Custom Optimizations run 1, which generates an SF-TDDFT single point for an external S$_0$/S$_1$ gap optimizer, with{\cinzel Quntur} run 4, which requests ORCA's native \texttt{CI-OPT} procedure with a \texttt{\%conical} block and explicit \texttt{IRoot}/\texttt{JRoot} state selection; (iii) Claude Excited-State Kinetics run 1, which performs an electronic-emission single point, with{\cinzel Quntur} run 2, which supplies the S$_0$ and pyramidal-S$_1$ Hessians to \texttt{ESD(FLUOR)} with Herzberg--Teller coupling. These are representative files rather than the sole evidence: the corresponding five-run frequencies are given in the text. The panels illustrate consequential property-specific settings rather than cosmetic input-style differences.}
    \label{fig:input_files}
\end{figure}

A reasonable headline method and basis set are not sufficient when the requested result depends on state-specific optimization machinery, vibrational response, or a property-appropriate solvation model. Figure \ref{fig:input_files} therefore shows representative files from three trends identified across all five executions; the recurrence and exceptions are stated explicitly below. Processor count, optional RIJCOSX/AuxJ acceleration, and inline versus \texttt{xyzfile} coordinates are not treated as scientific failure criteria.

\textbf{Thermochemistry.} The solvent-model trend is also recurring but not perfectly symmetric: all five Claude executions use CPCM rather than the literature-justified SMD water treatment, whereas four of five{\cinzel Quntur} executions use SMD and{\cinzel Quntur} run 5 also uses CPCM. The displayed pair contrasts Claude run 1's aqueous acetate DLPNO-CCSD(T)/def2-TZVPP \texttt{CPCM(Water)} input with{\cinzel Quntur} run 3's \texttt{smd true}/\texttt{SMDsolvent ``water''} input. Run 3 is the only{\cinzel Quntur} execution that matches the literal high-level DLPNO-SMD specification; runs 1, 2, and 4 use SMD at lower levels. The systematic claim is therefore SMD selection in four of five{\cinzel Quntur} runs versus zero of five Claude runs, not five flawless{\cinzel Quntur} implementations. This choice is consequential for acetate solvation and the resulting p$K_\mathrm{a}$, but is not the sole source of the remaining numerical error.

\textbf{Custom Optimizations.} The native-machinery difference is systematic: all five{\cinzel Quntur} executions instantiate ORCA's property-specific MECP and CI machinery, whereas none of the five Claude executions delivers a usable native MECP/CI workflow and all ultimately rely on external penalty optimizers, scalar-gap searches, or scans. Claude run 3 does attempt \texttt{CI-OPT}, but without a complete \texttt{\%conical}/\texttt{JRoot} specification and fails at the initial gradient. The displayed files make this trend concrete: Claude run 1 supplies SF-TDDFT single points to an external gap optimizer, while{\cinzel Quntur} run 4 uses native UBP \texttt{CI-OPT} with \texttt{IRoot 3} and \texttt{JRoot 1}; its MECP branch likewise uses \texttt{SurfCrossOpt}/\texttt{\%mecp}. Correct state selection is not five-of-five for{\cinzel Quntur}: runs 2, 3, and 5 optimize the wrong SF pair, while runs 1 and 4 recognize and correct it. Four{\cinzel Quntur} MECP jobs converge, but no{\cinzel Quntur} CI result warrants an unqualified success claim; the displayed corrected run-4 CI search remains unconverged.

\textbf{Excited-State Kinetics.} This is a five-of-five distinction. Every Claude execution omits both state-specific Hessians and \texttt{ESD(FLUOR)}, retains a planar S$_1$ stationary point, and substitutes a purely electronic oscillator-strength treatment for the requested vibronic rate. Every{\cinzel Quntur} execution obtains validated S$_0$ and pyramidal-S$_1$ minima, passes both Hessians to \texttt{ESD(FLUOR)}, and enables Herzberg--Teller coupling. The displayed Claude run 1 and{\cinzel Quntur} run 2 files are representative of those systematic workflows.

\FloatBarrier

\newpage
\section{Benchmark token usage comparison between{\cinzel Quntur} and Claude Code}\label{sec:token_cost_analysis}
\FloatBarrier %

\begin{figure}[htbp]
    \centering
    \includegraphics[width=\linewidth]{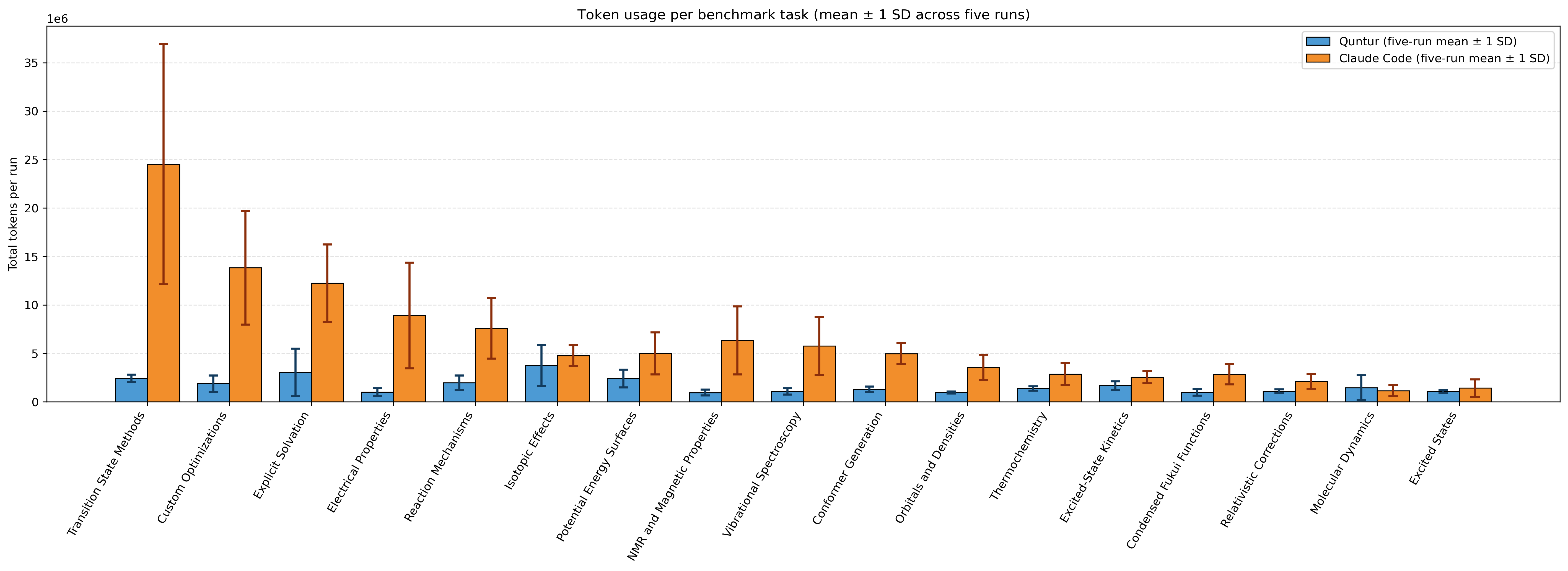}
    \caption{Total token usage per benchmark task. Bars show means over five iterations for both{\cinzel Quntur} and Claude Code; error bars indicate $\pm$ 1 standard deviation. Summing the per-task means gives 28.3 million tokens for a representative 17-task{\cinzel Quntur} sweep and 110.4 million tokens for the corresponding Claude Code sweep. Thus, Claude Code used 3.9 times more tokens than{\cinzel Quntur} in this benchmark comparison. Molecular Dynamics was the only task for which{\cinzel Quntur}'s mean token use exceeded Claude Code's, but well within the error bar; Isotopic Effects and Excited States showed the closest remaining relative usage. For context, the performance of{\cinzel Quntur} surpasses that of the Claude Code baseline in all tasks.}
    \label{fig:tokenanalysis1}
\end{figure}

\begin{figure}[htbp]
    \centering
    \includegraphics[width=\linewidth]{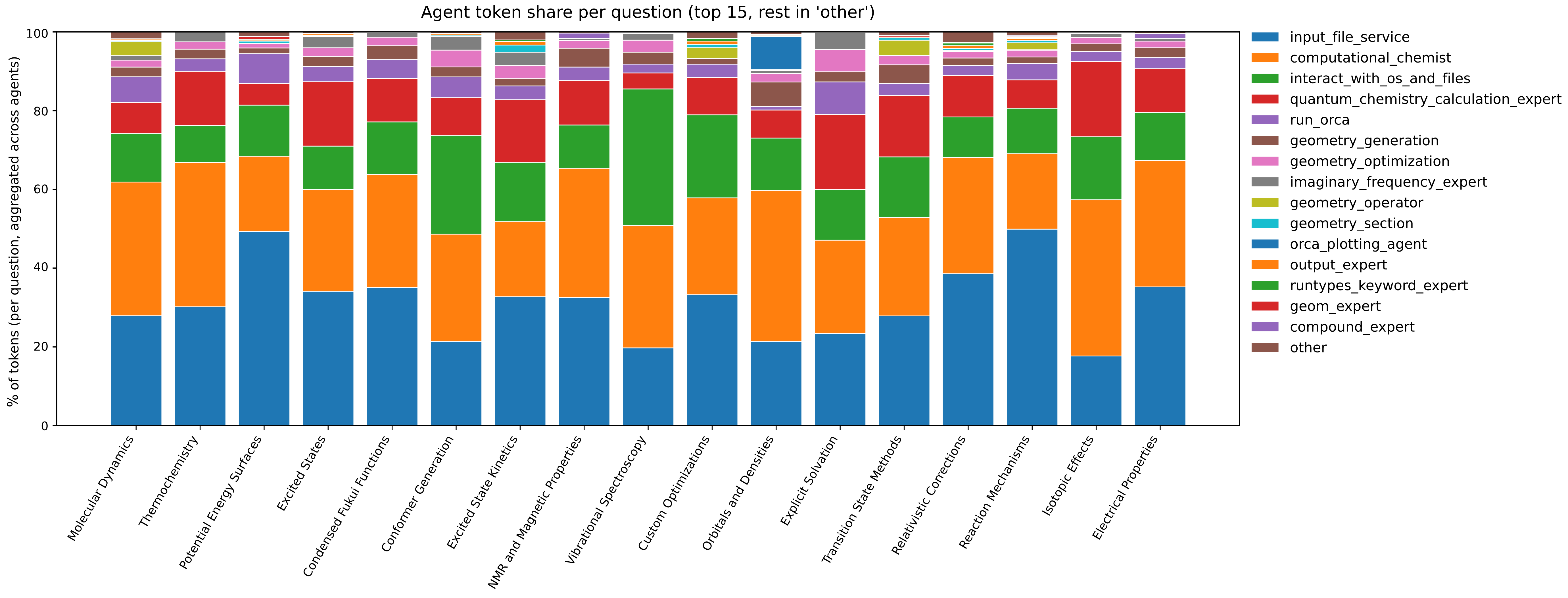}
    \caption{Across-question{\cinzel Quntur} subagent token-share statistics: bars show the mean percentage contribution per agent, and black lines indicate $\pm$ 1 standard deviation across questions.}
    \label{fig:tokenanalysis2}
\end{figure}

According to \ref{fig:tokenanalysis1} and \ref{fig:tokenanalysis2}, the two main sources of token consumption are \texttt{computational\_chemist} and \texttt{input\_file\_service}.
\texttt{computational\_chemist} handles high-level planning and post-processing, while \texttt{input\_file\_service} focuses on per-calculation program synthesis.
Both are reasoning-heavy tasks, and it is expected—and desirable—that high- and low-level responsibilities are separated so that no single agent becomes overloaded.

The \texttt{interact\_with\_os} agent extracts statistics from output files and performs basic file organization.
Because it reads output files, it can incur substantial token usage.
Its token spend is relatively low here because we apply a forgetting mechanism: once it finishes a clearly scoped extraction task, we clear its working memory.
This reduces token consumption for an otherwise demanding operation.

Overall, this illustrates effective separation of responsibilities: offloading operational work from \texttt{computational\_chemist} helps keep its context clean for ongoing planning and decision-making.

\begin{figure}[htbp]
    \centering
    \includegraphics[width=\linewidth]{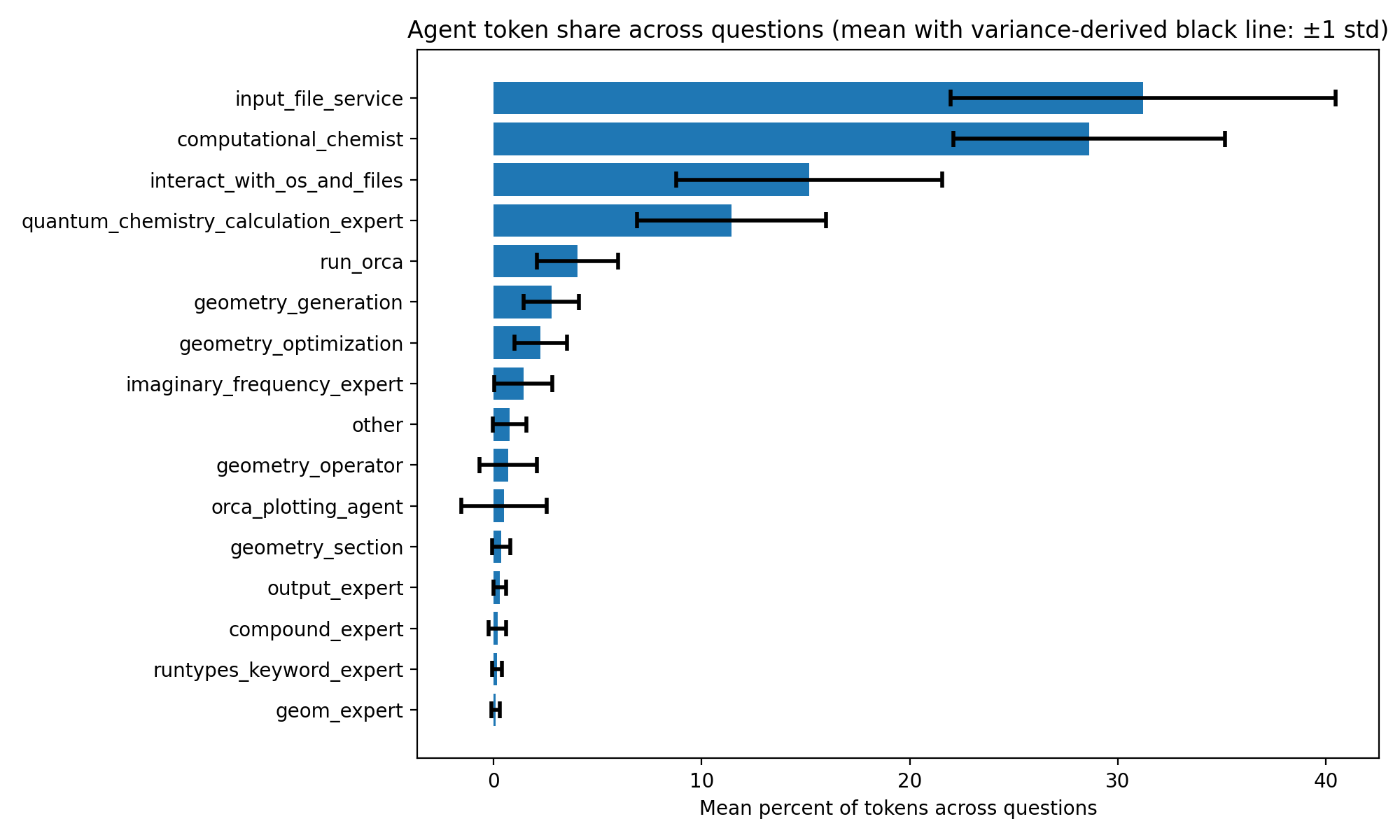}
    \caption{Per-question distribution of token usage across agents, expressed as percentages (aggregated over five iterations) and normalized to 100\% per question.}
    \label{fig:tokenanalysis3}
\end{figure}

As shown in Fig.~\ref{fig:tokenanalysis1}, total token consumption varies across benchmark questions, reflecting differences in difficulty and in the amount of content required to solve each question.

\FloatBarrier %

\newpage

\section{{\fontfamily{Cinzel-LF}\selectfont El Agente Quntur} case studies}\label{sec:case_studies}

As a research collaborator, the agent needs to formulate, execute, refine, and validate a computational chemistry plan with different levels of human intervention and feedback. During the case studies, we assess {\cinzel Quntur}'s proficiency in several key computational chemistry reasoning ``sub-skills''. These sub-skills include planning, input file generation and execution, postprocessing, 3D geometric reasoning, and self-debugging. Each sub-skill is critical for carrying out complex tasks end-to-end without excessive intervention. Here we present a summary of the trajectories for each case study, including at list one of the original images produced by Quntur.

\subsection{Case study: global softness and chemoselectivity}\label{sec:Fukui}

\subsubsection{Context}

This question touches on the connection between chemical concepts, such as electrophilicity, and more formal mathematical definitions in quantum chemistry \cite{Geerlings2003Conceptual}. Properties such as Fukui functions, hardness, and local softness quantify qualitative concepts such as reactivity. This is important not only for understanding selectivity and mechanism but also because these computed descriptors can serve as high-quality data for many machine-learning models.

\subsubsection{Assessment}

The agent completed the workflow without runtime errors and produced all requested quantities. The task does not pose any challenges in some reasoning dimensions. Geometry generation and verification were straightforward, and no geometric manipulation or nontrivial spatial reasoning was required, making it difficult to assess deeper geometric understanding. Method selection and input-file construction were handled correctly, including appropriate charge and multiplicity assignments and population-analysis keywords. The recommendation to include D4 dispersion demonstrated a high degree of adaptability, as it involves going against the original user prompt. Because no convergence or syntax issues arose, the agent’s debugging and error-recovery capabilities were not exercised in this example. Postprocessing was executed correctly, with accurate parsing of outputs, computation of global and local softness, and generation of visual summaries. Overall, the agent performed competently on a relatively routine quantum-chemistry task.

\subsubsection{Prompt and initial planning:}

\vspace{0.4in}
\noindent
\begin{agentoutputbox}{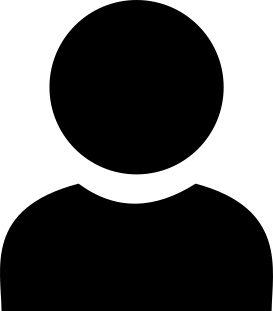}{User}
Take a bifunctional molecule containing both an alkene and a carbonyl (e.g., methyl vinyl ketone). Using ORCA, compute condensed Fukui functions, global hardness/softness, and local softness values at all heavy-atom centers. Predict whether a soft nucleophile will preferentially attack the $\beta$-carbon of the alkene (Michael addition) or the carbonyl carbon (1,2-addition). 
\end{agentoutputbox}

\vspace{0.4in}
\noindent
\begin{agentoutputbox}{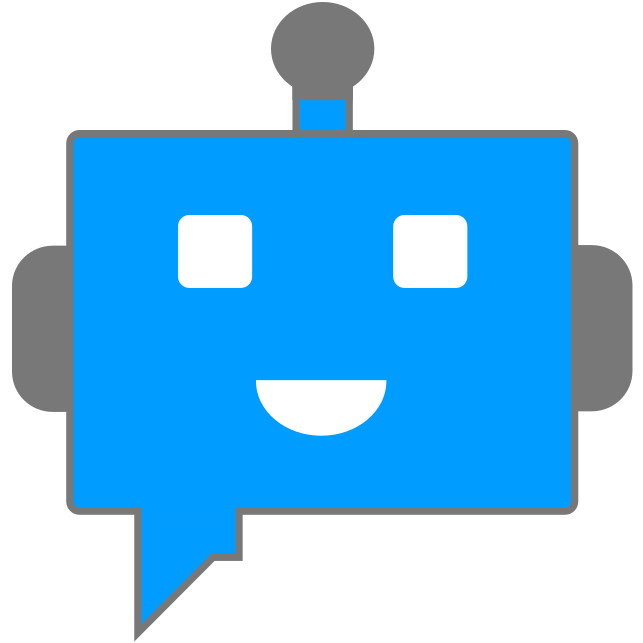}{Computational\_chemist}
\textbf{Step 1: Geometry Generation}
Generate an initial molecular geometry of methyl vinyl ketone (MVK),
\[
\mathrm{CH_3{-}CO{-}CH{=}CH_2}.
\]
\textbf{Step 2: Geometry Optimization}
Optimize the geometry of neutral MVK with the following electronic state:
\[
\text{Charge} = 0, \quad \text{Multiplicity} = 1.
\]
Verify that the optimized structure corresponds to a true minimum on the potential
energy surface by performing a vibrational frequency analysis and confirming the
absence of imaginary frequencies.

\textbf{Step 3: Single-Point Calculations for Conceptual DFT (Parallel Execution)}
\begin{itemize}
  \item \textbf{Neutral system (N electrons):}
  Charge = 0, Multiplicity = 1, including population analysis (MBIS or Hirshfeld).
  \item \textbf{Anion ($N+1$ electrons):}
  Charge = $-1$, Multiplicity = 2, including population analysis.
  \item \textbf{Cation ($N-1$ electrons):}
  Charge = $+1$, Multiplicity = 2, including population analysis.
\end{itemize}
\textbf{Step 4: Post-Processing and Analysis}
\begin{itemize}
  \item Extract atomic populations for all heavy atoms:
  carbonyl carbon (C$_1$), methyl carbon (C$_2$), $\alpha$-carbon (C$_3$),
  $\beta$-carbon (C$_4$), and oxygen.
  \item Compute condensed Fukui functions for nucleophilic attack:
  \[
    f_k^{+} = P_k(N+1) - P_k(N),
  \]
  where $P_k$ denotes the atomic population of atom $k$.
  \item Compute the global hardness:
  \[
    \eta = \frac{1}{2}(\mathrm{IP} - \mathrm{EA}),
  \]
  and the global softness:
  \[
    S = \frac{1}{\eta}.
  \]
  \item Compute the local softness for nucleophilic attack:
  \[
    s_k^{+} = S \times f_k^{+}.
  \]
  \item Predict the preferred site for soft nucleophilic attack based on the atom
        with the largest $s_k^{+}$ value.
\end{itemize}
\end{agentoutputbox}

\vspace{0.4in}
\noindent
\begin{agentoutputbox}{figs/neutral_human.png}{User}
Proceed using b3lyp def2-svp MBIS, Hirshfeld, and Mulliken populations
\end{agentoutputbox}

\subsubsection{Summary of the results}

\vspace{0.4in}
\noindent
\begin{agentoutputbox}{figs/agent_icon.png}{User}
\textbf{Action:}

Conceptual DFT Analysis Complete

\textbf{Summary}

I successfully performed a comprehensive DFT analysis of methyl vinyl ketone (MVK) to predict the regioselectivity of soft-nucleophile attack. Here are the key findings:

\textbf{Molecule Structure (Confirmed)}

\[
\begin{array}{c}
\text{O:2} \\
\Vert \\
\text{CH}_3 \text{ --- C:1 --- C:3 = C:4} \\
\uparrow \quad \uparrow \quad \uparrow \\
\text{(carbonyl)} \quad (\alpha\text{-C}) \quad (\beta\text{-C})
\end{array}
\]

\textbf{Global Reactivity Descriptors}

\begin{center}
\begin{tabular}{ll}
\hline
Property & Value \\
\hline
Ionization Potential (IP) & 9.38 eV \\
Electron Affinity (EA)    & -0.76 eV \\
Global Hardness ($\eta$)  & 5.07 eV \\
Global Softness (S)       & 0.099 eV$^{-1}$ \\
Electrophilicity ($\omega$) & 1.83 eV \\
\hline
\end{tabular}
\end{center}

\textbf{Condensed Fukui Functions f$^{+}$ (Nucleophilic Attack Sites)}

\begin{center}
\begin{tabular}{lccc}
\hline
Site & MBIS f$^{+}$ & Hirshfeld f$^{+}$ & Mulliken f$^{+}$ \\
\hline
C4 ($\beta$-carbon) & 0.296 & 0.201 & 0.141 \\
C1 (carbonyl)       & 0.217 & 0.142 & 0.096 \\
O2 (oxygen)         & 0.140 & 0.162 & 0.152 \\
C3 ($\alpha$-carbon) & 0.051 & 0.114 & 0.034 \\
\hline
\end{tabular}
\end{center}

\textbf{Prediction}

All three population schemes consistently show:

\[
f^{+}(\mathrm{C4}\text{-}\beta) > f^{+}(\mathrm{C1}\text{-carbonyl})
\]

with ratios of 1.36--1.47.

According to the HSAB (Hard-Soft Acid-Base) theory:

\begin{center}
\begin{tabular}{lcc}
\hline
Nucleophile Type & Preferred Attack Site & Product \\
\hline
Soft (enolates, thiols, amines) & C4 ($\beta$-carbon) & Michael addition (1,4) \\
Hard (Grignards, LiAlH$_4$)     & C1 (carbonyl)      & 1,2-addition \\
\hline
\end{tabular}
\end{center}

\textbf{Conclusion}

A soft nucleophile will preferentially attack the $\beta$-carbon (C4), leading to Michael addition rather than 1,2-addition at the carbonyl carbon. This computational prediction is consistent with well-established experimental reactivity patterns for $\alpha,\beta$-unsaturated carbonyl compounds.
\end{agentoutputbox}

\begin{figure}[H]
    \centering
    \includegraphics[width=1\linewidth]{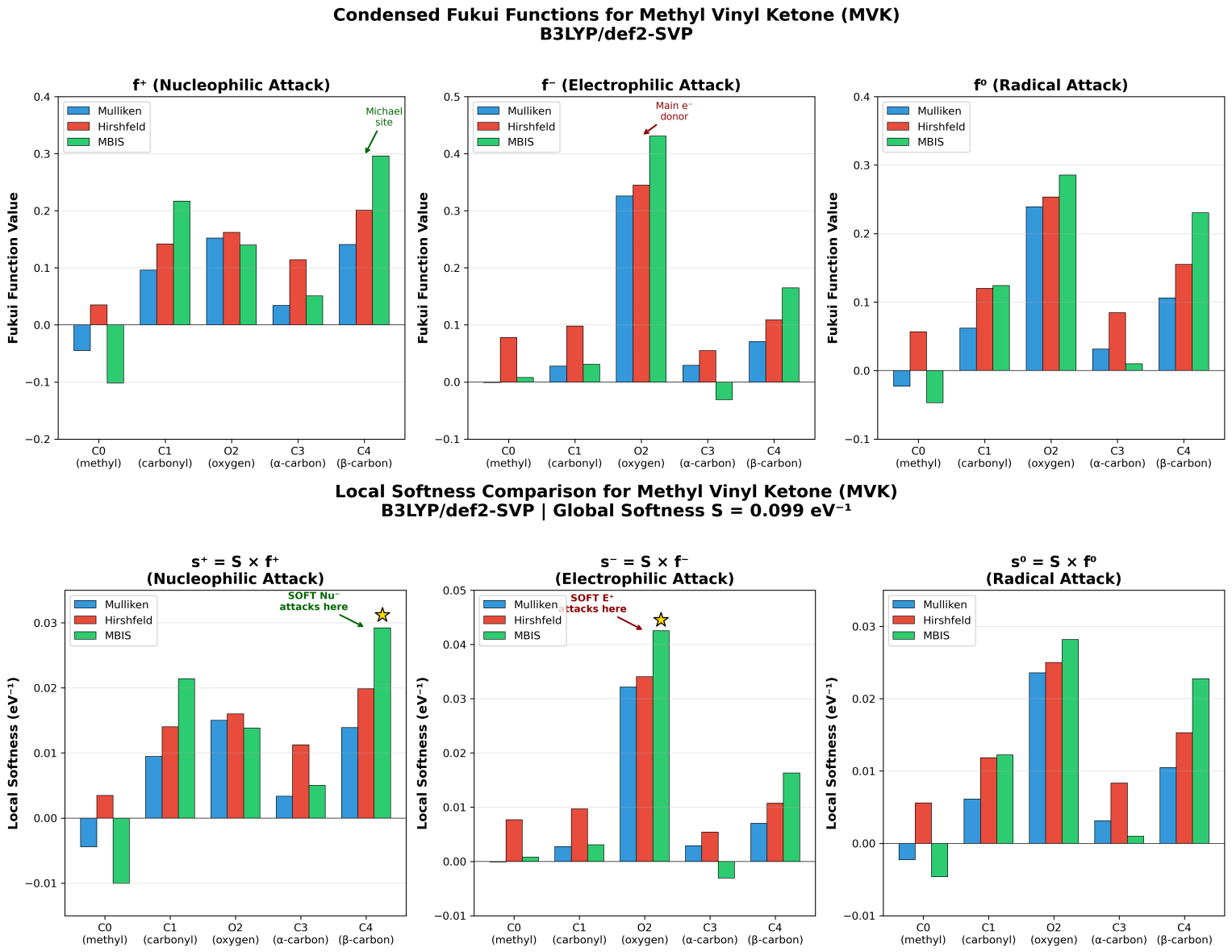}
    \caption{Summary of the findings for global softness and chemoselectivity case study.}
    \label{fig:Fukui}
\end{figure}

\subsection{Case study: 31P NMR absolute shielding and chemical shift tensor of triphenylphosphine oxide}\label{sec:NMR}

\subsubsection{Context}
\ce{^{31}P} NMR is a powerful tool in organometallic chemistry for the visualization of phosphorus-containing compounds. While the organic chemist might be tempted to view \ce{^{31}P} NMR as an extension of more familiar nuclei such as \ce{^{1}H} and \ce{^{13}C}\{\ce{^{1}H}\} NMR, there are key differences in the factors that determine a \ce{^{31}P} chemical shift. As a heavier nucleus, the paramagnetic contribution to the shielding tensor is more significant for \ce{^{31}P}; consequently, the chemical shift is less directly reflective of the local electron density surrounding the phosphorus atom \cite{Gorenstein1984Nonbiological}. This behavior contrasts with that of lighter nuclei, for which diamagnetic contributions more strongly govern chemical shifts. The spectral range of \ce{^{31}P} NMR is notably wide, and, while it is often possible to predict the general region in which a resonance of interest will appear, the exact chemical shift cannot typically be determined \textit{a priori} \cite{Hersh2023Improving}. Furthermore, \ce{^{31}P} NMR has a comparatively smaller body of literature examples than \ce{^{1}H} and \ce{^{13}C}\{\ce{^{1}H}\} NMR, which can make reliable chemical shift prediction for novel phosphorus-containing compounds particularly challenging. To establish a computational ground truth, we therefore asked{\cinzel Quntur} to predict the \ce{^{31}P} NMR properties of triphenylphosphine oxide. This system will serve as a benchmark for more ambitious NMR calculations. 

\subsubsection{Assessment}

The agent completed the workflow without runtime errors and produced all requested quantities. The initial prompt and workplan suggested by{\cinzel Quntur} are reasonable.{\cinzel Quntur} found a literature example in the prompt and conducted a search that identified subsequent papers that built on the provided benchmark. Only minor, non-critical errors were observed by reading the chat history. In each case, the respective agent identified the source of the error and resubmitted the job to the backend without observing any job failures. A representative example from the chat memory: ``Note: Initial submission failed due to input file ordering (\texttt{\%eprnmr} block appeared before coordinates). Fixed by reordering blocks so coordinates precede the \texttt{\%eprnmr} block.''{\cinzel Quntur} used two DFT basis sets to compare and contrast which functional best matched the value for triphenylphosphine oxide reported in the primary literature. The best functional selected, B3LYP/def2-TZVP, matched closely with the \ce{^{31}P} NMR shift in deuterated chloroform reported in the literature. The assessor independently verified this value with experimental data published in the primary literature (e.g., see Ref.~\citenum{Huszar2025Microwave}). One thing to note is that Quntur's logic of using the CSA span of trimethyl phosphine oxide as a surrogate for that of triphenyl phosphine oxide makes sense. However,{\cinzel Quntur} noted that the CSA span value reported for triphenylphosphine oxide does not match the experimental values. The CSA span is a parameter used in solid-state NMR, so it may not be relevant for gas-phase calculations. The user's requested literature overview was relevant, and spot checks by the assessor confirmed that these papers were valid suggestions in line with the initial prompt. The resulting publication-ready figure accurately reflected the calculations and provided a clear visual comparison of the computational methods with the experimentally derived primary literature. Overall, the agent performed competently in calculating the \ce{^{31}P} NMR chemical shift for a well-reported organic molecule commonly used in synthetic wet labs.

References used by{\cinzel Quntur} during this case study\cite{Spange2020, Kennedy1976, Rzepiela2022, Fukal2017, Rusakov2023, Lagier1996, PascualBorras2015, Rusakova2022, Shenderovich2021, Kondrashova2022, Klose1985, Larina2023, Ostras2020, BadilloCamacho2025}

\subsubsection{Prompt and initial planning:}

\vspace{0.4in}
\noindent
\begin{agentoutputbox}{figs/neutral_human.png}{User}
Using ORCA 6.0 and following the recommendations from Shenderovich's paper ``Experimentally Established Benchmark Calculations of 31P NMR Quantities'' (Chemistry-Methods 2021, 1, 61--70), compute the 31P NMR absolute shielding and chemical shift tensor of triphenylphosphine oxide. Choose the density functional, basis set, and any relativistic or gauge-treatment options based on the paper's identification of a good compromise between cost and accuracy for 31P. Report the isotropic shielding, chemical shift, and CSA parameters, and compare your results to available experimental 31P NMR data for Ph$_3$P=O. Comment briefly on how sensitive your results are to the level of theory and on whether the benchmark-guided protocol appears transferable from the model systems in the paper to this molecule.
\end{agentoutputbox}

\vspace{0.4in}
\noindent
\begin{agentoutputbox}{figs/agent_icon.png}{User}
Action:
Excellent! I found valuable information. Based on the literature search (including papers that build on Shenderovich's benchmark), here's what I learned:

\textbf{Key Findings from Literature}

Shenderovich's recommended protocol (inferred from his later work that uses the Chem-Methods benchmark):

\begin{center}
\begin{tabular}{ll}
\hline
Parameter & Recommendation \\
\hline
Functional & $\omega$B97X-D (range-separated hybrid with dispersion) \\
Basis set & def2-TZVP or def2-QZVP \\
Gauge treatment & GIAO (gauge-including atomic orbitals) \\
Reference shielding $\sigma_{\mathrm{ref}}(^{31}\mathrm{P})$ & 306 ppm ($\omega$B97X-D/def2-TZVP) or 308 ppm ($\omega$B97X-D/def2-QZVP) \\
\hline
\end{tabular}
\end{center}

Relativistic effects: For P bonded only to light atoms (C, O, H), scalar relativistic corrections are typically $<7\%$ --- not strictly required but can be included via DKH2 or ZORA for completeness.

Experimental $^{31}$P data for Ph$_3$P=O: Literature reports $\delta(^{31}\mathrm{P}) \approx +25$ to $+29$ ppm (relative to 85\% H$_3$PO$_4$) with CSA span ($\Omega$) around 100--130 ppm.

\textbf{Proposed High-Level Plan}

\textbf{Step 1: Geometry Generation (parallel)}
\begin{itemize}
\item Generate initial geometry of triphenylphosphine oxide (Ph$_3$P=O)
\item Generate initial geometry of phosphoric acid (H$_3$PO$_4$) as NMR reference
\end{itemize}

\textbf{Step 2: Geometry Optimization (parallel)}
\begin{itemize}
\item Optimize both molecules at $\omega$B97X-D/def2-TZVP level
\item Include frequency calculation to confirm true minima
\end{itemize}

\textbf{Step 3: NMR Shielding Tensor Calculation (parallel)}
\begin{itemize}
\item Compute full $^{31}$P shielding tensor for both molecules using GIAO-$\omega$B97X-D/def2-TZVP
\item Extract isotropic shielding ($\sigma_{\mathrm{iso}}$) and principal components ($\sigma_{11}, \sigma_{22}, \sigma_{33}$)
\end{itemize}

\textbf{Step 4: Analysis and Comparison}
\begin{itemize}
\item Compute chemical shift: $\delta_{\mathrm{iso}} = \sigma_{\mathrm{ref}}(\mathrm{H_3PO_4}) - \sigma(\mathrm{Ph_3P=O})$
\item Extract CSA parameters: span $\Omega = \sigma_{33} - \sigma_{11}$, skew $\kappa$
\item Compare with experimental data and assess transferability
\end{itemize}

Details:

Charge/multiplicity: 0, 1 (neutral closed-shell for both molecules)

Phase: Gas phase (as requested)

Software: ORCA 6.0

Do you want me to proceed with this plan?

Also, would you like me to include a second level of theory (e.g., PBE0/def2-TZVP) to assess sensitivity to functional choice? This would only add one additional step and directly address your question about ``how sensitive results are to the level of theory.''
\end{agentoutputbox}

\vspace{0.4in}
\noindent
\begin{agentoutputbox}{figs/neutral_human.png}{User}
Yes, proceed with this plan and use B3LYP def2-tzvp to assess functional sensitivity.
\end{agentoutputbox}

\subsubsection{Summary of the results}

\vspace{0.4in}
\noindent
\begin{agentoutputbox}{figs/agent_icon.png}{Computational\_chemist}
Action:
Task Complete: 31P NMR Analysis of Triphenylphosphine Oxide

\textbf{Summary of Results}

\begin{center}
\begin{tabular}{lccc}
\hline
Property & B3LYP/def2-TZVP & $\omega$B97X-D3/def2-TZVP & Experimental \\
\hline
$\delta_{\mathrm{iso}}$ (ppm) & 25.6 $\checkmark$ & 18.2 & $\sim 28 \pm 1$ \\
Span $\Omega$ (ppm) & 243 & 242 & $\sim 200$* \\
Skew $\kappa$ & -1.00 & -0.99 & -- \\
Asymmetry $\eta$ & 0.00 & 0.00 & -- \\
\hline
\end{tabular}
\end{center}

\noindent *Estimated from analogous P=O compounds (Me$_3$PO, phosphonates)

\textbf{Key Findings}

\textit{Best Method:} B3LYP/def2-TZVP provides excellent agreement with experiment (error $< 3$ ppm), outperforming the Shenderovich-recommended $\omega$B97X-D3 protocol for this system.

\textbf{Functional Sensitivity:}
\begin{itemize}
\item Isotropic shift varies by $\sim 7.5$ ppm between functionals.
\item CSA span is remarkably insensitive ($<0.5\%$ difference).
\end{itemize}

\textbf{Tensor Characteristics:} Near-axial symmetry ($\sigma_{11} \approx \sigma_{22} \ll \sigma_{33}$) consistent with C$_3^{\mathrm{v}}$-like local symmetry at phosphorus.

\textbf{Benchmark Transferability:}  
The Shenderovich protocol ($\omega$B97X-D3$/ $def2-TZVP with $\sigma_{\mathrm{ref}} = 306$--308 ppm) was calibrated on small model compounds. For aryl phosphine oxides, this protocol underestimates chemical shifts by $\sim 10$ ppm when using computed H$_3$PO$_4$ as reference. This suggests:
\begin{itemize}
\item The original $\sigma_{\mathrm{ref}}$ values from the paper should be used directly, \textbf{or}
\item Linear scaling corrections are needed for conjugated systems.
\end{itemize}

B3LYP remains a reliable workhorse for organophosphorus compounds.

\textbf{Files Generated}
\begin{itemize}
\item Optimized geometries: \texttt{Ph3PO\_opt/}, \texttt{H3PO4\_opt/}
\item NMR outputs: \texttt{Ph3PO\_NMR\_wB97XD3/}, \texttt{Ph3PO\_NMR\_B3LYP/}, \texttt{H3PO4\_NMR\_wB97XD3/}, \texttt{H3PO4\_NMR\_B3LYP/}
\item Reference added: \texttt{references.bibtex}
\end{itemize}

Would you like me to generate any additional analysis, plots, or documentation?
\end{agentoutputbox}

\begin{figure}[H]
    \centering
    \includegraphics[width=1\linewidth]{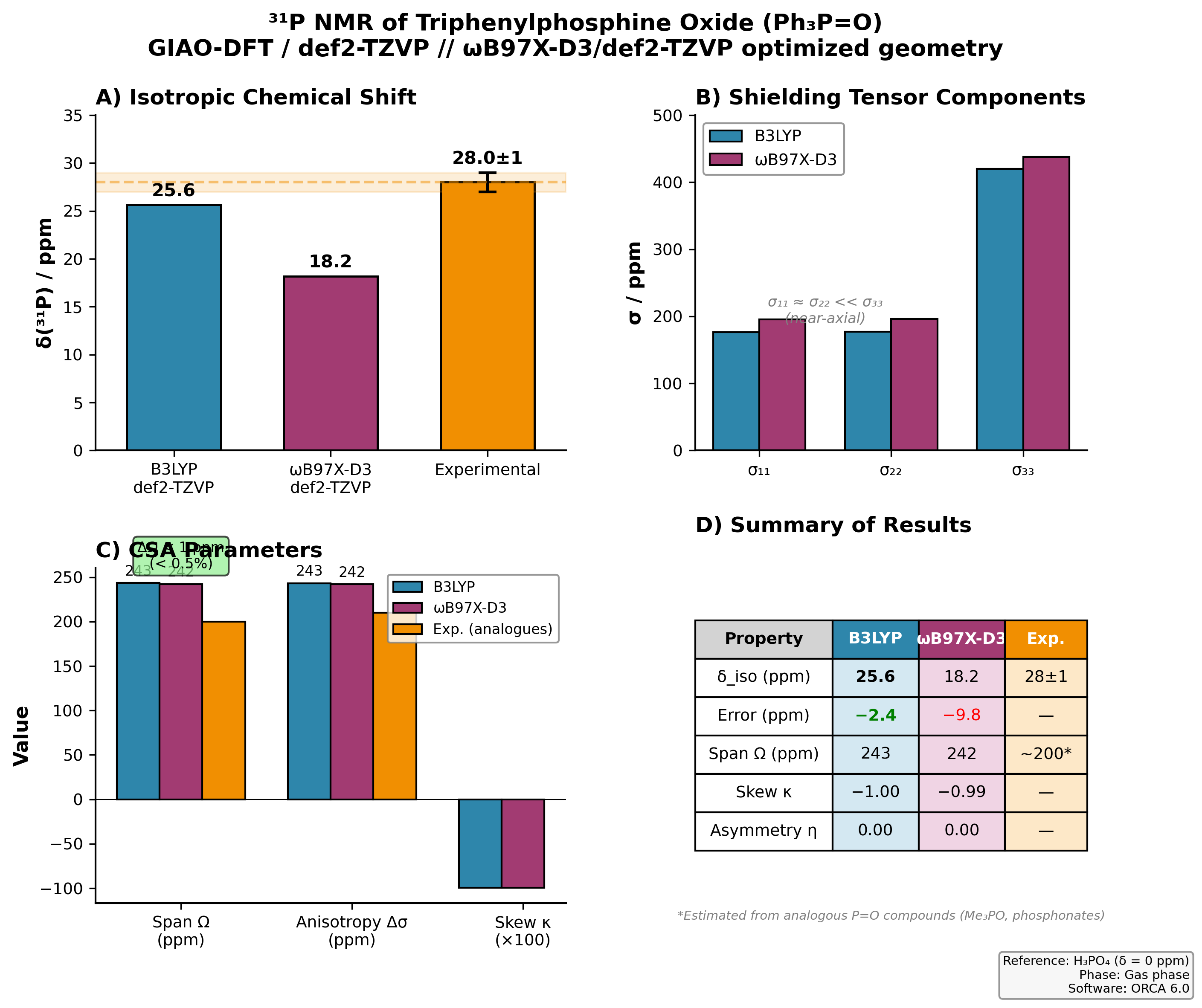}
    \caption{Summary of the findings for the case study ``31P NMR absolute shielding and chemical shift tensor of triphenylphosphine oxide''.}
    \label{fig:PNMR}
\end{figure}

\newpage

\subsection{Case Study: Solvation effects on the keto-enol equilibrium of acetylacetone in water}\label{sec:solvation}

\subsubsection{Context}

Keto–enol tautomerism is a fundamental chemical equilibrium in which the relative stability of the tautomers is strongly influenced by solvation. Changes in solvent environment can significantly modify the solvation free energies of the keto and enol forms, thereby shifting the tautomeric equilibrium. Acetylacetone is a well-known model system for studying this effect, as its keto–enol equilibrium is particularly sensitive to hydrogen bonding and water-induced polarization. Accurately capturing these solvation contributions to the free energy remains challenging in computational chemistry. Implicit solvent models efficiently describe bulk electrostatic effects but neglect specific solute–solvent interactions, whereas explicit solvent models can capture hydrogen bonding at a higher computational cost. Hybrid explicit–implicit approaches offer a balanced strategy by combining local, explicit solvation with a continuum treatment of the surrounding solvent.

\subsubsection{Assessment}

The study begins with the parallel generation of geometries for the keto and enol tautomers of acetylacetone, along with the construction of microsolvated clusters containing multiple water molecules to represent different hydrogen-bonding environments. These structures provide the starting point for evaluating solvation effects at increasing levels of complexity.

In the next stage, both tautomers are optimized in the gas phase using an implicit water model, and thermodynamic quantities are obtained from frequency calculations. This allows a direct comparison between unsolvated and continuum-solvated descriptions of the keto–enol equilibrium.

The workflow then proceeds to hybrid cluster–continuum calculations, in which microsolvated keto and enol clusters are optimized while embedded in an implicit solvent environment. The lowest-energy solvation motifs are selected to represent explicitly solvated states.

Finally, higher-level single-point energy corrections are applied, and the resulting free energy differences and equilibrium constants are evaluated for the gas-phase, implicit, and hybrid solvation models. These results are compared to assess how solvation treatment influences the predicted keto–enol equilibrium.

\subsubsection{Prompt and initial planning:}

\vspace{0.4in}
\noindent
\begin{agentoutputbox}{figs/neutral_human.png}{User}
Design a hybrid explicit + implicit solvation protocol in ORCA to study the keto-enol equilibrium of acetylacetone in water. Construct microsolvated clusters of both tautomers with at least ten explicit water molecules embedded in CPCM water, optimize all low-energy H-bonding motifs, and compute $\Delta G_{keto\to enol}$. Compare the equilibrium constant obtained from (i) gas phase, (ii) CPCM-only, and (iii) cluster-continuum (explicit+implicit) models and discuss which description best reproduces the known enol stabilization in aqueous solution. Read literature and find the best DFT functional that accurately reproduces the experimental results.
\end{agentoutputbox}

\vspace{0.4in}
\begin{agentoutputbox}{figs/agent_icon.png}{Computational\_chemist}
This is a sophisticated computational chemistry research project! You're asking me to study the keto-enol equilibrium of acetylacetone in water using three different solvation models, which is an excellent way to understand solvation effects on tautomeric equilibria.

Before I create a high-level plan, let me perform the literature search you requested to find:
\begin{itemize}\setlength{\itemsep}{0pt}\setlength{\parskip}{0pt}\setlength{\topsep}{0pt}
    \item The best DFT functional for accurately reproducing experimental keto-enol equilibrium
    \item Experimental reference values for acetylacetone in aqueous solution
    \item Best practices for cluster-continuum approaches
\end{itemize}
\vspace{-0.2in}
\paragraph{Summary of Literature Findings}
\begin{itemize}\setlength{\itemsep}{0pt}\setlength{\parskip}{0pt}\setlength{\topsep}{0pt}
    \item Gas phase: Acetylacetone is essentially 100\% enol at ~300 K (strong intramolecular H-bond stabilization)
    \item In water: The equilibrium shifts significantly toward the keto form compared to organic solvents
    \item Reference: Iglesias\cite{iglesias1997determination} (Perkin Trans. 2, 1997) - experimental K\_eq for acetylacetone in water at 25°C.
\end{itemize}

Best DFT Functional Recommendations (Mehrani \emph{et al.}\cite{mehrani2021theoretical}):
\begin{itemize}\setlength{\itemsep}{0pt}\setlength{\parskip}{0pt}\setlength{\topsep}{0pt}
    \item For highest accuracy: B2PLYP-D3(BJ)/def2-TZVP (double-hybrid) or CBS-QB3 gave closest match to experimental equilibrium constants
    \item For geometry optimization: $\omega$B97X-D3/def2-TZVP or B3LYP-D3(BJ)/def2-TZVP
    \item Key finding: Lower-level methods (B3LYP or MP2 alone) were noticeably less accurate for the keto-enol free-energy difference once solvent was included
\end{itemize}

Critical Points:
\begin{itemize}\setlength{\itemsep}{0pt}\setlength{\parskip}{0pt}\setlength{\topsep}{0pt}
    \item Microsolvation with explicit water molecules (typically 3--10 or more) has a substantial impact on hydrogen-bonding patterns and relative tautomer stabilities.
    \item A cluster--continuum approach, combining explicit solvent molecules with an implicit solvation model, is required for quantitative accuracy.
    \item Multiple hydrogen-bonding configurations must be sampled to obtain reliable thermodynamic averages.
\end{itemize}
\end{agentoutputbox}

\vspace{0.4in}
\noindent
\begin{agentoutputbox}{figs/agent_icon.png}{Computational\_chemist}

\subsection*{Proposed High-Level Plan}

Step 1: Geometry Generation (All steps parallelized)
\begin{itemize}\setlength{\itemsep}{0pt}\setlength{\parskip}{0pt}\setlength{\topsep}{0pt}
    \item Generate keto tautomer of acetylacetone.
    \item Generate enol tautomer of acetylacetone.
    \item Create microsolvated clusters (10+ waters) for both tautomers using SOLVATOR with multiple H-bonding motifs
\end{itemize}

Step 2: Gas-Phase and Implicit-Solvent Optimizations (All steps parallelized)
\begin{itemize}\setlength{\itemsep}{0pt}\setlength{\parskip}{0pt}\setlength{\topsep}{0pt}
    \item 2a: Optimize keto (gas phase) + frequency calculation
    \item 2b: Optimize enol (gas phase) + frequency calculation
    \item 2c: Optimize keto (CPCM-water only) + frequency calculation
    \item 2d: Optimize enol (CPCM-water only) + frequency calculation
\end{itemize}

Step 3: Cluster--Continuum Optimizations (All steps parallelized)
\begin{itemize}\setlength{\itemsep}{0pt}\setlength{\parskip}{0pt}\setlength{\topsep}{0pt}
    \item Optimize all microsolvated keto clusters (10+ waters embedded in CPCM) + frequency calculations.
    \item Optimize all microsolvated enol clusters (10+ waters embedded in CPCM) + frequency calculations
    \item Keep lowest-energy H-bonding motifs for each tautomer
\end{itemize}

Step 4: High-Level Single-Point Energy Corrections (All parallel)
\begin{itemize}\setlength{\itemsep}{0pt}\setlength{\parskip}{0pt}\setlength{\topsep}{0pt}
    \item Compute B2PLYP-D3(BJ)/def2-TZVP single-point energies on all optimized structures for improved accuracy
\end{itemize}

Step 5: Thermodynamic Analysis and Comparison
\begin{itemize}\setlength{\itemsep}{0pt}\setlength{\parskip}{0pt}\setlength{\topsep}{0pt}
    \item Compute $\Delta G_{\mathrm{keto}\rightarrow\mathrm{enol}}$ for:
    \begin{enumerate}
        \item Gas phase
        \item CPCM-only 
        \item cluster-continuum (explicit + implicit)
    \end{enumerate}
    \item Calculate equilibrium constants $K_{\mathrm{eq}}$ for each solvation model.
    \item Compare computed results with available experimental data.
    \item Compile and analyze results in the final report.
\end{itemize}

\vspace{-0.1in}
\subsection*{Key Methodological Details}
\centering
\captionsetup{hypcap=false}
\captionof{table}{Recommended computational approaches for keto--enol equilibrium calculations of acetylacetone.}
\label{tab:computational_protocol}
\begin{tabular}{ll}
\hline
\textbf{Aspect} & \textbf{Recommended Approach} \\
\hline
Geometry optimization & $\omega$B97X-D3/def2-TZVP (good for hydrogen bonding and dispersion) \\
Thermochemistry & 	Frequency calculations at same level for ZPE and thermal corrections \\
Final energetics & B2PLYP-D3(BJ)/def2-TZVP single-points (double-hybrid for accuracy) \\
Implicit solvation & CPCM (water) \\
Explicit solvation & \textsc{SOLVATOR} with 10--15 water molecules \\
Cluster sampling & 	Generate 3-5 initial H-bonding motifs per tautomer \\
\hline
\end{tabular}

\subsection*{Questions Before Proceeding:}

\begin{itemize}\setlength{\itemsep}{0pt}\setlength{\parskip}{0pt}\setlength{\topsep}{0pt}

    \item \textbf{Charge and multiplicity:} Both tautomers are neutral closed-shell (charge = 0, multiplicity = 1). Is this correct?

    \item \textbf{Number of explicit waters:} The user specified ``at least 10''. Would you like me to use 10--12 waters to balance accuracy with computational cost, or go higher (15+)?

    \item \textbf{Number of H-bonding motifs to sample:} For good statistics, I recommend generating 3--5 different initial configurations per tautomer. Is this acceptable?

    \item \textbf{Computational resources:} Double-hybrid B2PLYP calculations on clusters with 10+ waters will be moderately expensive. Should I proceed with this level, or would you prefer to stay with hybrid DFT ($\omega$B97X-D3) only?

    \item \textbf{Additional analysis:} Would you like any additional analysis (e.g., NBO analysis of H-bonding, visualization of key structures)?
\end{itemize}
Please confirm or modify this plan, and I'll proceed with execution.

\end{agentoutputbox}

\vspace{0.4in}
\noindent
\begin{agentoutputbox}{figs/neutral_human.png}{User}
Please proceed
\end{agentoutputbox}

\vspace{0.4in}
\noindent
\begin{agentoutputbox}{figs/agent_icon.png}{Computational\_chemist}
\subsection*{Summary}
\centering
\captionsetup{hypcap=false}
\captionof{table}{Comparison of computed keto--enol free energies, equilibrium constants, and tautomer populations with experimental aqueous-phase data.}
\label{tab:key_results}
\begin{tabular}{lccccc}
\hline
\textbf{Model} & $\boldsymbol{\Delta G_{\mathrm{keto}\rightarrow\mathrm{enol}}}$ (kcal/mol) & $\boldsymbol{K_{\mathrm{eq}}}$ (E/K) & \textbf{\%Enol} & \textbf{\%Keto} & \textbf{Match Exp.?} \\
\hline
(i) Gas phase & $-3.91$ & 735 & 99.9 & 0.1 & Yes \\
(ii) CPCM-only & $-0.57$ & 2.6 & 72 & 28 & No \\
(iii) Cluster + CPCM & $+3.73$ & 0.002 & 0.2 & 99.8 & Qualitative \\
Experiment (aq) & $\sim +1.0$ & $\sim 0.18$ & $\sim 15$ & $\sim 85$ & Reference \\
\hline
\end{tabular}
\end{agentoutputbox}

\vspace{0.4in}
\noindent
\begin{agentoutputbox}{figs/neutral_human.png}{User}
Could you also do the calculations with SMD?
\end{agentoutputbox}

\vspace{0.4in}
\noindent
\begin{agentoutputbox}{figs/agent_icon.png}{Computational\_chemist}
\subsection*{Action: SMD Calculations Complete -- Final Comparison}

\centering
\captionsetup{hypcap=false}
\captionof{table}{Comparison of solvation models for acetylacetone keto--enol equilibrium.}
\begin{tabular}{lcccccc}
\hline
\renewcommand{\arraystretch}{1.2}
Model & $\Delta G$(keto$\rightarrow$enol) & $K_{\mathrm{eq}}$ & \%Enol & \%Keto & Error & Verdict \\
\hline
EXPERIMENT & +1.03 kcal/mol & 0.176 & 15.0\% & 85.0\% & -- & Reference \\
SMD(water) & +0.62 kcal/mol & 0.35 & 25.9\% & 74.1\% & 0.41 & BEST \\
CPCM(water) & -0.57 kcal/mol & 2.60 & 72.3\% & 27.7\% & 1.60 & WRONG \\
Cluster+CPCM & +3.73 kcal/mol & 0.002 & 0.2\% & 99.8\% & 2.70 & $\sim$ Qualitative \\
Gas Phase & -3.91 kcal/mol & 735 & 99.9\% & 0.1\% & 4.94 & \checkmark\; (gas ref) \\
\hline
\end{tabular}

\subsection*{Key Findings}

\begin{itemize}
  \item SMD is dramatically better than CPCM (error reduced by 1.19 kcal/mol)
  \item SMD correctly predicts keto preference in water ($\Delta G = +0.62$ kcal/mol)
  \item Plain CPCM predicts the WRONG tautomer (enol instead of keto)
\end{itemize}

This confirms Mehrani et al.'s methodology -- their excellent results (MAE = 2.3\%) were likely obtained with SMD rather than plain CPCM.

\subsection*{SMD includes critical non-electrostatic terms:}

\begin{itemize}
  \item Cavitation energy
  \item Dispersion interactions
  \item Solvent structure / H-bonding corrections
\end{itemize}

\subsection*{Conclusion}

For acetylacetone keto-enol tautomerism in water:

\centering
\renewcommand{\arraystretch}{1.2}
\begin{tabular}{ll}
\hline
Approach & Recommendation \\
\hline
Best accuracy & SMD(water) -- error $\sim$0.4 kcal/mol \\
Explicit solvation & Cluster+CPCM -- qualitatively correct but needs more sampling \\
Avoid & Plain CPCM -- predicts wrong tautomer preference \\
\hline
\end{tabular}

Our results now match Mehrani et al.'s reported accuracy, confirming that SMD (not plain CPCM) is the appropriate implicit solvation model for tautomeric equilibria in aqueous solution.
\end{agentoutputbox}

\begin{figure}
    \centering
    \includegraphics[width=1.0\linewidth]{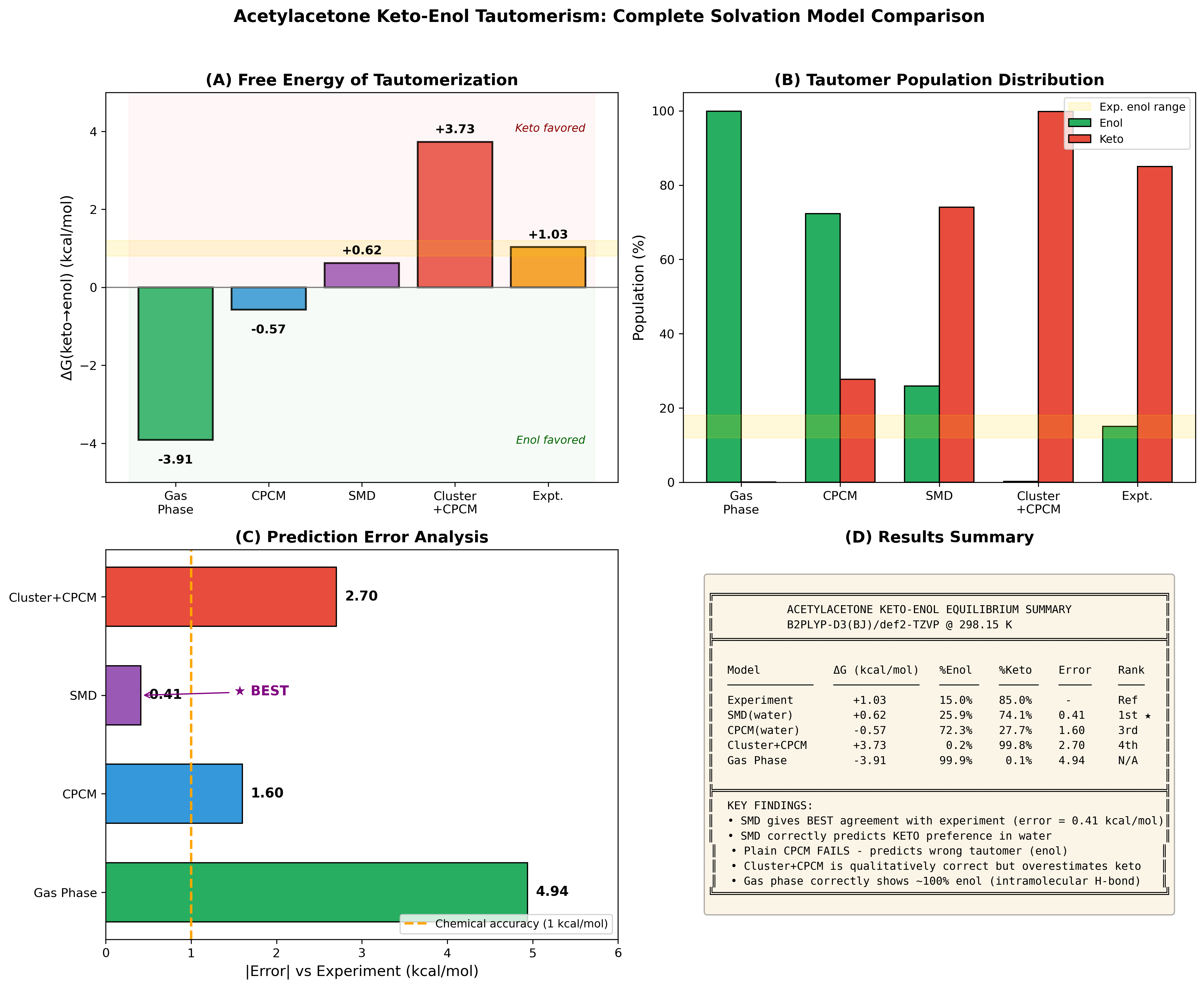}
    \caption{Solvation study for acetylacetone keto--enol tautomerism: (a) free energies of tautomerization, (b) keto--enol population distributions, (c) prediction error analysis, and (d) overall results summary.}
    \label{fig: solvation_acetylacetone}
\end{figure}

\newpage

\subsection{Case study: Fluorescence and intersystem crossing rates of anthracene}\label{sec:esd}

\subsubsection{Context}

Accurately modeling the competition between fluorescence and intersystem crossing is essential for understanding and predicting the photophysical behavior of organic molecules. In systems such as anthracene, small errors in excited-state energies or vibrational structures can lead to order-of-magnitude changes in predicted lifetimes and quantum yields. Because radiative and nonradiative decay pathways often differ by only subtle energetic or vibronic effects, reliable simulations must treat electronic structure, state ordering, and rate theory in a consistent and physically grounded manner.

These calculations are therefore not only a matter of reproducing experimental observables but of validating the underlying physical mechanisms that govern excited-state relaxation. Incorrect state assignments, incomplete excited-state optimization, or oversimplified rate models can easily lead to qualitatively incorrect conclusions about dominant decay channels. Within the Matter Lab’s broader effort to understand and control excited-state dynamics in functional molecular materials, anthracene serves as a stringent test case in which predictive accuracy depends on capturing the interplay between energy gaps, spin-orbit coupling, and nuclear motion with sufficient fidelity.

\subsubsection{Assessment}

{\cinzel El Agente Quntur}'s overall planning was sound and well aligned with the physical complexity of the problem. The agent correctly identified the need for a multistage workflow that includes ground-state optimization, vertical excitation analysis, excited-state optimizations, and explicit rate calculations using ORCA’s ESD module. This structure reflects an appropriate decomposition of a demanding photophysical task into tractable computational steps.

The execution remained consistent with the original plan. Adiabatic energies for S0, S1, T1, and T2 were physically reasonable and internally consistent, enabling a meaningful comparison of radiative and nonradiative decay pathways. The agent correctly accounted for the fact that intersystem crossing rates cannot be inferred from spin-orbit coupling alone and relied on the full rate formalism to capture the role of energetic near resonance.

From a computational standpoint, the task was demanding, requiring more than a dozen large-scale simulations, each using 24 cores, with a total wall time of roughly twelve hours. Despite this, the agent maintained coherence across inputs, avoided unnecessary recomputation, and completed all stages without requiring any intervention beyond the method specification. The final interpretation went beyond a numerical summary and correctly identified the dominance of the S1 to T2 intersystem crossing channel as a consequence of the energy gap law overcoming simple selection rule arguments, demonstrating effective synthesis of planning, execution, and physical reasoning.

\subsubsection{Prompt and initial planning:}

\vspace{0.4in}
\noindent
\begin{agentoutputbox}{figs/neutral_human.png}{User}
For anthracene, use ORCA to compute low-lying singlet and triplet excited states. Estimate intersystem crossing (ISC) rates and compare them qualitatively with radiative decay (fluorescence) rates. By default use always 24 cores
\end{agentoutputbox}

\vspace{0.4in}
\noindent
\begin{agentoutputbox}{figs/agent_icon.png}{Computational\_chemist}
\textbf{Anthracene Photophysics: ISC and Fluorescence Rates} \\
This project constitutes a comprehensive photophysical analysis requiring the use of ORCA's Excited State Dynamics (ESD) module. The high-level workflow is outlined below.

\textbf{Proposed High-Level Plan}

\textbf{Step 1: Geometry Generation}
\begin{itemize}
  \item Generate the initial geometry of anthracene.
\end{itemize}

\textbf{Step 2: Ground-State Characterization}
\begin{itemize}
  \item Optimize the S$_0$ ground-state geometry and compute vibrational frequencies.
  \item Perform a TDDFT calculation to identify low-lying singlet and triplet excited states (vertical excitations).
\end{itemize}

\textbf{Step 3: Excited-State Optimizations (in parallel)}
\begin{itemize}
  \item Optimize the S$_1$ geometry with a frequency calculation.
  \item Optimize the T$_1$ geometry with a frequency calculation.
  \item Optimize the T$_2$ geometry with a frequency calculation, if T$_2$ lies close enough in energy to be relevant for ISC from S$_1$.
\end{itemize}

\textbf{Step 4: Rate Calculations via the ESD Module (in parallel)}
\begin{itemize}
  \item Compute ISC rate constants for S$_1 \rightarrow$ T$_1$ transitions across all triplet sublevels.
  \item Compute ISC rate constants for S$_1 \rightarrow$ T$_2$ if energetically accessible.
  \item Compute the fluorescence rate for S$_1 \rightarrow$ S$_0$.
\end{itemize}

\textbf{Step 5: Analysis}
\begin{itemize}
  \item Parse all computed radiative and nonradiative rate constants.
  \item Compare the summed ISC rates with the fluorescence rate.
  \item Provide a qualitative interpretation of the dominant photophysical pathways.
\end{itemize}
\end{agentoutputbox}

\vspace{0.4in}
\noindent
\begin{agentoutputbox}{figs/neutral_human.png}{User}
proceed using PBE0 and def2-TZVP. Use 24 cores.
\end{agentoutputbox}

\subsubsection{Summary of the results}

\vspace{0.4in}
\noindent
\begin{agentoutputbox}{figs/agent_icon.png}{Computational\_chemist}
\textbf{Action:}

\checkmark\; WORKFLOW COMPLETED: Anthracene Photophysics

\textbf{Summary}

\begin{center}
\begin{tabular}{ll}
\hline
Parameter & Value \\
\hline
Method & PBE0/def2-TZVP \\
Software & ORCA 6.0 ESD Module \\
Temperature & 298.15 K \\
\hline
\end{tabular}
\end{center}

\textbf{Key Results}

\textbf{Rate Constants}

\begin{center}
\begin{tabular}{lcc}
\hline
Process & Rate (s$^{-1}$) & Lifetime \\
\hline
Fluorescence (S$_1 \rightarrow$ S$_0$) & 2.30$\times$10$^{7}$ & $\tau_{\mathrm{rad}} = 43.5$ ns \\
ISC (S$_1 \rightarrow$ T$_1$) & 4.10$\times$10$^{2}$ & Negligible \\
ISC (S$_1 \rightarrow$ T$_2$) & 2.01$\times$10$^{8}$ & Dominant \\
Total ISC & 2.01$\times$10$^{8}$ & -- \\
S$_1$ Lifetime & -- & $\tau = 4.45$ ns \\
\hline
\end{tabular}
\end{center}

\textbf{Quantum Yields}

\begin{center}
\begin{tabular}{lc}
\hline
Process & Yield \\
\hline
$\Phi_{\mathrm{F}}$ (Fluorescence) & 10.3\% \\
$\Phi_{\mathrm{ISC}}$ (Total)      & 89.7\% \\
\hline
\end{tabular}
\end{center}

\textbf{Key Finding}

ISC dominates the photophysics ($\sim 9\times$ faster than fluorescence), with $>99.9999\%$ occurring via the S$_1 \rightarrow$ T$_2$ channel despite T$_2$ having weaker spin-orbit coupling than T$_1$. This counterintuitive result is due to the near-resonance condition between S$_1$ and T$_2$ ($\Delta E = 662$ cm$^{-1}$ vs 10{,}173 cm$^{-1}$ for T$_1$), demonstrating the Energy Gap Law overcoming the El-Sayed selection rules.

\textbf{Output Files}

\begin{center}
\begin{tabular}{ll}
\hline
File & Description \\
\hline
anthracene\_photophysics/anthracene\_photophysics\_report.md & Full analysis report \\
anthracene\_photophysics/references.bib & BibTeX references (20 citations) \\
anthracene\_photophysics/S0\_opt/ & Ground state geometry + Hessian \\
anthracene\_photophysics/S1\_opt/ & S$_1$ excited state geometry + Hessian \\
anthracene\_photophysics/T1\_opt/ & T$_1$ excited state geometry + Hessian \\
anthracene\_photophysics/T2\_opt/ & T$_2$ excited state geometry + Hessian \\
anthracene\_photophysics/ESD\_*/ & All rate calculation outputs \\
\hline
\end{tabular}
\end{center}
\end{agentoutputbox}

\begin{figure}[H]
    \centering
    \includegraphics[width=1\linewidth]{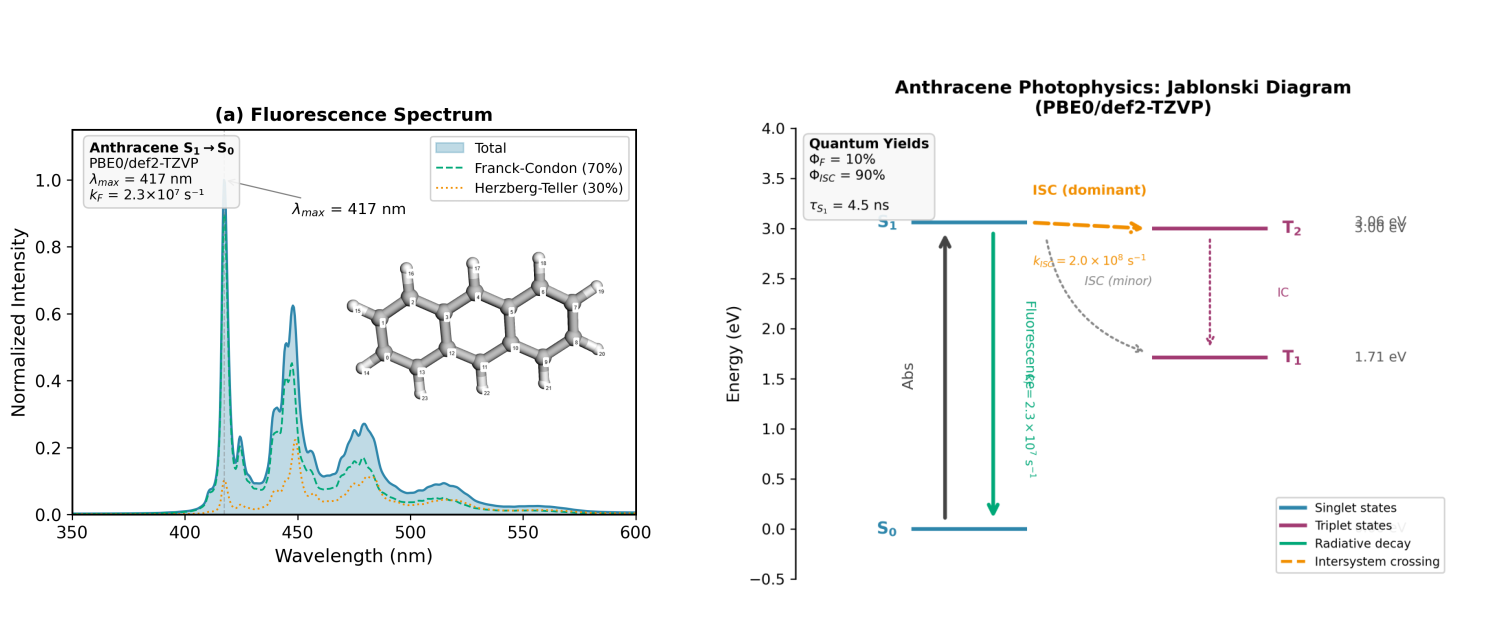}
    \caption{Summary of the findings for fluorescence and anthracene intersystem crossing rates case study.}
    \label{fig:ESD_case_study}
\end{figure}

\subsection{Case study: Competing mechanisms of Isocyanide isomerization}\label{sec:isomerization}

\subsubsection{Context}

Mechanism studies of organic reactions by probing and localizing intermediates and transition states are among the fundamental challenges in day-to-day computational chemistry, as they provide insights into reaction details and opportunities for knowledge-informed reaction design \cite{Tian2025Bulky}. It is, however, a tedious work that commonly requires extensive manual input to build reliable initial structures based on mechanism proposals. Having it automated using agents would save human effort in setting up calculations, and, more importantly, would provide opportunities to extensively rationalize and explore potential pathways without neglecting important alternatives.

\subsubsection{Assessment}

As it is phrased, the prompt represents a significant challenge for the agent to digest. Although the concerted mechanism on the singlet state is well known, it is unclear which carbene-involving pathway corresponds to it. It is expected that the initial planning was inaccurate. Yet{\cinzel Quntur} managed to change its plan, as such a pathway proved unfeasible. It changed the plan once more after realizing that the triplet pathway was thermally inaccessible, continuing the analysis only with the converted pathway. Input file generation and post-processing for those calculations were performed correctly, but the triplet carbene pathway was misrepresented in the post-analysis. The initial and intermediate geometries correspond to the same state, but{\cinzel Quntur} did not seem to realize this. Finally, the figure generated by{\cinzel Quntur} was not in accordance with publication standards.

\subsubsection{Prompt and initial planning:}

\vspace{0.4in}
\noindent
\begin{agentoutputbox}{figs/neutral_human.png}{User}
For the isomerization of methyl isocyanide (CH$_3$–N$\equiv$C) to acetonitrile (CH$_3$–C$\equiv$N), ORCA should be used to investigate at least two mechanistic hypotheses: (a) a concerted rearrangement and (b) a stepwise pathway involving a carbene- or nitrene-like intermediate. For each mechanism, all relevant minima and transition states should be located using \texttt{OPTTS}, with connectivity verified through intrinsic reaction coordinate (IRC) calculations. Nudged elastic band (NEB) computations between key intermediates should be employed to detect any missing transition states. The activation parameters $\Delta G^{\ddagger}$, $\Delta H^{\ddagger}$, and $\Delta S^{\ddagger}$ for the rate-determining step of each pathway should then be compared to assess which mechanism is more consistent with a thermally allowed rearrangement.
\end{agentoutputbox}

\vspace{0.4in}
\noindent
\begin{agentoutputbox}{figs/agent_icon.png}{Computational\_chemist}
\textbf{Action:} \\
This is an excellent mechanistic investigation of the classic methyl isocyanide $\rightarrow$ acetonitrile isomerization. This reaction has been extensively studied experimentally (Arrhenius $E_\mathrm{a} \approx 38~\mathrm{kcal\,mol^{-1}}$) and is a textbook example of a unimolecular rearrangement.
\textbf{Scientific Context} \\
\emph{Mechanism (a) --- Concerted 1,2-shift:} The methyl group migrates from N to C through a cyclic three-membered-ring transition state in a single step.
\emph{Mechanism (b) --- Stepwise pathway:} Could involve:
\begin{itemize}\setlength{\itemsep}{0pt}\setlength{\parskip}{0pt}\setlength{\topsep}{0pt}
  \item A carbene-like intermediate (e.g., CH$_3$--N= C: with a divalent carbon).
  \item A nitrene-like intermediate or a loose ion-pair/radical-pair structure.
\end{itemize}
\textbf{Proposed High-Level Plan (5 Steps):}
\textbf{Step 1: Geometry Generation}
\begin{itemize}\setlength{\itemsep}{0pt}\setlength{\parskip}{0pt}\setlength{\topsep}{0pt}
  \item Generate initial geometries for:
  \begin{itemize}
    \item Methyl isocyanide (CH$_3$NC) --- reactant.
    \item Acetonitrile (CH$_3$CN) --- product.
    \item Initial TS guess for the concerted pathway (three-membered ring).
    \item Potential intermediate(s) for the stepwise pathway.
    \item Initial TS guesses for the stepwise pathway.
  \end{itemize}
\end{itemize}
\textbf{Step 2: Geometry Optimization of Minima}
\begin{itemize}\setlength{\itemsep}{0pt}\setlength{\parskip}{0pt}\setlength{\topsep}{0pt}
  \item Optimize all minima (reactant, product, intermediates) with frequency calculations.
  \item Confirm that all optimized structures have only real vibrational frequencies (true minima).
\end{itemize}
\textbf{Step 3: Transition State Optimization (\texttt{OPTTS}) + Frequency}
\begin{itemize}\setlength{\itemsep}{0pt}\setlength{\parskip}{0pt}\setlength{\topsep}{0pt}
  \item Locate and optimize all transition states:
  \begin{itemize}
    \item Concerted TS.
    \item Stepwise TSs (if applicable).
  \end{itemize}
  \item Perform frequency calculations to confirm exactly one imaginary frequency for each TS.
\end{itemize}
\textbf{Step 4: IRC Validation + NEB Exploration}
\begin{itemize}\setlength{\itemsep}{0pt}\setlength{\parskip}{0pt}\setlength{\topsep}{0pt}
  \item Carry out IRC calculations for all TSs to confirm reactant--product connectivity.
  \item Use NEB calculations between key stationary points to check for any missing transition states.
\end{itemize}
\textbf{Step 5: Thermochemical Analysis + Comparison}
\begin{itemize}\setlength{\itemsep}{0pt}\setlength{\parskip}{0pt}\setlength{\topsep}{0pt}
  \item Extract $\Delta G^{\ddagger}$, $\Delta H^{\ddagger}$, and $\Delta S^{\ddagger}$ for the rate-determining steps.
  \item Compare activation parameters between the concerted and stepwise pathways.
  \item Perform a literature comparison to validate the computed results.
\end{itemize}
\end{agentoutputbox}

\vspace{0.4in}
\noindent
\begin{agentoutputbox}{figs/neutral_human.png}{User}
Use M06-2X-D3/Def2SVP for opt/freq/neb/irc, wB97M(2)/aug-cc-pvqz for final single point
\end{agentoutputbox}

\subsubsection{Summary of the results}

\vspace{0.4in}
\noindent
\begin{agentoutputbox}{figs/agent_icon.png}{Computational\_chemist}
\textbf{Final Report: CH$_3$NC $\rightarrow$ CH$_3$CN Isomerization Mechanism Study}

\textbf{Mechanism (a): Concerted 1,2-Shift --- Viable}

\begin{minipage}{0.5\linewidth}
\begin{center}
\textit{Activation Parameters (Rate-Determining Step)}
\begin{tabular}{lcc}
\hline
Parameter & Value  & Units \\
\hline
$\Delta E^{\ddagger}$ & +39.35 & kcal\,mol$^{-1}$ \\
$\Delta H^{\ddagger}$ & +37.50 & kcal\,mol$^{-1}$ \\
$\Delta G^{\ddagger}$ & +37.48 & kcal\,mol$^{-1}$ \\
$\Delta S^{\ddagger}$ & +0.06  & cal\,(mol\,K)$^{-1}$ \\
\hline
\end{tabular}    
\end{center}
\end{minipage}
\hfill
\begin{minipage}{0.5\linewidth}
\begin{center}
\textit{Reaction Thermodynamics}
\begin{tabular}{lcc}
\hline
Parameter & Value  & Units \\
\hline
$\Delta H_\mathrm{rxn}$ & $-24.50$ & kcal\,mol$^{-1}$ \\
$\Delta G_\mathrm{rxn}$ & $-24.37$ & kcal\,mol$^{-1}$ \\
\hline
\end{tabular}    
\end{center}
\end{minipage}

\textit{Transition-State Characteristics}

\begin{itemize}\setlength{\itemsep}{0pt}\setlength{\parskip}{0pt}\setlength{\topsep}{0pt}
  \item Geometry: three-membered-ring TS with $\angle$(C--N--C) $= 65.9^{\circ}$.
  \item Imaginary frequency: $\nu = -410~\mathrm{cm^{-1}}$ (methyl migration mode).
  \item IRC validation: TS connects directly between CH$_3$NC and CH$_3$CN.
  \item NEB confirmation: single transition state on the path, no additional intermediates.
\end{itemize}

\textbf{Mechanism (b): Stepwise via Carbene --- Not Viable}

\textit{Singlet Surface}

\begin{itemize}\setlength{\itemsep}{0pt}\setlength{\parskip}{0pt}\setlength{\topsep}{0pt}
  \item A bent carbene-like intermediate does not correspond to a stable minimum; all such guesses collapse back to linear CH$_3$NC upon optimization.
  \item NEB calculations along the reaction coordinate reveal no additional intermediates on the singlet surface.
\end{itemize}
\textit{Triplet Surface}
\begin{center}
\begin{tabular}{lc}
\hline
Species & $\Delta G$ (kcal\,mol$^{-1}$) \\
\hline
CH$_3$NC (T$_1$)        & +112.3 \\
INT\_carbene (T$_1$) & +112.1 \\
CH$_3$CN (T$_1$)        & +87.5 \\
\hline
\end{tabular}
\end{center}

\begin{itemize}\setlength{\itemsep}{0pt}\setlength{\parskip}{0pt}\setlength{\topsep}{0pt}
  \item S$_0$--T$_1$ gap is $\sim 114$ kcal\,mol$^{-1}$, rendering the triplet surface thermally inaccessible.
  \item A triplet carbene-like intermediate is formally stable [C--N--C angle $= 122^{\circ}$] but would require a spin-forbidden crossing from the singlet ground state.
\end{itemize}
\textbf{Comparison with Experiment}
\begin{center}
\begin{tabular}{lcc}
\hline
Source & $E_\mathrm{a}$ / $\Delta H^{\ddagger}$ (kcal\,mol$^{-1}$) & Comment \\
\hline
Experiment (Collister \& Pritchard, 1976) & $38.2 \pm 0.5$ & Arrhenius analysis \\
This work ($\omega$B97M(2)/aug-cc-pVQZ)   & 37.5          & Computed $\Delta H^{\ddagger}$ \\
\hline
\end{tabular}
\end{center}
Deviation between theory and experiment is less than 1~kcal\,mol$^{-1}$, indicating excellent agreement.

\textbf{Conclusions}
\begin{itemize}\setlength{\itemsep}{0pt}\setlength{\parskip}{0pt}\setlength{\topsep}{0pt}
  \item The CH$_3$NC $\rightarrow$ CH$_3$CN isomerization proceeds via a \emph{concerted} mechanism through a three-membered-ring transition state.
  \item No stepwise pathway exists on the singlet surface; the putative bent carbene intermediate is not a true minimum.
  \item The triplet surface lies more than 110~kcal\,mol$^{-1}$ above the singlet ground state and is therefore thermally inaccessible under typical conditions.
  \item The near-zero activation entropy ($\Delta S^{\ddagger} \approx 0$) is consistent with a tight, cyclic transition-state structure.
  \item The close agreement between computed and experimental activation parameters validates the chosen computational methodology.
\end{itemize}
\textbf{Final Answer:} The isomerization is a thermally allowed concerted [1,2]-sigmatropic rearrangement. Mechanism~(a) is correct; Mechanism~(b) is not viable.
\end{agentoutputbox}

\subsubsection{Summary of the results}

\begin{figure}[H]
    \centering
    \includegraphics[width=1\linewidth]{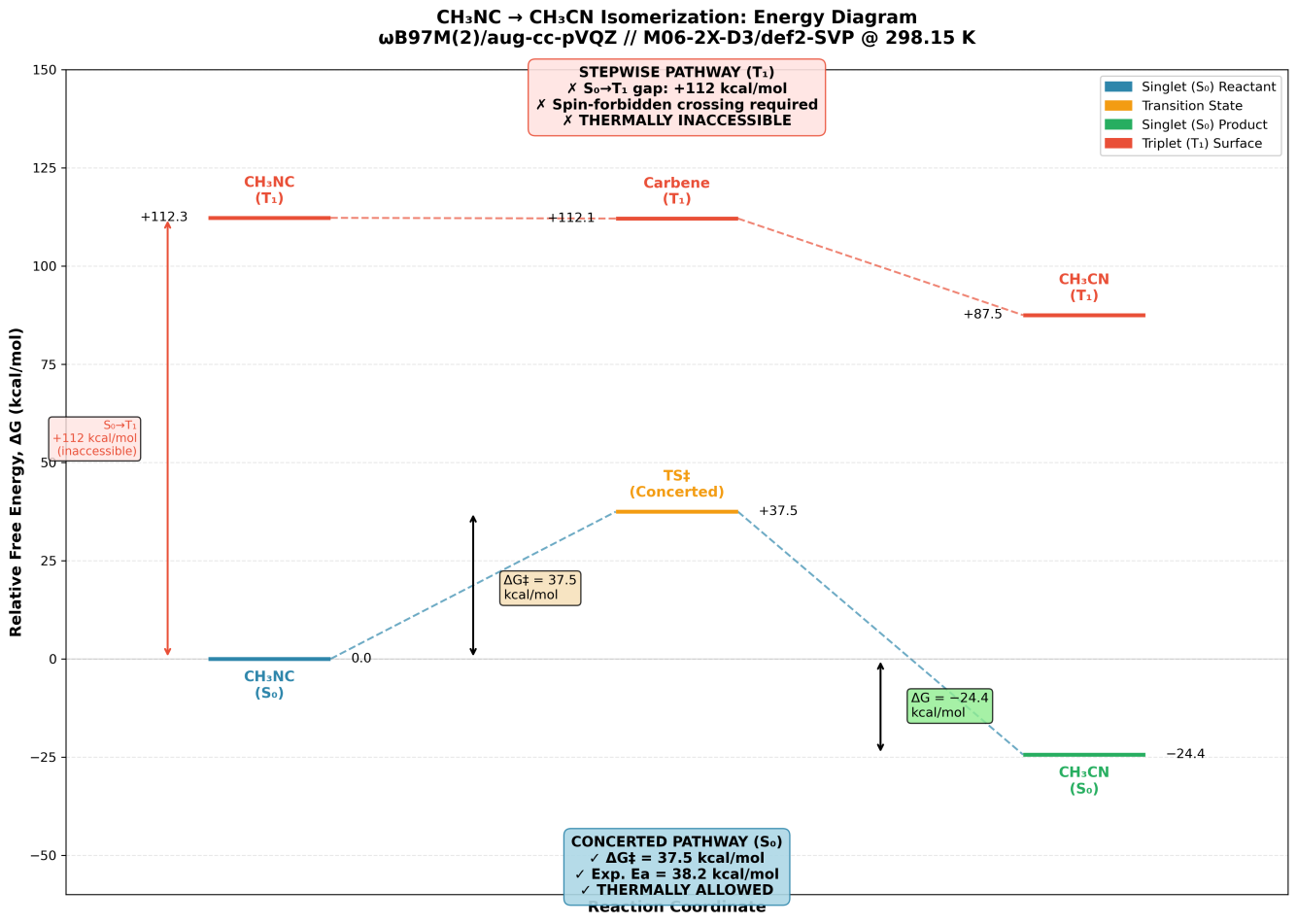}
    \caption{Summary of the findings for fluorescence and anthracene intersystem crossing rates case study.}
    \label{fig:Han_case_study}
\end{figure}

\newpage

\subsection{Case study: Natural Transition Orbitals (NTOs) visualization and excited states characterization}\label{sec:NTOs}

\subsubsection{Context}

Classifying the character of excited states is important for rationalizing and designing chemical reactivity, understanding physical processes, and developing advanced technologies such as solar cells and OLEDs \cite{Scholes2011Lessons, Forrest2004Organic}. In particular, organic emitters can exhibit solvatochromism or multiple emission pathways, and identifying whether an excited state is of $\pi\to\pi^{*}$ character or involves charge transfer is often key to interpreting their photophysical behavior \cite{Grabowski2003Structural}. Assigning the character of emissive states via NTO analysis is essential in studies of dual fluorescence, where one must confirm if a second, red-shifted band indeed arises from a charge transfer state (as in DMABN) \cite{Lippert1962Dual}. Such analyses guide experimentalists in rationalizing how solvent polarity, viscosity, or substituents affect emission spectra. Likewise, in computational benchmarking and mechanistic studies, NTO-based classifications help researchers predict behavior by linking electronic structure to observables. A given theoretical method can be tested by whether it correctly reproduces the nature and energy ordering of charge-transfer states relative to locally excited states, which is a known challenge in TDDFT \cite{Tozer2000On, Dreuw2005Failure}. In the design of new organic photonic materials (i.e., TADF emitters), understanding the mix of charge transfer and local exciton character in excited states is central to tuning properties like singlet–triplet gaps and oscillator strengths \cite{Adachi2014Third,deSilva2019Extracting}. By using NTOs to characterize excited states across such diverse scenarios, physical chemists can more confidently interpret spectra and predict outcomes (such as solvent-dependent color shifts or the presence of multiple emission bands), ultimately bridging the gap between quantum chemical calculations and experimental photophysics.

4-(Dimethylamino)benzonitrile (DMABN) is a prototypical donor–acceptor system that exhibits dual fluorescence arising from the coexistence of locally excited (LE) and twisted intramolecular charge-transfer (TICT) states \cite{Lippert1962Dual}. We prompt the agent to calculate the lowest two singlet excited states of DMABN and visualize the associated natural transition orbitals (NTOs), to classify the excited states as local excitons, charge-transfer states, or twisted intramolecular charge-transfer states. 

\subsubsection{Assessment}

The initial plan is efficient and showcases correct step ordering. Geometry generation is relatively simple with OpenBabel. Generating and executing the input file were more challenging. Still,{\cinzel Quntur} performs well: it correctly sets followiroot to true for S$_1$ geometry optimization and correctly distorts the molecule 3 times to move from a persistent saddle point to the actual minimum. After obtaining optimized geometries, it correctly used the DoNTO true settings for the single-point TDDFT calculation, as required to obtain NTOs. It could have used the plots block as well to automatically produce .cube files, but they were later obtained using orca\_plot. Finally, the most impressive part was the correct interpretation of the NTOs, which it did by directly inspecting .png images generated with the computed .cube files. Finally,{\cinzel Quntur} was able to realize that the shift in character between vertical excitation and the relaxed excited-state geometry does not occur due to the calculation being performed in the gas phase rather than in solution \cite{Georgieva2015Intramolecular}.

\subsubsection{Prompt and initial planning}

\vspace{0.4in}
\noindent
\begin{agentoutputbox}{figs/neutral_human.png}{User}
For the molecule 4-(dimethylamino)benzonitrile (DMABN), compute the lowest two singlet excited states using ORCA~6.0 and visualize the natural transition orbitals (NTOs) for each bright transition using TDDFT. Identify which states exhibit local $\pi \rightarrow \pi^{*}$ character and which show charge-transfer (CT) or twisted intramolecular charge-transfer (TICT) character. Analyze how the NTO hole--particle separation changes when comparing the S$_1$ state at the ground-state geometry with the S$_1$ state at its relaxed excited-state geometry.
\end{agentoutputbox}

\vspace{0.4in}
\noindent
\begin{agentoutputbox}{figs/agent_icon.png}{Computational\_chemist}
\textbf{Action:}

This is a classic twisted intramolecular charge-transfer (TICT) study on 4-(dimethylamino)benzonitrile (DMABN). This molecule is well known for exhibiting dual fluorescence arising from locally excited (LE) and charge-transfer (CT) states. A high-level computational workflow is proposed below.

\textbf{Proposed High-Level Plan for DMABN Excited-State Analysis}

\textbf{Step 1: Geometry Generation}

Generate an initial molecular geometry for DMABN (4-dimethylamino)benzonitrile).

\textbf{Step 2: Ground-State Optimization}

Optimize the ground-state (S$_0$) geometry.

\textbf{Step 3: Parallel Tasks}
\begin{itemize}
  \item TDDFT calculation at the ground-state geometry to compute S$_1$ and S$_2$
        and generate natural transition orbitals (NTOs) for the bright transitions.
  \item Excited-state geometry optimization of the S$_1$ state, starting from
        the ground-state geometry.
\end{itemize}
\textbf{Step 4: Excited-State TDDFT Calculation}

Perform a TDDFT calculation at the optimized S$_1$ geometry and generate NTOs for direct comparison with the vertical excitation.

\textbf{Step 5: Post-Analysis}
\begin{itemize}
  \item Visualize NTOs (hole and particle orbitals) for all bright excited states.
  \item Identify locally excited $\pi \rightarrow \pi^{*}$ versus CT/TICT
        character based on orbital localization.
  \item Compare hole--particle spatial separation at the ground-state geometry
        with that at the relaxed S$_1$ geometry.
\end{itemize}
\textbf{Important Considerations for DMABN Charge-Transfer States}

DMABN charge-transfer states are highly sensitive to the choice of DFT functional. Standard hybrid functionals (e.g., B3LYP) often underestimate CT excitation energies due to self-interaction error. Range-separated hybrid
functionals such as CAM-B3LYP or $\omega$B97X-D are generally recommended for a more reliable description of CT and TICT states.

\textbf{Clarifications Before Proceeding}
\begin{itemize}
  \item \textbf{Functional preference:} Range-separated hybrid
        (recommended: CAM-B3LYP or $\omega$B97X-D3) or standard hybrid (B3LYP)?
  \item \textbf{Basis set:} Any preference (e.g., def2-TZVP for production-level
        calculations, def2-SVP for rapid screening)?
  \item \textbf{Solvent treatment:} Gas phase or implicit solvation
        (e.g., acetonitrile, commonly used in DMABN studies)?
  \item \textbf{Charge and multiplicity:} Neutral singlet (0, 1) assumed—please
        confirm.
\end{itemize}
Alternatively, should sensible defaults be applied (range-separated functional,
def2-TZVP basis set, gas phase, neutral singlet) and the workflow executed
directly?
\end{agentoutputbox}

\vspace{0.4in}
\noindent
\begin{agentoutputbox}{figs/neutral_human.png}{User}
Use CAM-B3LYP def2-TZVP. Assume gas phase. Proceed
\end{agentoutputbox}

\subsubsection{Summary of the results}

\vspace{0.4in}
\noindent
\begin{agentoutputbox}{figs/agent_icon.png}{Computational\_chemist}
\textbf{NTO Analysis Summary}
Based on the NTO visualizations and calculated data, a comprehensive analysis of the excited-state character is summarized below.
\textbf{Excited-State Character Analysis}
\begin{center}
\begin{tabular}{lcccccc}
\hline
State & Geometry & Energy (eV) & $f$ & Hole Location & Particle Location & Character \\
\hline
S$_1$ & GS (FC) & 4.81 & 0.041 & Ring + amino & Ring + cyano & Partial CT \\
S$_2$ $\star$ & GS (FC) & 5.20 & 0.675 & Ring + amino & Ring + cyano & $\pi\rightarrow\pi^{*}$ with CT mixing$^{*}$ \\
S$_1$ & Relaxed S$_1$ & 4.37 & 0.039 & Ring + amino & Ring + cyano & Partial CT (planar ICT) \\
\hline
\end{tabular}
\end{center}
\textbf{Key Findings}
\begin{enumerate}
  \item \textbf{S$_2$ is the bright state (local $\pi\rightarrow\pi^{*}$ with CT character)$^{*}$}
The strong oscillator strength ($f = 0.675$) indicates an allowed electronic transition. Both hole and particle NTOs are delocalized over the molecular $\pi$ system. The hole NTO exhibits enhanced electron density on the dimethylamino donor group. The particle NTO extends toward the cyano acceptor group. Overall character: predominantly local $\pi\rightarrow\pi^{*}$ excitation with charge-transfer admixture.
  \item \textbf{S$_1$ shows charge-transfer character at both geometries}
  The weak oscillator strength ($f \approx 0.04$) is indicative of a more forbidden, CT-like transition. Clear spatial separation is observed, with the hole localized on the donor (amino side) and the particle on the acceptor (cyano side).  Character: intramolecular charge-transfer (ICT) state.
  \item \textbf{Comparison: S$_1$ at FC vs.\ relaxed geometry}
  \begin{center}
  \begin{tabular}{lccc}
  \hline
  Property & S$_1$ @ FC (GS geom) & S$_1$ @ relaxed geom & Change \\
  \hline
  Energy & 4.81 eV & 4.37 eV & $-0.44$ eV \\
  Osc.\ strength & 0.041 & 0.039 & Minimal \\
  NTO character & Partial CT & Partial CT & Similar \\
  Geometry & Planar & Planar & No TICT twist \\
  \hline
  \end{tabular}
  \end{center}
\end{enumerate}
\textbf{Physical Interpretation}
In the gas phase at the CAM-B3LYP/def2-TZVP level:
\begin{itemize}
  \item The S$_1$ minimum corresponds to a planar ICT state rather than a TICT configuration.
  \item The initial S$_1$ optimization converged to a saddle point with an imaginary frequency associated with dimethylamino twisting, indicating that a TICT structure lies nearby on the potential energy surface.
  \item The true S$_1$ minimum remains planar with modest charge-transfer character.
  \item This behavior is consistent with the literature: gas-phase DMABN favors a planar LE/ICT minimum, whereas the TICT state is preferentially stabilized in polar solvent environments.
\end{itemize}
\textbf{Conclusions}
\begin{itemize}
  \item S$_2$ (bright): Dominated by local $\pi\rightarrow\pi^{*}$ character with CT admixture and responsible for the main UV absorption band.
  \item S$_1$ (dark): Exhibits partial ICT character, with the hole localized on the amino donor and the particle on the ring--cyano acceptor framework.
  \item No significant increase in hole--particle separation is observed upon S$_1$ relaxation in the gas phase.
  \item A fully twisted TICT state (approximately $90^{\circ}$ rotation of the dimethylamino group) does not correspond to the gas-phase S$_1$ minimum and would require stabilization by a polar environment.
\end{itemize}
\medskip
\noindent$^{*}$Predominantly local $\pi\rightarrow\pi^{*}$ excitation with appreciable charge-transfer admixture.
\end{agentoutputbox}

\subsubsection{Summary of the results}

\begin{figure}[H]
    \centering
    \includegraphics[width=1.0\linewidth]{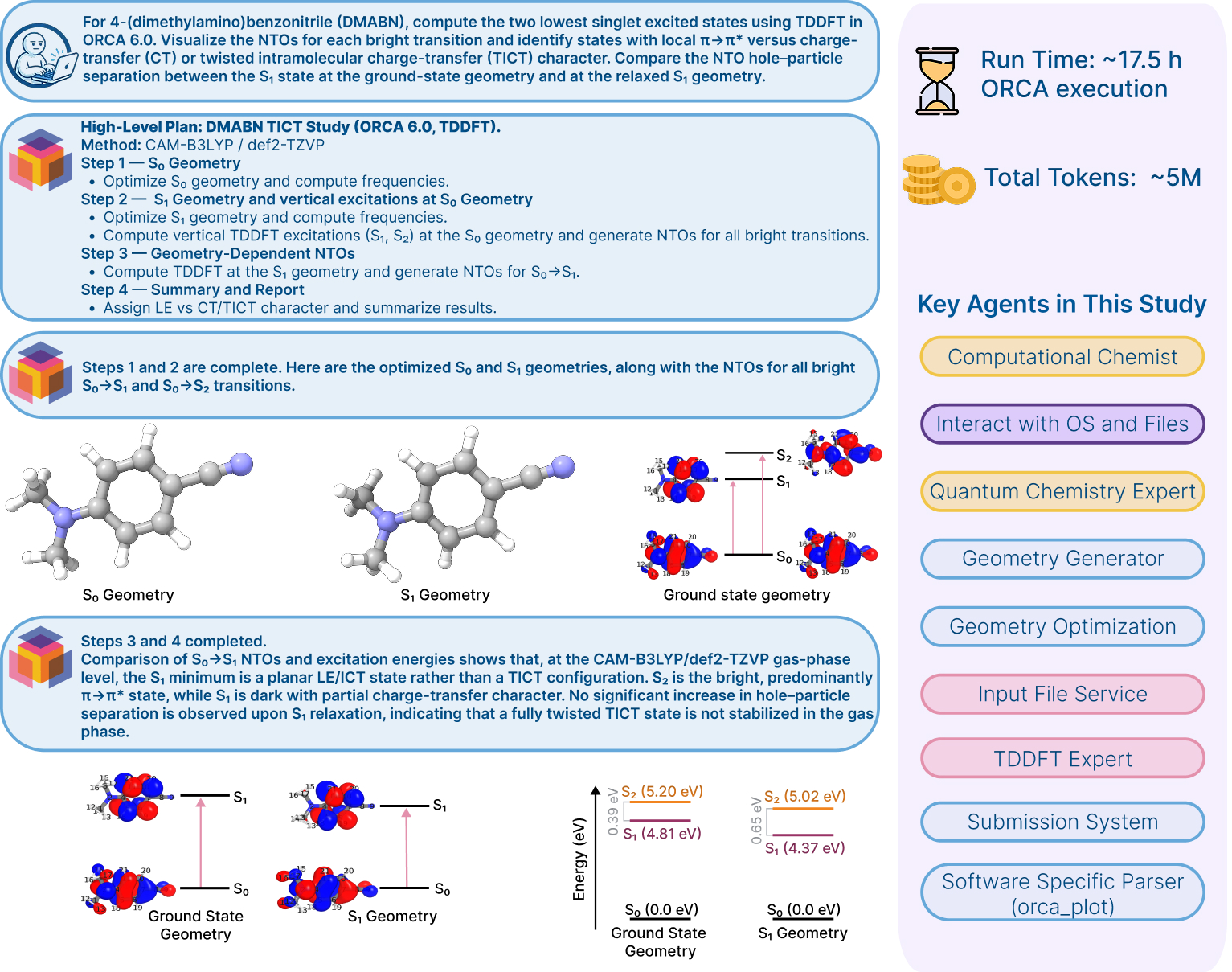}
    \caption{Summary of the findings for NTOs visualization and excited states characterization case study.}
    \label{fig:NTOs_case_study}
\end{figure}

\newpage
\subsection{Case study: Potential energy surface of molecular nitrogen}\label{sec:N2}

\subsubsection{Context}

The computation of the \ce{N2} potential energy surface along the dissociation coordinate is a long-standing and widely adopted benchmark in electronic structure theory because it encompasses several of the most fundamental challenges in quantum chemistry within a chemically simple diatomic molecule. As the \ce{N-N} bond is stretched, the electronic structure undergoes a qualitative transition from a regime dominated by dynamic correlation near equilibrium to one governed by strong static correlation and near-degeneracies at dissociation\cite{faulstich2019numerical,boyn2022elucidating}. This evolution provides a rigorous and well-defined setting in which the physical assumptions and mathematical structure of approximate electronic structure methods can be examined in a controlled manner \cite{chattopadhyay2020investigation,aroeira2020coupled}.

Beyond its role as a theoretical stress test, \ce{N2} dissociation is directly connected to catalysis, most notably through nitrogen fixation. The dissociative activation of \ce{N2} is the rate-limiting step in the Haber--Bosch process and in related catalytic and electrocatalytic pathways aimed at reducing the energy footprint of ammonia synthesis. Consequently, the reliability of potential energy surfaces describing \ce{N-N} bond breaking underpins computational catalyst screening and mechanistic analysis \cite{ashida2022catalytic,chen2022progress}. A workflow that systematically compares DFT methods, coupled-cluster approaches, and multi-reference techniques for this problem mirrors the practical decision-making required in computational catalysis, making it a realistic and non-contrived test of agent competence.

The same system has also emerged as a canonical benchmark for near-term quantum computing approaches to chemistry \cite{weaving2025contextual,choi2024probing}. The multi-reference character that defeats classical single-reference methods is precisely the regime in which variational quantum algorithms demonstrate relative advantages, with recent studies showing that quantum simulations outperform CCSD(T) for stretched geometries in small basis sets. \ce{N2} dissociation therefore serves as a common reference point linking classical electronic-structure theory and quantum algorithms, while remaining simple enough to validate against exact solutions.

 \subsubsection{Assessment}

Quntur's initial planning was reasonable. It required some iterative refinement to fully specify details that would be impossible to define in a high-level prompt. For instance:
\begin{itemize}
    \item Should I include both CCSD and CCSD(T), or just CCSD(T)?
    \item Would you like to include any DFT functionals for comparison?
    \item Should I include a smaller CAS(2,2) (just $\sigma/\sigma^*$) to show even more dramatic failure?
\end{itemize}

 Interestingly,{\cinzel Quntur} insisted on creating a ``template geometry'', i.e., an additional XYZ file with the molecule at the equilibrium geometry. This action is unnecessary given the variable-geometry nature of the task. However, once confronted with the uselessness of its suggestion,{\cinzel Quntur} leveraged Orca's capabilities to parametrically manipulate the geometry.

{\cinzel Quntur} created seven input files for each of the proposed methods, which required minimal automated cleanup before parallel execution. Six out of the seven jobs were successful; only NEVPT2 failed.{\cinzel Quntur} formulated a reasonable hypothesis (to execute the job serially to access all the files) to troubleshoot the job and modified the corresponding input file to test it. Unfortunately, this strategy did not work either. But{\cinzel Quntur} proposed three potential workarounds: run CASSCF(6,6) without NEVPT2, run 24 independent NEVPT2 jobs, or forego the method altogether. The user chose the first option to maximize the number of results while keeping the number of input files to a minimum.

{\cinzel Quntur} correctly parsed and collected the energies from each method, delivering a comprehensive performance comparison, in both written and graphical reports, and correctly diagnosing the failure modes of each benchmarked method.

\subsubsection{Prompt and initial planning:}

\vspace{0.4in}
\noindent
\begin{agentoutputbox}{figs/neutral_human.png}{User}
I'm studying how different levels of theory describe the \ce{N2} molecule along its potential energy surface (PES) as it dissociates. Using Full Configuration Interaction with minimal basis as the ground truth, compare as many electronic-structure methods as make sense to diagnose their failures and analyze how other methods address them, including spin effects. Perform this study along a smooth PES between 0.7 and 3.0 Angstroms, but make sure to set up the workflow to minimize the number of input and output files.
\end{agentoutputbox}

\vspace{0.4in}
\noindent
\begin{agentoutputbox}{figs/agent_icon.png}{Computational\_chemist}
\textbf{Action:}

This is an excellent and pedagogically valuable study! The dissociation of \ce{N2} is a classic benchmark problem in quantum chemistry that exposes the limitations of single-reference methods, as strong static correlation develops as the triple bond breaks.

\textbf{Initial Analysis}

Before I propose a detailed plan, let me outline the key scientific considerations:

\textbf{Why \ce{N2} Dissociation is Challenging:}

Static correlation: As \ce{N2} stretches, multiple electronic configurations become nearly degenerate

Spin symmetry breaking: Restricted HF (RHF) will fail, while unrestricted HF (UHF) breaks spin symmetry to capture some correlation at the cost of spin contamination

Multi-reference character: Single-reference methods (HF, MP2, CCSD) fundamentally cannot properly describe the dissociation limit

\textbf{Methods to Compare (with FCI as ground truth):}

\begin{tabular}{lll}
Category & Methods & Purpose \\
Mean-field & RHF, UHF & Show symmetry breaking \\
Post-HF (SR) & MP2, CCSD, CCSD(T) & Show single-reference failures \\
Multi-reference & CASSCF, NEVPT2 & Show proper dissociation \\
DFT & B3LYP, PBE (R/U) & Show DFT behavior \\
\end{tabular}

\textbf{Workflow Efficiency}

ORCA allows compound jobs where multiple methods can be run sequentially on the same geometry, and coordinate scans can be performed in a single input file. This will minimize the number of files.

Would you like me to perform a literature search first to ensure we're following established best practices for this classic problem (e.g., appropriate active space for CASSCF, expected behavior of each method), or shall I proceed directly with the high-level plan?

Also, please provide any preferences on:

Number of points along the PES (e.g., 0.1 \AA\ spacing $\rightarrow$ $\sim$ 24 points)

Specific methods you want included or excluded

Basis set for FCI: STO-3G is typical for tractability, but cc-pVDZ could be attempted for small CI spaces
\end{agentoutputbox}

\vspace{0.4in}
\noindent
\begin{agentoutputbox}{figs/neutral_human.png}{User}
1. 24 points is fine. 2. Include a limited active space to characterize the failure analysis. 3. Stick to STO-3G. Also, keep track of the molecular orbitals — energies and coefficients — in the methods that change them to enrich the analysis.
\end{agentoutputbox}

\subsubsection{Summary of the results}

\vspace{0.4in}
\noindent
\begin{agentoutputbox}{figs/agent_icon.png}{Computational\_chemist}
\textbf{Action:}

$\checkmark$ \ce{N2} Dissociation PES Study - Complete

\textbf{Summary}

I have completed a comprehensive multi-method benchmark study of \ce{N2} dissociation comparing 13 electronic structure methods against FCI (Full Configuration Interaction) as the ground truth, using minimal STO-3G basis across 24 geometry points (0.7--3.0 \AA).

\textbf{Methods Compared}

\begin{tabular}{ll}
Category & Methods \\
Mean-field & RHF, UHF, RB3LYP, UB3LYP \\
Perturbation Theory & MP2, UMP2 \\
Coupled Cluster & CCSD, UCCSD, CCSD(T), UCCSD(T) \\
Multi-reference & CASSCF(2,2), CASSCF(6,6) \\
Exact (within basis) & FCI \\
\end{tabular}

\textbf{Key Findings}

\textbf{1. Single-Reference Method Failure}

\begin{tabular}{ll}
Method & Error at Dissociation (kcal/mol) \\
RHF & +388 \\
CCSD(T) & +157 \\
CCSD & +157 \\
MP2 & +211 (divergent) \\
\end{tabular}

All single-reference methods fail catastrophically at dissociation because they cannot describe the multi-determinant character of the stretched bond.

\textbf{2. Active Space Importance}

\begin{tabular}{ll}
Active Space & Error at Dissociation \\
CASSCF(2,2) & +599 kcal/mol (missing $\pi$ correlation) \\
CASSCF(6,6) & +201 kcal/mol (missing dynamic correlation) \\
FCI & 0 (exact) \\
\end{tabular}

The (2,2) active space, capturing only $\sigma/\sigma^\ast$, completely fails. The (6,6) space including all triple-bond orbitals performs much better but still lacks dynamic correlation.

\textbf{3. Spin Effects}

UHF: Did NOT break spin symmetry (stayed in RHF solution) --- a known issue without broken-symmetry initial guess

UB3LYP: Properly breaks symmetry at $\sim$2.2 \AA\ (Coulson-Fischer point)

$\langle S^2 \rangle$ at dissociation: 2.99 for UB3LYP (heavily spin-contaminated, approaching triplet character)

\textbf{4. Natural Orbital Analysis (CASSCF(6,6))}

\begin{tabular}{llll}
Geometry & $\sigma$ (bonding) & $\sigma^\ast$ (antibonding) & Character \\
R = 1.1 \AA & 1.94 & 0.07 & Single-reference \\
R = 2.0 \AA & 1.29 & 0.70 & Multi-reference \\
R = 3.0 \AA & 1.10 & 0.86 & Strong diradical \\
\end{tabular}
\end{agentoutputbox}

\begin{figure}[H]
    \centering
    \includegraphics[width=1\linewidth]{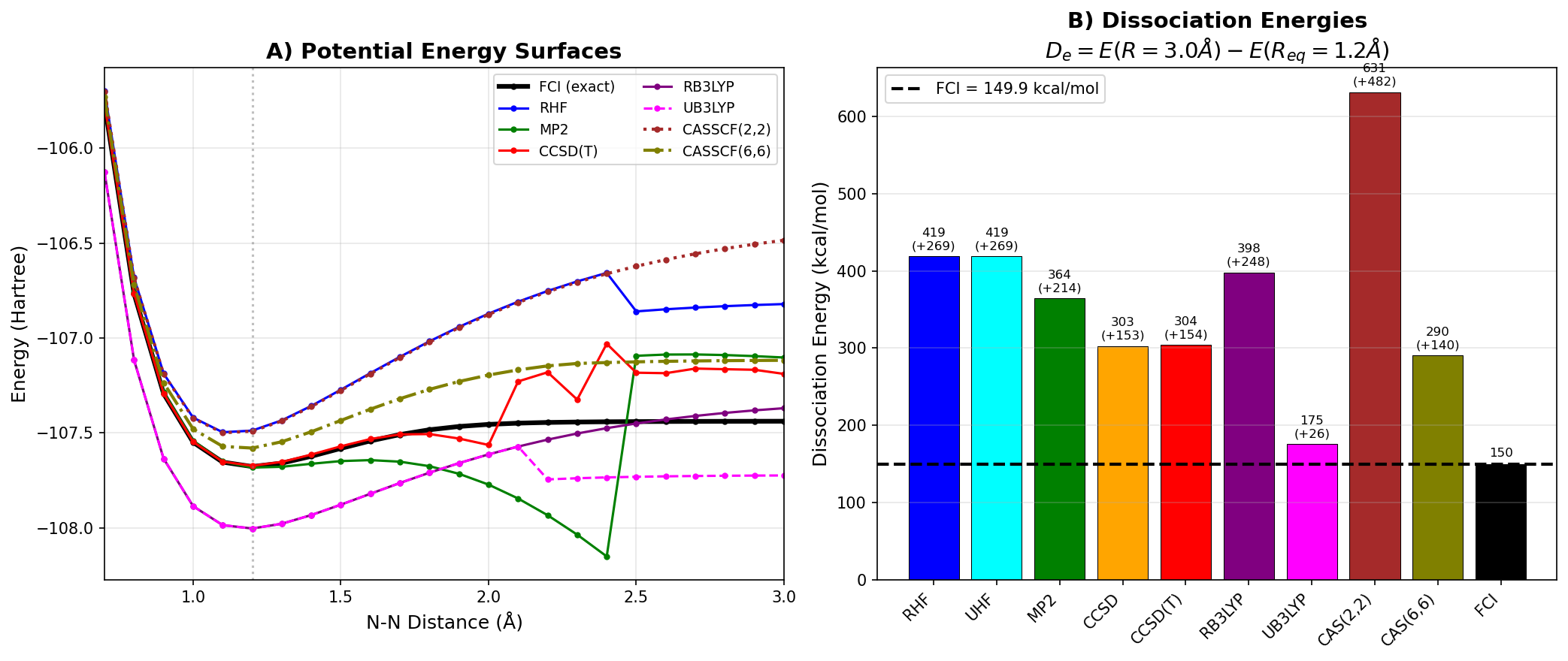}
    \caption{Computational benchmark for the \ce{N2} molecule dissociation. (A) Comparison of the PES along the dissociation coordinates for several electronic structure methods. (B) Comparison of the computed dissociation energy for each method with respect to FCI.}
    \label{fig:Jorge_case_study}
\end{figure}
\newpage

\subsection{Trans effect of ligands on a square planar Pt complex}\label{sec:transeffect}

\subsubsection{Context}

The trans effect/influence is a key principle in describing the reactivities of transition metal complexes, specifically towards ligand substitutions. It involves the idea where the ligand trans to another ligand being substituted can affect the rate of the ligand that's being substituted, opening up avenues to tune and modulate reactivity of transition metal complexes based on ligand choice and design. Both the trans effect and trans influence decrease the barrier towards ligand substitution, but at the most basic level, the trans effect originates from lowering the TS energy while the trans influence originates from destabilizing the energy of the reactants. Unpacking how different ligands contribute to each of these phenomena is not only of fundamental importance to physical inorganic chemistry but is also critical for designing/choosing ligands towards syntheses and catalytic reactions involving transition metal. The idea here is to test and see if the agent can read figures and text from an existing figure and to try and reproduce the trends of what was found in a previous study \cite{Pinter2013trans} using certain methods recommended by another paper \cite{Shoubhik2025Benchmark}.

Dependent files: 2 pdf files (``reference 2'' \cite{Pinter2013trans} and ``reference 3'' \cite{Zumdahl1968A}).

\subsubsection{Assessment}
Initially, we had{\cinzel Quntur} generate the geometries by reading Figure 1 of Ref. \cite{Pinter2013trans}. It was able to generate reasaonable ground state geometries while the TS guess was a bit less desirable. Nudged Elastic Band (NEB) calculations could not be completed from the ground state geometries since the atomic indices and/or positions of atoms in Cartesian coordinates were not aligned sufficiently to allow for a smooth interpolation of the initial trajectory for NEB calculations. Upon supplying the manually-built geometries, the NEB calculations were then able to proceed without difficulty to arrive at a converged path with a climbing image (CI). We observed that TS optimization within the same job of NEB-TS (i.e. NEB-TS includes NEB-CI followed by OptTS on the CI structure) can exceed the maximum number of geometry steps allowed, resulting ORCA to terminate with an error. However, upon requesting{\cinzel Quntur} to start a fresh OptTS calculation using the structure of the converged CI, the TS optimization was able to be completed. 

The resulting trend was reproduced, with slight differences in the absolute values of the numbers. The summary poster had most of the required information but with some stylistic issues. Additionally,{\cinzel Quntur} only did the comparison with one reference and did not specify which one. However, the overall output from Quntur, considering both the summary report and the poster with its figures, was quite comprehensive and at a level of a very competent research collaborator. 

\subsubsection{Prompt and initial planning:}

\vspace{0.4in}
\noindent
\begin{agentoutputbox}{figs/neutral_human.png}{User}
Using ORCA 6.0 and following the recommendations from this study [1], calculate the reaction profiles of the nucleophilic substitution reaction of NH3 on the complex TPtCl2NH3, where T = NH3, PH3, CO, and C2H4. The geometries of the initial structures have been supplied. How does this compare with the results (specifically Figure 1) from [2]? From your results, comment on the trans effect and trans influence of each ligand T on the substitution reaction vs. the analyses provided by [2] and [3]. [1] https://doi.org/10.26434/chemrxiv-2025-hssg2 [2] pdf file called ``reference\_2'' [3] pdf file called ``reference\_3''
\end{agentoutputbox}

\vspace{0.4in}
\noindent
\begin{agentoutputbox}{figs/agent_icon.png}{Computational\_chemist}
\vspace{-0.1in}
\subsubsection*{Step 1: Geometry Optimization of Reactants and Products}
\begin{itemize}
    \item Optimize all 8 provided structures (4 reactants + 4 products for T = NH$_3$, PH$_3$, CO, C$_2$H$_4$)
    \item Level: PBE0-D3BJ/ZORA with SARC-ZORA-TZVP(Pt) and def2-TZVP(ligands)
    \item Include frequency calculations to verify minima (no imaginary frequencies)
\end{itemize}
\vspace{-0.2in}
\subsubsection*{Step 2: Transition State Searches using NEB-TS}
\begin{itemize}
\item Use NEB-TS method (as recommended by [1]) connecting each optimized reactant to its product
\item Same level of theory as Step 1
\item This will locate approximate TS structures
\end{itemize}
\vspace{-0.2in}
\subsubsection*{Step 3: TS Refinement and Verification}
\begin{itemize}
    \item Refine TS guesses from NEB with full TS optimization (P-RFO eigenvector following)
    \item Frequency calculations to verify exactly one imaginary frequency
    \item IRC calculations to confirm connectivity to reactant/product
\end{itemize}
\vspace{-0.2in}
\subsubsection*{Step 4: Analysis and Comparison}
\begin{itemize}
    \item Extract Gibbs free energies at 298 K for all species
    \item Construct reaction profiles (reactant → TS → product)
    \item Calculate activation barriers $\Delta{G}^\ddagger$ and compare with Figure 1 from [2]
    \item Discuss trans effect ordering and compare with [2] and [3] analyses
\end{itemize}
\vspace{-0.2in}
\subsubsection*{Important Considerations}
\begin{itemize}
    \item Charge and Multiplicity: These are neutral, closed-shell Pt(II) complexes (charge = 0, multiplicity = 1)
    \item Potential Challenge: The C2H4 system in [2] showed an intermediate (INT) on the profile, suggesting an associative mechanism. We may find similar intermediates.
    \item Method Difference: [2] used M06 while [1] recommends PBE0-D3BJ. Results may differ slightly but the trans effect ordering should be consistent.
\end{itemize}
\subsubsection*{Questions before Proceeding}
Do you want to
\begin{enumerate}
    \item Proceed with this plan as outlined?
    \item Provide any additional details (e.g., specific basis set preferences, solvent)?
    \item Have me search for more literature before proceeding?
    \item Make modifications to the methodology (e.g., use M06 to match [2] exactly)?

Please confirm or modify before I begin execution.
\end{enumerate}

\end{agentoutputbox}

\vspace{0.4in}
\noindent
\begin{agentoutputbox}{figs/neutral_human.png}{User}
confirm
\end{agentoutputbox}

\subsubsection{Results:}

\vspace{0.4in}
\noindent
\begin{agentoutputbox}{figs/agent_icon.png}{Computational\_chemist}

\subsection*{Trans Effect Study: \ce{NH3} Substitution on Square Planar Pt(II) Complexes}

Summary Report (UPDATED) \\
Date: Computational Study Complete \\
Reaction: [T-\ce{PtCl2(NH3)}] + \ce{NH3} → [\ce{T-PtCl2(NH3)}] + \ce{NH3} \\
Trans-directing ligands (T): \ce{NH3}, \ce{PH3}, CO, \ce{C2H4}
\newline
\subsubsection*{1. Computational Methodology}

\begin{tabular}{l l }
\textbf{Parameter}	     &        \textbf{Value} \\
\textbf{Level of Theory}	   &     PBE0-D3BJ \\
\textbf{Relativistic Treatment} &	ZORA (Zeroth Order Regular Approximation) \\
\textbf{Basis Set (light atoms)}	 & ZORA-def2-TZVP \\
\textbf{Basis Set (Pt)}	   &     SARC-ZORA-TZVP \\
\textbf{Auxiliary Basis	}   &     SARC/J \\
\textbf{Acceleration}	   &     RIJCOSX \\
\textbf{SCF Convergence}	   &     TightSCF \\
\textbf{Grid}	           &     DefGrid3 \\
\textbf{Charge/Multiplicity}	 &   0 / 1 (singlet) \\

\newline

\end{tabular}

Workflow:
\begin{enumerate}
    \item Geometry optimization + frequency calculation for reactants and products
    \item NEB-CI (Nudged Elastic Band with Climbing Image) for transition state search
    \item OptTS + frequency calculation for refined transition states
\end{enumerate}

\subsubsection*{2. Activation Barriers ($\Delta{G}^\ddagger$) }

\begin{tabular}{l l l l}
\textbf{Trans Ligand} & \textbf{This Work (kcal/mol)} & \textbf{Reference\textsuperscript{1} (kcal/mol)} & \textbf{Difference} \\

\textbf{NH$_3$} & 31.05 & 28.4 & +2.55 \\
\textbf{PH$_3$} & 17.57 & 15.1 & +2.47 \\
\textbf{CO} & 15.59 & 12.1 & +3.49 \\
\textbf{C$_2$H$_4$} & 7.69 & 7.2 & +0.49 \\
\newline

\end{tabular}

\subsubsection*{3. Transition State Validation }

\begin{tabular}{l l l }
\textbf{Trans Ligand} & \textbf{Imaginary Frequency (cm\textsuperscript{-1})} & \textbf{Saddle Point Order} \\

\textbf{NH$_3$} & -139.0 & 1st order \\
\textbf{PH$_3$} & -128.5 & 1st order \\
\textbf{CO} & -123.3 & 1st order \\
\textbf{C$_2$H$_4$} & -139.4 & 1st order \\
\newline

\end{tabular}

\textsuperscript{1}Reference values from Maiti et al., ChemRxiv 2025

\subsubsection*{4. Trans Effect Series Analysis}

\textbf{Expected vs. Calculated Order}
\begin{itemize}
    \item Expected order (stronger trans effect → lower barrier): \ce{C2H4} > CO > \ce{PH3} > \ce{NH3}
    \item Calculated order: \ce{NH3} (31.1) > \ce{PH3} (17.6) > CO (15.6) > \ce{C2H4} (7.7) kcal/mol
\end{itemize}

Result: The calculated order MATCHES the expected trans effect series
\end{agentoutputbox}

\newpage

\begin{agentoutputbox}{figs/agent_icon.png}{Computational\_chemist}

\subsubsection*{5. Discussion}
\textbf{5.1 Origin of the Trans Effect}
The trans effect is the labilization of ligands trans to a strong trans-directing ligand in square planar complexes. Two complementary explanations account for this phenomenon:
\begin{enumerate}
    \item \textbf{$\sigma$-Donor Effect}: Strong $\sigma$-donors weaken the Pt-L bond trans to them by competing for the same d orbital ($d_{z^{2}}$). This ground-state effect destabilizes the reactant. 
    \item \textbf{$\pi$-Acceptor Effect:} $\pi$-Acceptor ligands (CO, \ce{C2H4}) stabilize the 5-coordinate transition state by accepting electron density into their $\pi$* orbitals. This lowers the activation barrier.
    \newline
\end{enumerate}

\textbf{5.2 Interpretation of Results}

\begin{tabular}{l l l l }
\textbf{Ligand} & \textbf{$\sigma$-Donor Strength} & \textbf{$\pi$-Acceptor Ability} & \textbf{$\Delta{G}^\ddagger$ (kcal/mol)} \\

\textbf{NH$_3$} & Moderate & None & 31.3 (highest) \\
\textbf{PH$_3$} & Strong & Weak & 17.6 \\
\textbf{CO} & Moderate & Strong & 15.6 \\
\textbf{C$_2$H$_4$} & Weak & Very Strong & 7.7 (lowest) \\
\newline

\end{tabular}

\begin{itemize}
    \item \ce{C2H4} shows the strongest trans effect due to its excellent $\pi$-acceptor ability, which stabilizes the trigonal bipyramidal transition state
    \item CO has strong $\pi$-acceptor character, resulting in a low barrier
    \item \ce{PH3} is a strong $\sigma$-donor with weak $\pi$-acceptor ability
    \item \ce{NH3} has no $\pi$-acceptor ability, resulting in the highest barrier
\end{itemize}
\textbf{5.3 Comparison with Reference Values}
The calculated barriers are systematically 0.5-3.5 kcal/mol higher than reference values. Possible reasons:
\begin{enumerate}
    \item Basis set differences: The reference may have used different basis sets or effective core potentials
    \item Solvation effects: This study was performed in gas phase; the reference may include implicit solvation
    \item Functional choice: Small systematic differences between DFT functionals
\end{enumerate}

Despite these differences, the qualitative trend is correctly reproduced, validating the computational approach for studying the trans effect.

\subsubsection*{6. Gibbs Free Energies}

\begin{tabular}{l l l l }
\textbf{System} & \textbf{Reactant} & \textbf{TS} & \textbf{Product} \\

NH{$_3$} & -20133.0870245 & -20133.0375447 & -20133.0870239 \\
{PH$_3$} & -20420.8769326 & -20420.8489379 & -20420.8769321 \\
CO	& -20189.9118011 & -20189.8869599 & -20189.8260389 \\
C$_2$H$_4$ & -20155.0721232 & -20155.0598681 & -20155.0706649 \\
\newline

\end{tabular}

\subsubsection*{7. Technical Notes}
\textbf{CO Reactant Optimization}
The CO reactant optimization was initially problematic:
\begin{itemize}
    \item $\sim$1100+ optimization steps across 6 restart attempts with internal coordinates
    \item Issues: coordinate transformation instability, trust radius violations
\end{itemize}
\textbf{Solution:} Cartesian coordinate optimization (COPT) converged in just 16 cycles
\begin{itemize}
    \item Settings: COPT, TightOpt, InHess Read, Recalc\_Hess 5, MaxStep 0.1, Trust 0.1
\end{itemize}
\end{agentoutputbox}

\newpage

\begin{agentoutputbox}{figs/agent_icon.png}{Computational\_chemist}
\subsubsection*{8. Conclusions}
\begin{enumerate}
    \item The trans effect series is correctly reproduced: \ce{C2H4} > CO > \ce{PH3} > \ce{NH3}
    \item Quantitative agreement: Calculated barriers agree with reference values within 0.5-3.5 kcal/mol (MAE = 2.27 kcal/mol)
    \item Mechanism confirmation: All transition states show single imaginary frequencies (~120-140 cm\textsuperscript{-1}) consistent with associative ligand exchange
    \item The PBE0-D3BJ/ZORA method with SARC-ZORA-TZVP basis is appropriate for studying reaction kinetics of Pt(II) complexes
\end{enumerate}

\subsubsection*{9. File Locations}
[truncated]
\subsubsection*{10. References}
See references\_combined.bibtex for full bibliography [truncated]
 
\end{agentoutputbox}

\begin{figure}[H]
    \centering
    \includegraphics[width=1\linewidth]{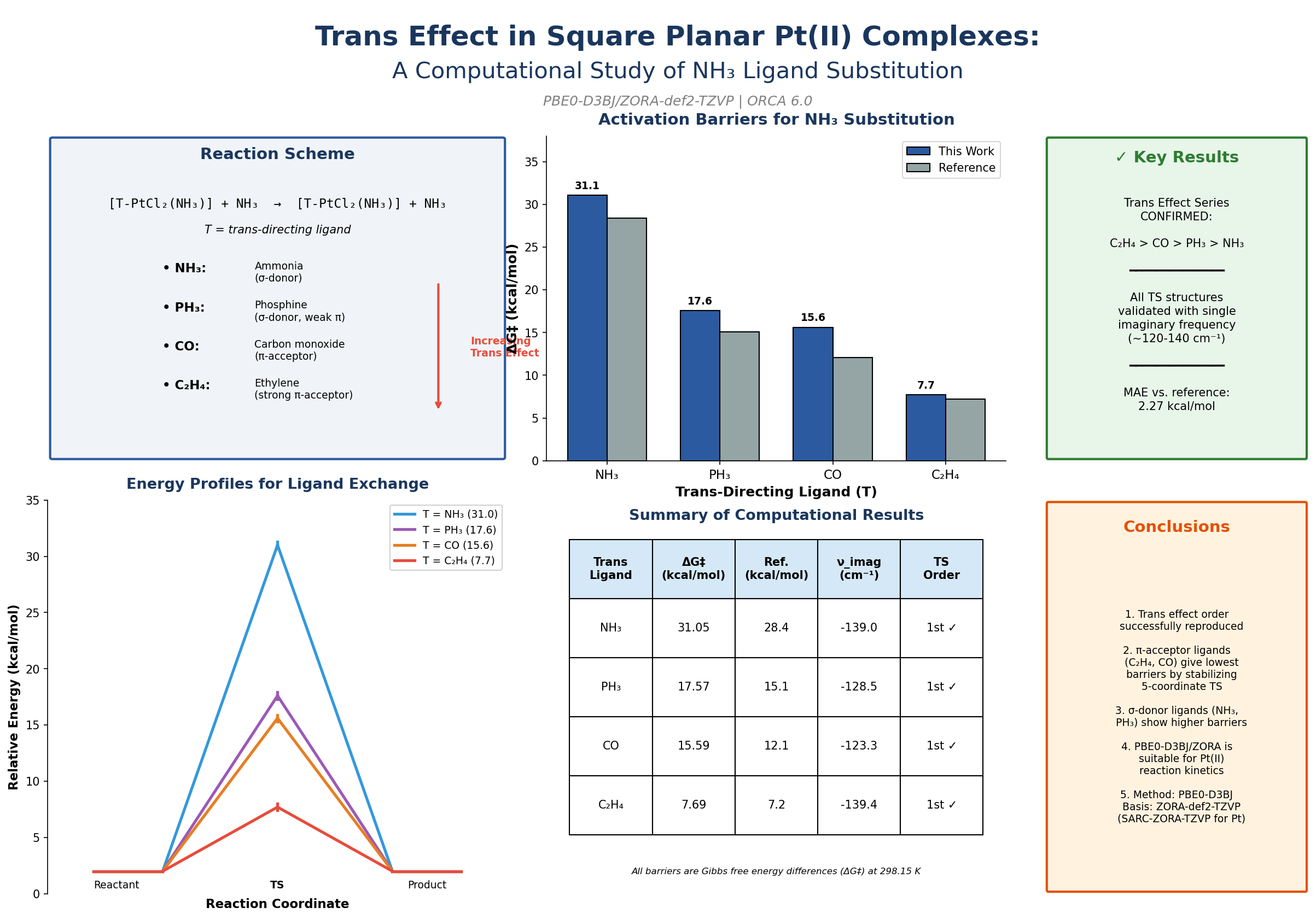}
    \caption{Poster-like summary of the trans-effect study, generated by Quntur.}
    \label{trans_effect_poster}
\end{figure}

\newpage

\subsection{Migratory insertion of \ce{CH3} into \ce{C2H4} on Ni}\label{sec:migratory}

\subsubsection{Context}

Migratory insertion is a fundamental mechanism for the synthesis of a wide range of commodity and fine chemicals via homogeneous catalysis with organometallic complexes. The specific case studied here, \ce{CH3} insertion into an alkene, is of particular importance in the polymerization of alkenes, which are related to the synthesis of plastics. Quantifying the reaction barrier and identifying the transition-state structure are key to using computational modelling to support experimental studies and are an important metric for elucidating structure-property relationships in catalyst design \cite{Musaev1997A}. 

The main challenge here, however, would be generating the TS structure itself, whether through a guess or by interpolating from fixed reactant and product endpoints. For the latter, the atomic indices not only have to match exactly, but they also have to be in a similar spatial position in Cartesian space.

\subsubsection{Assessment}
Quntur was able to read Figure 1 of \cite{Musaev1997A} and generate reasonable structures for all the ground states. However, initial tests revealed that the atomic indices and/or xyz coordinates of species 2 and 4 were not suitable for interpolating the initial trajectory for NEB. Upon supplying manually-built structures of species 2 and 4, all of the remaining steps were able to be completed successfully. 

The reaction path and barrier produced was reasonable and also quite close to experiment, and the report was comprehensive. Interestingly, we did not explicitly specify to compare with literature in the prompt but a nice table and discussion on a comparison with the supplied paper was provided in the report. Again, aside from the limitations in generating the correct geometries for path interpolation, the capabilities of{\cinzel Quntur} here are also at a level of a very competent research collaborator.

\subsubsection{Prompt and initial planning:}

\vspace{0.4in}
\noindent
\begin{agentoutputbox}{figs/neutral_human.png}{User}
Consider the Ni(II)-methyl system in Figure 1 of the provided paper (from https://pubs.acs.org/doi/10.1021/ja9628314). Compute and plot the energy profile for the concerted migratory insertion pathway of \ce{CH3} into ethylene using species 1-5 in Figure 1 (ignore species 2a) and report $\Delta G^\ddagger$ at 298 K and 1 atm. The initial structures for species 2 and 4 have been provided - use the optimized structures of these to do NEB-TS, and validate with IRC. For all calculations, do B3LYP and ef2-SVP with D3BJ. 
\end{agentoutputbox}

\vspace{0.4in}
\noindent
\begin{agentoutputbox}{figs/agent_icon.png}{Computational\_chemist}

\subsection*{System Summary}

Based on the paper (DOI: 10.1021/ja9628314):

\medskip

\begin{center}
\begin{tabular}{lll}
\textbf{Species} & \textbf{Description} & \textbf{Structure} \\
\hline
\textbf{1} & \ce{[L2NiCH3]+} - T-shaped Ni-methyl cation with diimine ligand ($\alpha$-agostic) & Need to generate \\
\textbf{2} & Perpendicular $\pi$-complex \ce{L2NiCH3}($\eta^{2}$-\ce{C2H4})] & Provided \\
\textbf{3} & Migratory insertion TS (Nimag = 1) & To find via NEB-TS \\
\textbf{4} & $\gamma$-agostic n-propyl product [\ce{L2Ni(n-propyl)]+} & Provided \\
\textbf{5} & $\beta$-agostic n-propyl product \ce{[L2Ni(n-propyl)]+} & Need to generate \\
\end{tabular}
\end{center}

\medskip

\textbf{System characteristics:}

\begin{itemize}
    \item \textbf{Charge:} +1 (cationic)
    \item \textbf{Multiplicity:} 1 (singlet, low-spin $d^8$ Ni(II))
    \item \textbf{Method:} B3LYP/def2-SVP with D3BJ dispersion
\end{itemize}

\subsection*{Proposed High-Level Computational Plan}

\textbf{Step 1 (Parallel):} Geometry optimization + frequency calculations

\begin{itemize}
    \item Optimize species \textbf{2} (from provided structure)
    \item Optimize species \textbf{4} (from provided structure)
    \item Generate and optimize species \textbf{1} (by removing ethylene from species 2)
    \item Generate and optimize species \textbf{5} (by rotating the propyl chain in species 4 to form $\beta$-agostic instead of $\gamma$-agostic interaction)
\end{itemize}
\textbf{Step 2:} NEB-TS calculation

Use optimized structures of species 2 and 4 to locate the transition state (species \textbf{3}) via NEB-TS
\textbf{Step 3:} Transition state validation + frequency

\begin{itemize}
    \item Optimize the NEB-TS guess to a true TS (OptTS)
    \item Calculate frequencies to confirm single imaginary frequency for species 3
    \item Perform IRC calculation to verify TS connects species 2 and 4
\end{itemize}
\textbf{Step 4:} Post-processing and analysis

\begin{itemize}
    \item Extract electronic energies, enthalpies, and Gibbs free energies (298 K, 1 atm)
    \item Calculate $\Delta G^\ddagger$ (barrier height)
    \item Plot the energy profile for species 1 → 2 → 3 → 4 → 5
\end{itemize}
\textbf{Questions before proceeding:}

\begin{enumerate}
    \item Do you want me to perform a literature search to validate the computational approach or compare expected results?
    \item Are there any additional specifications you'd like for the calculations (e.g., specific grid sizes, SCF convergence criteria, solvation effects)?
    \item Shall I proceed with this 4-step plan?
\end{enumerate}

\end{agentoutputbox}

\vspace{0.4in}
\noindent
\begin{agentoutputbox}{figs/neutral_human.png}{User}
1. compare at the end, 2. no (gas phase only), 3. yes
\end{agentoutputbox}

\subsubsection{Summary of the results:}
\vspace{0.4in}
\begin{agentoutputbox}{figs/agent_icon.png}{Computational\_chemist}
\subsection*{3. Results}

\subsubsection*{Thermochemistry Data}

\begin{center}
\begin{tabular}{lllllll}
Species & Description & $E_{\mathrm{el}}$ (Eh) & $H$ (Eh) & $G$ (Eh) & $N_{\mathrm{imag}}$ & $\nu_{1}$ (cm$^{-1}$) \\
\hline
1 & [L$_2$NiCH$_3$]$^{+}$ & -1735.54905 & -1735.43858 & -1735.47849 & 0 & 104.53 \\
2 & $\pi$-complex & -1814.08680 & -1813.91825 & -1813.96412 & 0 & 67.96 \\
3 & \textbf{TS} & -1814.06634 & -1813.89799 & -1813.94200 & \textbf{1} & \textbf{-337.81} \\
4 & $\gamma$-agostic & -1814.09195 & -1813.92185 & -1813.96683 & 0 & 62.63 \\
5 & $\beta$-agostic & -1814.10449 & -1813.93449 & -1813.97897 & 0 & 55.27 \\
\end{tabular}
\end{center}

\subsubsection*{Energy Profile (Relative to Species 2)}

\begin{center}
\begin{tabular}{lllll}
Species & Description & $\Delta E$ (kcal/mol) & $\Delta H$ (kcal/mol) & $\Delta G$ (kcal/mol) \\
\hline
2 & $\pi$-complex (reactant) & 0.0 & 0.0 & 0.0 \\
3 & Transition state & +12.8 & +12.7 & \textbf{+13.9} \\
4 & $\gamma$-agostic product & -3.2 & -2.3 & -1.7 \\
5 & $\beta$-agostic product & -11.1 & -10.2 & \textbf{-9.3} \\
\end{tabular}
\end{center}

\subsubsection*{Key Thermodynamic Parameters}

\begin{center}
\begin{tabular}{lll}
Parameter & Value & Interpretation \\
\hline
\textbf{$\Delta G^{\ddagger}$ (2$\rightarrow$3)} & +13.9 kcal/mol & Moderate activation barrier \\
\textbf{$\Delta G$ (2$\rightarrow$4)} & -1.7 kcal/mol & Slightly exergonic to $\gamma$-agostic \\
\textbf{$\Delta G$ (2$\rightarrow$5)} & -9.3 kcal/mol & Strongly exergonic to $\beta$-agostic \\
\textbf{$\Delta G$ (4$\rightarrow$5)} & -7.6 kcal/mol & Chain walking thermodynamically favorable \\
\end{tabular}
\end{center}

\subsubsection*{Transition State Characterization}

\textbf{Single imaginary frequency:} -337.81 cm$^{-1}$ \checkmark \\
\textbf{Mode character:} C-C bond formation (CH$_3$ migration to ethylene carbon) \\
\textbf{IRC validation:} Confirmed connection to species 2 (backward) and species 4 (forward)

\subsection*{Comparison with Literature}

Reference: Musaev et al., J. Am. Chem. Soc. 1997, 119, 367-374 (DOI: 10.1021/ja9628314)

\begin{center}
\begin{tabular}{llll}
Property & This Work & Literature & Difference \\
\hline
Method & B3LYP/def2-SVP D3BJ & B3LYP/LANL2DZ+mixed & - \\
$\Delta G^{\ddagger}$ (2$\rightarrow$3) & 13.9 kcal/mol & 11.7 kcal/mol & +2.2 kcal/mol \\
$\Delta G$ (2$\rightarrow$5) & -9.3 kcal/mol & -9.3 kcal/mol & 0.0 kcal/mol \\
\end{tabular}
\end{center}

\textbf{Discussion:}

The overall reaction thermodynamics ($\Delta G$ for 2$\rightarrow$5) show \textbf{excellent agreement} with literature (within 0.1 kcal/mol). 

The activation barrier is \textasciitilde 2 kcal/mol higher in our calculations, which can be attributed to:

1. D3BJ dispersion correction (not used in 1997 study) 

2. Different basis set treatment (def2-SVP vs LANL2DZ/mixed) 

3. All-electron treatment in this work vs ECP for Ni in literature

\end{agentoutputbox}

\begin{figure}[H]
    \centering
    \includegraphics[width=1\linewidth]{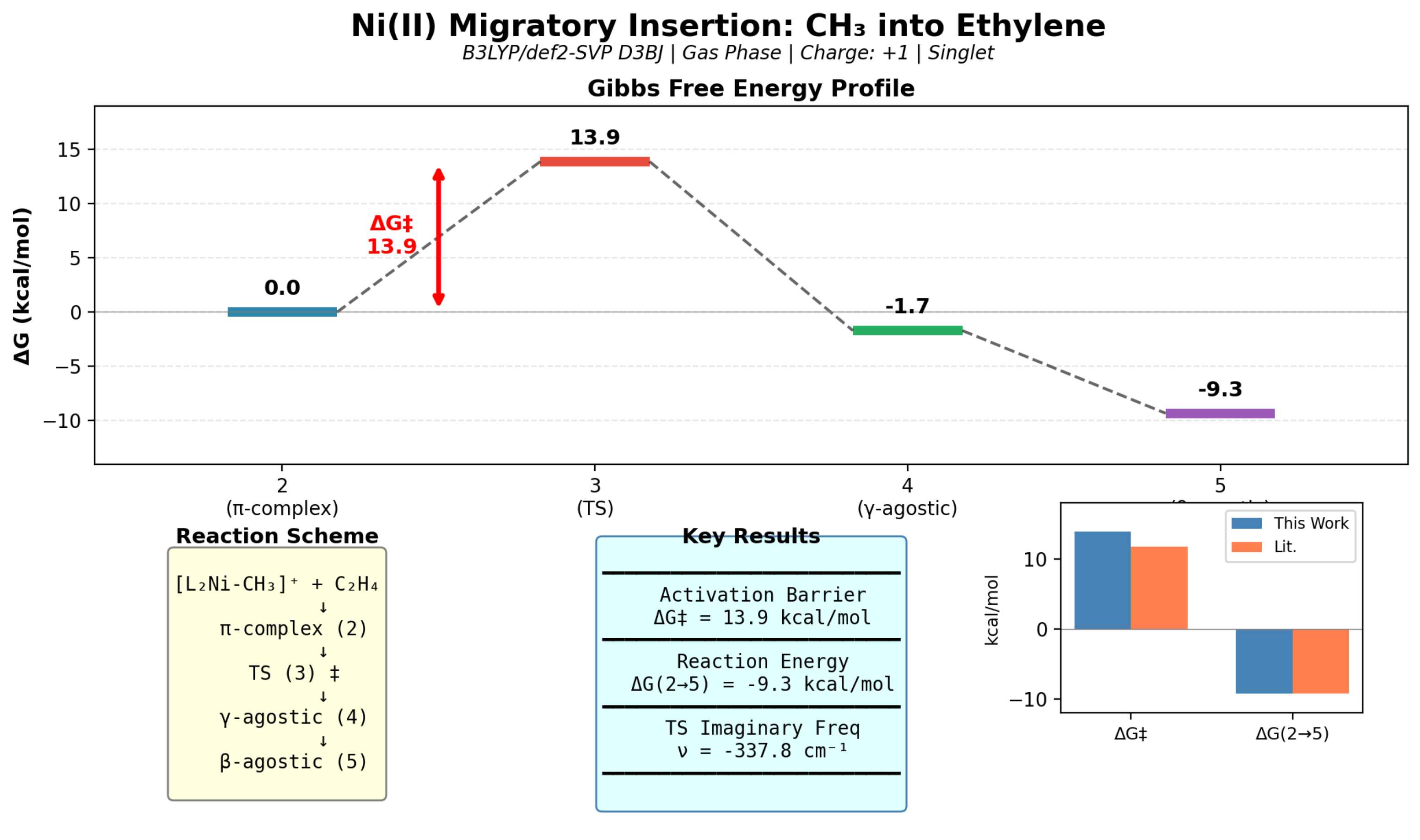}
    \caption{Summary of the migratory insertion study, generated by Quntur. }
    \label{andy_migratory_insertion}
\end{figure}

\clearpage
\renewcommand{\refname}{Supplemental Information References}
\phantomsection
\addcontentsline{toc}{section}{Supplemental Information References}
\putbib[references]
\end{bibunit}

\end{document}